\documentclass[structabstract,longauth]{aa}

\usepackage{graphicx}
\usepackage{lscape}
\usepackage{amsmath}
\usepackage{txfonts}
\usepackage{url}
\usepackage[usenames,dvipsnames,svgnames,table]{xcolor}
\usepackage{ulem}
\usepackage{tabularx}
\usepackage{longtable}
\usepackage{multirow}
\usepackage{subcaption} 

\usepackage{array}    
\usepackage{booktabs} 

\usepackage{hyperref}

\usepackage{xcolor}

\newcommand\logten{\ensuremath{\log_{10}}}

\begin{document} 

\title{\textit{\textbf{Gaia}} Data Release 1}
\subtitle{The variability processing \& analysis\\ and its application to the south ecliptic pole region}

\author{
L.~Eyer\inst{\ref{inst1}},
N.~Mowlavi\inst{\ref{inst1}},
D.W.~Evans\inst{\ref{inst2}},
K.~Nienartowicz\inst{\ref{inst3}},
D.~Ord\'o\~nez\inst{\ref{inst4}},
B.~Holl\inst{\ref{inst4}},
I.~Lecoeur-Taibi\inst{\ref{inst4}},
M.~Riello\inst{\ref{inst2}},
G.~Clementini\inst{\ref{inst22}},
J.~Cuypers\inst{\ref{inst5}},
J.~De~Ridder\inst{\ref{inst6}},
A.C.~Lanzafame\inst{\ref{inst7},\ref{inst8}},
L.M.~Sarro\inst{\ref{inst9}},
J.~Charnas\inst{\ref{inst4}},
L.P.~Guy\inst{\ref{inst4}},
G.~Jevardat~de~Fombelle\inst{\ref{inst3}},
L.~Rimoldini\inst{\ref{inst4}},
M.~S\"{u}veges\inst{\ref{inst4}},
F.~Mignard\inst{\ref{inst14}},
G.~Busso\inst{\ref{inst2}},
F.~De Angeli\inst{\ref{inst2}},
F.~van Leeuwen\inst{\ref{inst2}},
P.~Dubath\inst{\ref{inst4}},
M.~Beck\inst{\ref{inst1}},
%
J.J.~Aguado\inst{\ref{inst9}},
J.~Debosscher\inst{\ref{inst6}},
E.~Distefano\inst{\ref{inst8}},
J.~Fuchs\inst{\ref{inst15}},
P.~Koubsky\inst{\ref{inst15}},
T.~Lebzelter\inst{\ref{inst21}},
S.~Leccia\inst{\ref{inst12}},
M.~Lopez\inst{\ref{inst10}}, 
A.~Moitinho\inst{\ref{inst18}},
S.~Regibo\inst{\ref{inst6}},
V.~Ripepi\inst{\ref{inst12}},
M.~Roelens\inst{\ref{inst1}},
L.~Szabados\inst{\ref{inst11}},
B.~Tingley\inst{\ref{inst23}},
V.~Votruba\inst{\ref{inst15}},
S.~Zucker\inst{\ref{inst19}},
%
C.~Aerts\inst{\ref{inst6}},
F.~Barblan\inst{\ref{inst1}},
S.~Blanco-Cuaresma\inst{\ref{inst1}},
M.~Grenon\inst{\ref{inst1}},
A.~Jan\inst{\ref{inst4}},
D.~Lorenz\inst{\ref{inst21}},
B.~Miranda\inst{\ref{inst18}},
S.~Morgenthaler\inst{\ref{inst20}},
C.~Ordenovic\inst{\ref{inst14}},
L.~Palaversa\inst{\ref{inst1}},
A.~Prsa\inst{\ref{inst16}},
M.I.~Ruiz-Fuertes\inst{\ref{inst4}},
%
R.I.~Anderson\inst{\ref{inst13}},
H.E.~Delgado\inst{\ref{inst9}},
Y.~Dzigan\inst{\ref{inst19}},
R.~Hudec\inst{\ref{inst15}},
A.~Jonckheere\inst{\ref{inst5}},
P.~Klagyivik\inst{\ref{inst11}},
A.~Kutka\inst{\ref{inst15}},
M.~Moniez\inst{\ref{inst24}},
J.-M.~Nicoletti\inst{\ref{inst20}},
P.~Park\inst{\ref{inst1}},
E.~Van~Hemelryck\inst{\ref{inst5}},
M.~Varadi\inst{\ref{inst11}},
%
A.~Kochoska\inst{\ref{inst16}},
A.F.~Lanza\inst{\ref{inst8}},
M.~Marconi\inst{\ref{inst12}},
G.~Marschalko\inst{\ref{inst11}},
S.~Messina\inst{\ref{inst8}},
I.~Musella\inst{\ref{inst12}},
I.~Pagano\inst{\ref{inst8}},
G.~Sadowski\inst{\ref{inst17}},
M.~Schultheis\inst{\ref{inst14}}
%
%
}
\authorrunning{Eyer et al.}
\institute{Department of Astronomy, University of Geneva, Ch. des Maillettes 51, CH-1290 Versoix, Switzerland\label{inst1}\\
\email{Laurent.Eyer@unige.ch}
\and
Institute of Astronomy, University of Cambridge, Madingley Road, Cambridge CB3 0HA, UK\label{inst2}
\and 
SixSq, Rue du Bois-du-Lan 8, CH-1217 Geneva, Switzerland\label{inst3}
\and 
Department of Astronomy, University of Geneva, Ch. d'Ecogia 16, CH-1290 Versoix, Switzerland\label{inst4}
\and
Royal Observatory of Belgium, Ringlaan 3, B-1180 Brussels, Belgium\label{inst5}
\and
Institute of Astronomy, KU Leuven, Celestijnenlaan 200D, 3001 Leuven, Belgium\label{inst6}
\and
Universit\`a di Catania, Dipartimento di Fisica e Astronomia, Sezione Astrofisica, Via S. Sofia 78, I-95123 Catania, Italy\label{inst7} 
\and
INAF-Osservatorio Astrofisico di Catania, Via S. Sofia 78, I-95123 Catania, Italy\label{inst8}
\and
Dpto. Inteligencia Artificial, UNED, c/ Juan del Rosal 16, 28040 Madrid, Spain\label{inst9} 
\and
Departamento de Astrof\'isica, Centro de Astrobiolog\'ia (INTA-CSIC), PO Box 78, E-28691 Villanueva de la Ca\~nada, Spain\label{inst10}
\and
Konkoly Observatory, Research Centre for Astronomy \& Earth
   Sciences, Hungarian Academy of Sciences, H-1121 Budapest, Konkoly Thege ut 15-17, Hungary\label{inst11}
\and
INAF-Osservatorio Astronomico di Capodimonte, Via Moiariello 16, 80131, Napoli, Italy\label{inst12}
\and
Department of Physics and Astronomy, Johns Hopkins University, 3701 San Martin Dr, Baltimore, MD 21218, USA\label{inst13}
\and
Laboratoire Lagrange, Observatoire de la Cote d'Azur, Boulevard de l'Observatoire, CS 34229, F-06304 Nice Cedex 04, France\label{inst14}
\and
Academy of Sciences of the Czech Republic, Fricova 298, 25165 Ondrejov, Czech Republic\label{inst15}
\and
Villanova University, 800 W Lancaster Ave 19085 Villanova,  United States\label{inst16}
\and
Institut d'Astronomie et d'Astrophysique (IAA), Université Libre de Bruxelles (ULB), CP 226, Boulevard du Triomphe, B-1050 Brussels, Belgium\label{inst17}
\and
FCUL, Campo Grande, Edif. C8, 1749-016 Lisboa, Portugal\label{inst18}
\and
Tel Aviv University, Tel Aviv 69978, Israel\label{inst19}
\and
EPFL SB MATHAA STAP, MA B1 473 (Bâtiment MA), Station 8, CH-1015 Lausanne, Switzerland\label{inst20}
\and
University of Vienna, Department of Astrophysics, Tuerkenschanz\-strasse 17, A1180 Vienna, Austria\label{inst21}
\and
INAF - Osservatorio Astronomico di Bologna, Via Ranzani n. 1, 40127 Bologna, Italy\label{inst22}
\and
Stellar Astrophysics Centre, Aarhus University, Department of Physics and Astronomy, 120 Ny Munkegade, Building 1520, DK-8000 Aarhus C, Denmark\label{inst23}
\and
Laboratoire de l'Accélérateur Linéaire Université Paris-Sud, Bâtiment 200,
91898 Orsay, France\label{inst24}}

\date{Received ; accepted} 

\abstract
   {The ESA {\it Gaia} mission provides a unique time-domain survey for more than one billion sources brighter than $G$=20.7~mag.
   {\it Gaia} offers the unprecedented opportunity to study variability phenomena in the Universe thanks to multi-epoch $G$-magnitude photometry in addition to astrometry, blue and red spectro-photometry, and spectroscopy (the latter for sources brighter than magnitude $G \approx$ 17).
   Within the {\it Gaia} Data Processing and Analysis Consortium (DPAC), Coordination Unit 7 (CU7) has the responsibility to detect variable objects, classify them, derive characteristic parameters for specific variability classes, and provide global descriptions of variable phenomena.
   }
%
   {In this article, we describe the variability processing and analysis that we plan to apply to the successive data releases, and we present its application to the $G$-band photometry results of the first 14 months of \textit{Gaia} operations that comprises 28~days of Ecliptic Pole Scanning Law (EPSL) and 13~months of Nominal Scanning Law (NSL).
   We also present the general properties of the \textit{Gaia} sampling law.
   The data processing and analysis in this first \textit{Gaia} Data Release is applied to Cepheids and RR~Lyrae variables, restricted to sources around the South Ecliptic Pole (SEP).
   }
%
   {
   The methods used in the pipeline are taken from classical Statistics, Data Mining and Time Series analysis. 
   We also developed original procedures to deal with the specific properties of the {\it Gaia} data, as well as additional software tools needed to handle and visualize the data.
   The results are validated by comparing them with data available in the literature that cover the relevant regions of the sky.
   }
%
   {
   Out of the 694 million, all-sky, sources that have calibrated $G$-band photometry in this first stage of the mission, about 2.3~million sources that have at least 20 observations are located within 38 degrees from the SEP.
   We detect about 14\% of them as variable candidates, among which the automated classification identified 9347 Cepheid and RR~Lyrae candidates using a combination of three classifiers.
   Additional visual inspections and selection criteria led to the publication of 3194 Cepheid and RR~Lyrae stars, described in \citet{DPACP-13}.
   Under the restrictive conditions for DR1, the completenesses of Cepheids and RR~Lyrae stars are estimated at 67\% and 58\%, respectively, numbers that will significantly increase with subsequent Gaia data releases.
   Their properties, together with time series and associated statistics, are available in the \textit{Gaia} DR1 catalogue.
   }
%
   {
   Data processing within the {\it Gaia} DPAC is essentially iterative, the quality of the data and the results being improved at each iteration.
   The results presented in this article show a glimpse of the exceptional harvest that is to be expected from the \textit{Gaia} mission for variability phenomena. 
   }


\maketitle

\section{Introduction
\label{sec:introduction}}
The Universe contains many celestial objects whose observed quantities are varying on a human time scale \citep[see][]{EyerMowlavi2008}.
Therefore multi-epoch surveys have been a productive line of research, revealing immensely interesting physical phenomena and being used to understand the Universe as a whole as well as its constituents.
The number of photometric multi-epoch surveys has been rising quickly during these last two decades, as they benefit from technological developments (CCDs, computer performances).
They received a boost from scientific ``hot'' topics such as microlensing (e.g. OGLE, \citealt{Udalski1992OGLE}; EROS, \citealt{EROS1999}; MACHO, \citealt{Alcock1993MACHO}) exo-planetary transits (e.g. SuperWasp, \citealt{Pollacco2006WASP}; HAT, \citealt{Bakos2004HAT,Bakos2013HAT}; CoRoT, \citealt{Baglin2006COROT}, \citealt{Barge2008COROT}; Kepler and K2, \citealt{Borucki2010Kepler}), transient searches (Catalina, \citealt{Drake2009CRTS,Djorgovski2011CRTS}; PTF, \citealt{Law2009PTF}), and Near Earth Objects (LINEAR \citealt{2000Icar..148...21S}; Pan-STARRS \citealt{2002SPIE.4836..154K}).
However, among the many surveys, the ESA {\it Gaia} spacecraft is performing a truly unique and exceptional time-domain survey \cite[see for example][]{Eyeretal2012, Eyeretal2013, Eyeretal2015}.
There are several reasons for this: (1)~{\it Gaia} will achieve an exquisite astrometric accuracy, (2)~{\it Gaia} observes more than one billion objects from about 2 to 20.7 mag, (3)~{\it Gaia} repeatedly observes with one set of instruments the entire sky, (4)~{\it Gaia} delivers near-simultaneous white light $G$-magnitude photometry, blue and red spectro-photometry, and spectroscopy (the latter for sources brighter than magnitude $\sim$ 17 in $G$-band), see \citet{DPACP-1}.

Thanks to the previous experience with the {\it Hipparcos} mission \citep{perryman1997Hipparcos}, software development to analyse large data sets has been recognised as a major challenge of the {\it Gaia} mission.
Consequently, the {\it Gaia} Data Processing and Analysis Consortium (DPAC) was formed, with subdivisions into Coordination Units (CUs) and Data Processing Centres (DPCs), see \citet{Mignardetal2008}.
One of the identified tasks of the Consortium is dedicated to variable celestial objects.
The task has been assigned to the Variability Processing and Analysis Coordination Unit 7 (CU7), coordinated at the University of Geneva, with the responsibility to detect variable objects, classify them, derive characteristic parameters for specific variability classes, and give global descriptions of variable phenomena.
CU7 is composed of 80 members, 9 of whom are located in Geneva.
The DPC associated to CU7 is also located in Geneva.
Participation of an industrial partner in Switzerland, SixSQ, is also acknowledged.

CU7 interacts with other CUs within DPAC, each having its specific responsibilities.
We mention CU4 responsible for processing object sources, and in particular binary stars, CU5 which calibrates the photometric data and is our major source of information, CU6 which processes spectra, deriving in particular radial velocities, and CU8 which extracts astrophysical parameters from available \textit{Gaia} data.
The data processing and analysis results are made available by the ESA and \textit{Gaia} DPAC through \textit{Gaia} Data Releases. 

In parallel to those variability processing and analysis tasks performed by CU7 on well-calibrated \textit{Gaia} data, a team led by Cambridge analyses the \textit{Gaia} data in near-real time to detect transient phenomena that require quick follow-up by the scientific community in order not to lose science.
On the detection of such phenomena, the Cambridge team issues so-called Science Alerts to inform the community.
We refer to \citet{HodgkinWyrzykowskiBlagorodnova_etal13} and \citet{Wyrzykowski16} for additional details.

For the first \textit{Gaia} Data Release \citep{Brown16}, the photometric data calibrated by CU5 \citep{DPACP-12} was used to analyse time series in a restricted region of the sky (SEP region).
The SEP region analysed covers part of the Large Magellanic Cloud that has been monitored by various ground-based surveys, such as OGLE and EROS, and which are used to validate our variability analysis results.
Thanks to this EPSL phase, we reach up to almost 200 measurements per object in less than a month, which is representative of what will be achieved by the NSL in other regions of the sky at the end of the five-year mission.
Although the EPSL has a denser and more regular sampling and in general more prone to aliasing than the NSL, these data provide a small showcase for the whole mission.

This article focusses on the automated processing and analysis performed in CU7, and is organized as follows.
\begin{itemize}
  \item Sect.~\ref{sec:overviewVarProcessing} presents a general overview of the variability processing and analysis tasks, i.e. the methods that we use or plan to use to tackle the global analysis of the {\it Gaia} data.
Our goal is to systematically detect variable objects, classify them, derive characteristic parameters for specific variability classes, and give global descriptions of variable phenomena.
This approach was previously tested on other photometric, radial velocity, and astrometric surveys ({\it Kepler}; OGLE; EROS; {\it Hipparcos}; CORAVEL, \citealt{Baranne1979CORAVEL}; SDSS, \citealt{York2000SDSS}) and simulated {\it Gaia} data.
  \item Sect.~\ref{sec:theData} describes the two different data sets which were processed for the \textit{Gaia} DR1, including the definition of the sky region for this release, the transformations of times, the conversion of measured fluxes in magnitudes, and the constraints applied to time series.
  \item Sect.~\ref{sec:C0sepResults} presents the results of the processing of preliminary \textit{Gaia} photometry including the first 3 months of operations: from crossmatch with literature to statistics describing and validating the data, followed by the detection of variability, characterisation of variable objects, and their classification in variability types, focussing on Cepheids and RR~Lyrae stars.
  \item Sect.~\ref{sec:C1SepResults} describes the results of the pipeline applied to the \textit{Gaia} data 
from the first 14 months of operations, limited to a subset of Cepheids and RR~Lyrae stars identified in Sect.~\ref{sec:C0sepResults}.
An overview of the published tables and their contents is included. 
  \item Sect.~\ref{sec:completeness} presents an assessment of the completeness of the identified variable sources with respect to other surveys in the same area.
  \item Sect.~\ref{sec:gvs-data} compares the distributions of parameters of the published \textit{Gaia} variable stars with the literature.
\end{itemize}

In addition to Sections~\ref{sec:overviewVarProcessing} to \ref{sec:theData} mentioned above, Sect.~\ref{sec:conclusions} concludes the paper, and four appendices are available: 
Appendix~\ref{sec:Timesampling} discusses various properties of the time sampling of the nominal five-year mission and highlights what makes it interesting for the detection and characterisation of variable objects,
Appendix~\ref{Appendix:classification} related to classification,
Appendix~\ref{Appendix:GVS} related to Global Variability Studies, and Appendix~\ref{Appendix:acronyms} summarizing the list of acronyms used in the paper.

\section{Overview of variability processing and analysis
\label{sec:overviewVarProcessing}}
\begin{figure}
\centering
\includegraphics[width=0.47\textwidth]{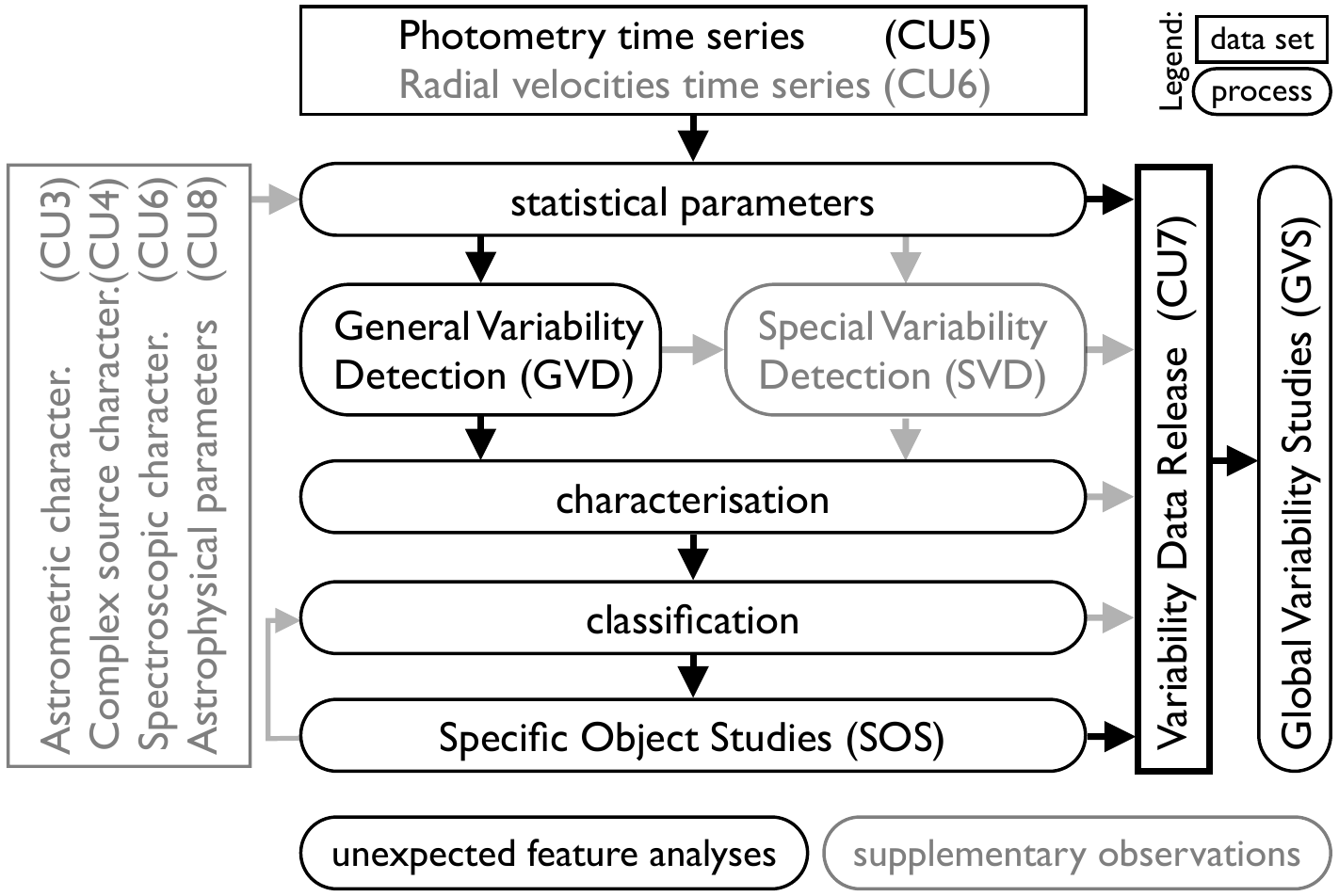}
\caption{
  General processing diagram for the CU7 variability processing and analysis.
  The connection with other CUs is described in Sect.~\ref{sec:overviewVarProcessing} and the various processing modules in Sect.~\ref{sec:dataprocessingModules}. Elements that have not been applied in this release are drawn in grey.}
\label{fig:generalProcessing}
\end{figure}

The general processing diagram adopted for CU7 is shown in Fig.~\ref{fig:generalProcessing}.
To effectively deal with the unprecedented amount of data available, we adopt an iterative process for variability processing.
We progressively reduce the size of the data set to process while increasing the complexity of processing and analysis of the subsets.
First we detect variability, then characterise this variability and define attributes, which are used to classify the objects.
For specific objects (i.e. variability types) we evaluate the general classification results and determine further parameters.
This step is also a validation of the classification, which is fed back to the same named module.
The global variability studies use all the available information to check and describe the variability phenomena.
There are furthermore two additional work-packages. (1)~Supplementary observations validate the results of our analysis using or performing new observations (not necessary up to now since existing public surveys were sufficient to validate the results of our analysis).
(2)~The unexpected feature analysis work-package is a special activity to analyse any unexpected behaviour, which may come either from physical phenomena, from the spacecraft or the data reduction.
The different processes mentioned in Fig.~\ref{fig:generalProcessing} will be explained in more details in Sect.~\ref{sec:dataprocessingModules}.

The input for CU7 and its associated Data Processing Centre in Geneva (DPCG) includes data received from other processing units: CU 3, 4, 5, 6 and 8.
In turn, these CUs depend on our results to validate their processing.
In addition, some of these CUs (CU4 and 8) rely on our data products, the variability catalogue, to perform their processing.
CU7 is thus an essential component of the {\it Gaia} mission: validating calibrations performed within DPAC, feeding other CUs with data for further analyses, 
and populating the {\it Gaia} catalogues with the results of our variability analysis.

\subsection{Data processing flow
\label{sec:DPCG_DataProcessingFlow}}
The data processing performed for a single Data Reduction Cycle (DRC) aims at producing all results for a single \textit{Gaia} data release.
Such processing typically spans over multiple months and requires the collaboration of many members that are geographically scattered.
The first \textit{Gaia} Data Release (\textit{Gaia} DR1) thus saw the first operational DRC (DRC1) for the combined DPCG and CU7 teams. 

\begin{figure}
\centering
\includegraphics[width=0.47\textwidth]{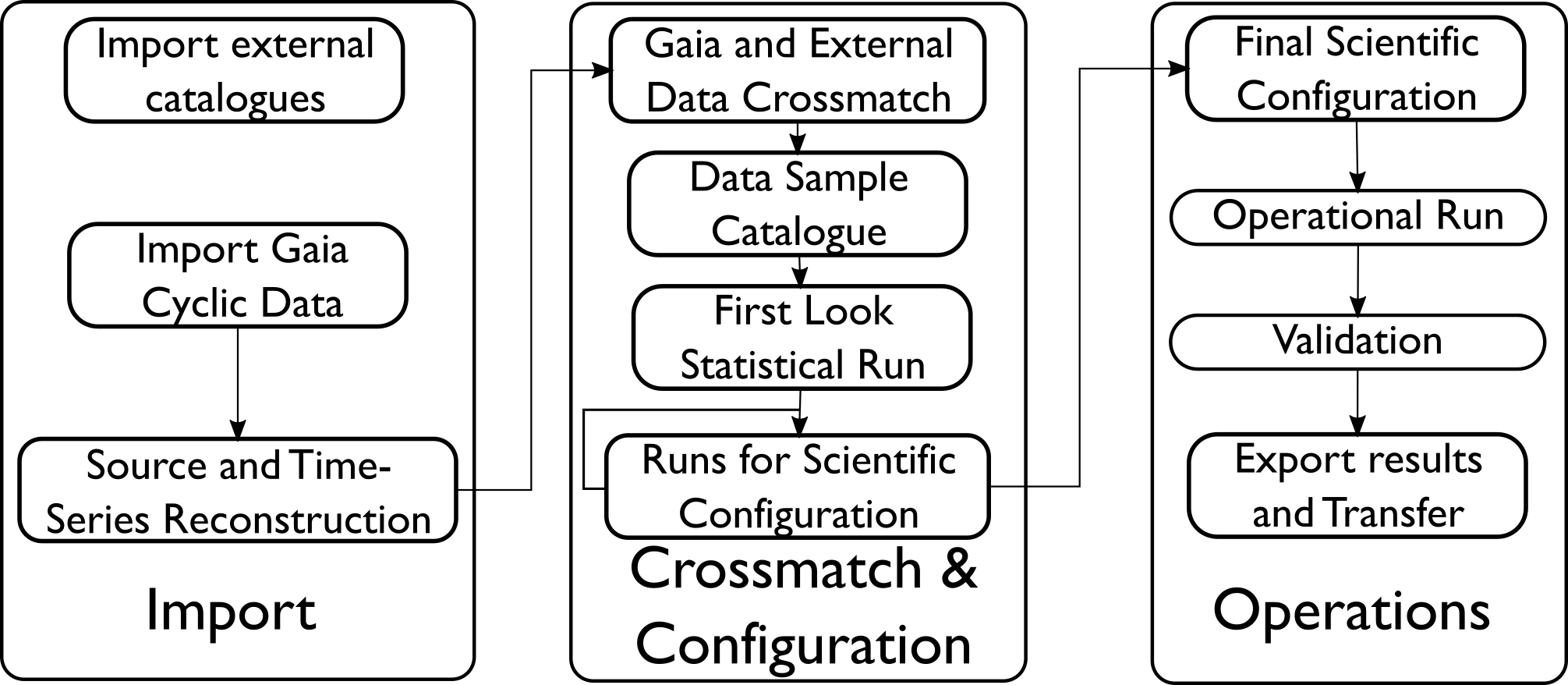}
\caption{High level activity plan for DPCG data reduction cycle (Import, Crossmatch \& Configuration, Operations). 
}
\label{dpcg_drc}
\end{figure}

The main activities of the data processing are in chronological order: data import in our database, crossmatch of sources with external catalogues (from literature), scientific configuration and testing of CU7 modules, final processing and export of the data to the other CUs. 
These activities are performed iteratively, since they imply additional verification and validation tasks aimed at detecting errors and improving the results.
The organization of these activities is shown in Fig.~\ref{dpcg_drc}. 

The DRC1 presents the opportunity to test our methodology on a moderately large data set in preparation for processing the amount and variety of data expected by the end of mission.
For DRC1 the iterations were all done over the whole dataset.
This was possible because of the small size of the dataset, which had a short processing time (less than an hour).
Future \textit{Gaia} datasets will contain a data volume requiring increasingly long processing times (expected processing time for the second DRC is measured in weeks on the same infrastructure).

\subsubsection{System infrastructure}

The atomicity of source-by-source processing allows for a perfectly parallel usage of our computing cluster that is gradually expanded.
It is currently composed of 57 quad core nodes, totalling 456 simultaneous processes, each process having 4 GB of RAM.
The batch system execution is orchestrated via DPCG-developed middleware that allows stream-processing for real-time aggregation of results and their instant visualization.
The data storage is a PostgreSQL 9.5 \citep{pgsql} database with a number of extensions enabling Java, Gnu R, Javascript server-side integration, semi-structured data, advanced indexing techniques like spatial indexing or functional indexing as well as partitioning to be executed within the database engine.
To structure the data in an efficient manner, the incoming \textit{Gaia} data from the Main Data Base (MDB) at DPCE (Data Processing Centre ESAC) is stored in partitioned tables.
The database is hosted on a single node with 40 cores and 256 GB of RAM, with optimised directly attached storage volumes for this exercise.
We are developing a parallel database solution based on Postgres-XL \citep{pgxl} to achieve horizontal scalability in forthcoming Data Processing Cycles. 

\subsubsection{Data preparation: data import, reconstruction and filtering, external catalogues, external-crossmatch}

After receiving all sources and photometry from DPAC's internal MDB scattered over thousands of \textit{Gaia} files, the data is first imported in parallel into the DPCG database.
Using hybrid SQL-Java queries, the sources and time series are reconstructed from individual transits.
Reconstruction is the process of time-ordering of photometric transits from the input data into times-series assigned to sources, which are then grouped into catalogues.
Because of the high number of measurements and unknown distributions of the input data, this expensive Input-Output operation step is necessary as \textit{Gaia} data is not naturally ordered by sources or FoV transits that make up the time series. 

During reconstruction, the observation epoch tagged with the on-board time is first transformed into a Julian Epoch and then transferred to the Solar system barycentre
(see Sect.~\ref{sec:time_and_conversions}).
Time checks during reconstruction allow for detection and rejection of fundamentally invalid sources.
Specifically we check that the transits that are associated with a given source do not violate the \textit{Gaia} scanning law sampling in the data we receive.
An example would be a source with two consecutive transits with a time interval shorter than that predicted between the preceding and following fields of view of the spacecraft.

Simultaneously, external catalogues are imported, with their attributes (including classification types) and photometry if needed, to be used as basis for algorithm configurations and to perform comparative studies of results.
After the \textit{Gaia} catalogue creation, sources from external catalogues are crossmatched against the \textit{Gaia} data with the help of quad-tree spatial indexing using the Postgres q3c library \citep{2006ASPC..351..735K}, besides other metrics. 
This crossmatch uses the method described in Sect.~\ref{sect:crossmatch} and provides datasets for use in tests, validations and classifier training. 

\subsubsection{Initial configuration, time series operators, first look run}
 
The pipeline is highly configurable, with hundreds of parameters that must be balanced for performance constraints and scientific return, defined on a run basis.
Initial configuration parameters are derived from configurations tuned during the pre-DRC1 development phase based on synthetic data, other surveys and \textit{Gaia} Operational Rehearsals.
An important part of the configuration preparation is to decide what time series operators are needed by various scientific modules of the system and what their configuration should be. 
Typical time series operators perform flux to magnitude conversion, outlier removal, error cleaning on input time series and create derived (transformed and/or filtered) time series suitable to be digested by specific algorithms. 
Chaining of the time series operators creates a hierarchy of derived time series that is used as required by the scientific modules analyses without losing provenance of the input used to get results.

Once sources with photometry are reconstructed and cleaned, DPCG executes a First Look statistical analysis run to gain some understanding of the fundamentals of the DRC data, deriving global distributions of tens of parameters that we visualize using an interactive application with a drill-down capability to see individual time series.

The results from the Crossmatch catalogues runs allow for a scientific performance evaluation of the pipeline. 
The configuration parameters of every algorithm are set by CU7 members iteratively, until we reach satisfactory results on the dataset scale (i.e. shape of chosen distributions) and at source level (i.e. classification obtained, periods found) that allow us to mark the end of the Configuration phase.

\subsubsection{Operations phase, Validation and Export}

When a final algorithm configuration for the dataset is produced, DPCG enters the operational phase, followed by the validation and completeness checks.
Then results are exported to \textit{Gaia} MDB format, validated again and transferred to DPCE for integration into the MDB repository hosted there. 
For DRC1, DPCG exported to DPCE not only scientific results per source but also the reconstructed $G$-band time series in flux and magnitude.

\subsection{Data processing modules
\label{sec:dataprocessingModules}}
This section describes the various data processing modules in the variability pipeline that are used or will be used for producing the quantities to be found in the successive {\it Gaia} Data Releases.
These modules follow the processing steps outlined in Fig.~\ref{fig:generalProcessing}.

\subsubsection{Statistical parameters}

The first step in the processing chain is the computation of a number of basic descriptive, inferential and correlation statistics of all time series. 
These statistics provide a first general overview of the data and their distribution and are used to determine whether variability is present in 
a time series of {\it Gaia} observations. Descriptive statistics computed on the temporal evolution of the time series include:
the number of observations, time duration of the time series, mean observation time and min/max time difference between successive
observations. Parameters that characterise the light curve include measures of the minimum, maximum, range, mean, median, variance, skewness, kurtosis,  point-to-point scatter, inter-quartile range (IQR), median absolute deviation, 
and signal-to-noise ratio. 
Weighted and unweighted measures, biased and unbiased by sample size 
\citep[for central moments, see][]{2014A&C.....5....1R}, as well as robust estimates (when applicable) are computed and compared.  
Several inferential test statistics are computed on the time series including the Kolmogorov-Smirnov test
for equality of continuous distributions, 
\citep{Kolmogorov,Smirnov}, 
the Ljung-Box test 
for randomness, 
\citep{LjungBox}, the Abbe hypothesis test \citep{vonNeumann1941, vonNeumann1942},  the chi-squared and Stetson test statistics \citep{1996PASP..108..851S}.  Some of these measures are used in the classification of a time series as constant or variable. 
All these measures are available in the {\it Gaia} catalogue.  \label{DataProcessing:Statistics}

\subsubsection{Variability detection}






The separation between variable and non-variable objects can be described in terms of significant levels of light-curve variations compared to selected thresholds in a set of parameters.
We first identify the properties of sources whose brightness can be considered constant. 
Such `constant' {\it Gaia} objects define a reference with respect to which the variability of other objects is assessed. 
Different types of variability may require different methods and, in the most difficult cases, one has to resort to some non-standard specialised methods \citep[as for the detection of exoplanetary transits,][and trends in time series, \citealt{2006ASPC..349...15E}]{Dzigan_Zucker:2011}.


For such reasons, variability detection in the variability processing and analysis is organized in two broad classes of methods:
one, the general variability detection (GVD) class, scans the {\it Gaia} data using general standard variability detection algorithms;
the other, the special variability detection (SVD) class, searches the {\it Gaia} data for some specific type of variability not generally detectable using standard methods.


\paragraph{General Variability Detection.}

The GVD processing step aims to detect efficiently variability in {\it Gaia} data  while at the same time coping with challenges, such as the different number of measurements from one source to another, the dependence of the measurement error on the brightness, the presence of spurious outlying values, etc.
The GVD task can be best accomplished by adopting multiple criteria targeting different and complementary aspects of variability (e.g., see Sect.~\ref{sec:GVD_run}).
Such criteria can be used by hypothesis tests based on p-value thresholds \citep[e.g.,][]{laurentThesis,2006ASPC..349...15E}.
However, with such a method, it is important that the estimated uncertainties are reliable and that parameter distributions are unaffected by possible data artefacts.
Currently in the {\it Gaia} data, the quoted uncertainties on the fluxes do not represent correctly the true error yet \citep{DPACP-11} and the presence of spurious outliers is not negligible, rendering the approach difficult.
An alternative approach is to use the values of variability parameters and complementary information from other CUs
as attributes for a dedicated classifier to automatically combine the different criteria and detect variable objects.
The classifier method is chosen for processing the initial {\it Gaia} data because of its ability to adapt to attribute distributions as represented in real data and provide an optimal separation between constant and variable objects. 
Typically, a classifier is trained with a similar number of constant and variable objects, preferably drawn from the same region of the sky where variability detection is applied (for the present paper, near the South Ecliptic Pole).
The list of classification attributes is optimised according to the features that are recognisable in the data (e.g., the value of skewness is limited or even counterproductive if  spurious outliers are frequent in the data). 
The selection of sources and attributes for training the classifier and the results of variability predictions are described in Sect.~\ref{sec:GVD_run}.

\paragraph{Special Variability Detection.}

The classes targeted by SVD include short time-scale and small amplitude periodic variables as well as variability induced by planetary transits and solar-like (magnetic) activity.
SVD methods can be rather demanding in terms of computing resources, 
and their application to the whole {\it Gaia} dataset may not fit in the timeframe imposed by the cyclic {\it Gaia} analysis.
These methods are therefore applied to some subset of the {\it Gaia} dataset, preselected based on prior information on the sources.

The short time-scale variability method aims at the detection and characterisation of variability on time-scales from a few tens of a second up to 12 hours by exploiting {\it Gaia}'s peculiar sampling and its per-CCD photometry (see Sect.~\ref{sect:ccd}).
The per-CCD photometry gives access to the highest sampling rate with 9 consecutive measurements separated by 4.85~seconds. 
This work package includes detection and characterisation of general short time-scale variability, short-period signals (e.g. roAp, $\delta$~Scuti, $\beta$~Cephei, ZZ~Ceti, AM~CVn stars), and stochastically variable phenomena.
The algorithm is based on the variogram calculation with given lags \citep{1999A&AS..136..421E}, the analysis of linear photometric trends per-CCD on one transit passage, and Bartlett's method \citep{Bartlett:1948,Bartlett:1950}.

The search for small amplitude periodic variables is carried out using the periodogram.
The aim is to identify periodic light variations with very poor signal-to-noise ratio ($\sim$1) 
and amplitudes down to the milli-magnitude level. 

For the planetary transit detection two methods are adopted: the classical Box-fitting Least Squares method  \citep[BLS,][]{Kovacs_etal:2002}, and a method optimised to search for transits in data with long time baseline and sparse sampling like in the case of {\it Gaia} \citep{Tingley:2011}. 
These methods balance effectiveness and computational cost and one or both will be used at different stages of the mission.
The expected yield of the planetary transit detection is discussed in \cite{Dzigan_Zucker:2012} and is estimated from a few hundreds to a few thousands.
{\it Gaia} will provide $G_{\rm BP}$ and $G_{\rm RP}$ magnitudes that are nearly simultaneous with the $G$-band photometry. Although less precise than the {\it G}-band photometry, they can be used as independent confirmation of the transits.
Furthermore the radial velocities derived from {\it Gaia} will allow us to eliminate some false detections (like grazing binaries) which may mimic exoplanetary transits.
The results of the planetary transit detection are used in the Directed Follow-Up strategy \citep[DFU,][]{Dzigan_Zucker:2011,Dzigan_Zucker:2013}, which is the base for the {\it Gaia} Transiting Exoplanets (GATE) Initiative \citep{Zucker_etal:2015}.
Such an initiative, combined with the {\it Gaia} Science Alerts resources, aims at combining the results of the {\it Gaia} planetary transit detection with carefully scheduled ground-based photometric observations, using Bayesian considerations.

The main method for detecting solar-like activity is based on the colour-magnitude variability correlation expected for the rotational modulation due to spots and faculae in late-type stellar atmospheres.
This method was originally applied with UBV photometry by \cite{Messina_etal:2006} and, in the {\it Gaia} case, makes use of the FoV $G$ magnitude and $G_{\rm BP} - G_{\rm RP}$ colour variations.
Flare candidates are identified from outliers in the $G$ vs. $G_{\rm BP} - G_{\rm RP}$ variation correlations and  specific patterns (rise and/or decay) in the per-CCD photometry.
Solar-like variable candidates are subsequently analysed by Specific Object Studies packages to determine rotational periods \citep{distefano_etal:12} and flare parameters.

The results from SVD methods are not included in the first {\it Gaia} data release.
 \label{DataProcessing:Variability}

\subsubsection{Characterisation}
The Characterisation work package analyses the time series identified as variable by the preceding general and special variability detection modules.
For periodic objects, the variability behaviour is characterised using a classical Fourier decomposition to find the simplest possible description of the observed variability. 
Each source is scanned for periodicity and the most significant period is retained in order to fit a Fourier series model. 

The period search process comprises a large number of period search methods and run-time options. 
Well known methods for frequency analysis, such as those based on least squares harmonic analysis \citep{Deeming1975,Lomb1976,Scargle1982,2009A&A...496..577Z}, phase dispersion minimization \citep{Jurkevich1971,Stellingwerf1978}, string length methods \citep{LK1965,Dworetsky1983}, and a few others \citep[FAMOUS,][]{Mignard2005,Kuiper1960} are all available. 
A number of methods are compared in \citet{NicolettiThesis}.
For the analysis of the \textit{Gaia} DR1 data, the only method used was one based on a generalised least squares harmonic analysis \citep{2009A&A...496..577Z}. The option to de-trend the time series prior to period search is available but was not used in this study so far.

For {\it Gaia} DR1, we have searched for frequencies above $2/\Delta T$, where $\Delta T$ is the length of the time interval of the data series, with steps of $0.05/\Delta T$, with no gain in finer steps. The upper limit of the frequency range was chosen depending on the desired output: short period variables can be found if the value is high (e.g. 30~d$^{-1}$), but this could result in many unwanted aliases \citep{1999A&AS..135....1E} or parasite frequencies.
If one searches only for longer period variables, one can limit the search to smaller frequency values.

False alarm probability (FAP) can be used to indicate the probability that the highest maximum (or lowest minimum in some methods) resulting from a period search that occurs by chance and not by a true variable phenomenon.
While different approaches and recommendations for the analysis of {\it Gaia} data can be found in \cite{Suveges2015}, we have computed FAP values following \citet{Baluev2008} for GDR1; subsequent data releases may use different methods.

The general model of variability that we employ to fit a time series of {\it Gaia} observations is given by a Fourier decomposition of $N_f$ independent frequencies $f$ with up to $N_h$ harmonics each, and polynomial terms up to degree $N_p$ (for low-order trends):
\begin{equation}
y = \sum_{n=1}^{N_{f}} \sum_{k=1}^{N_{h}(n)} 
    \left[a_{n,k} \sin(2\pi k f_{n} t) +  b_{n,k} \cos(2\pi k f_{n} t)\right] +  \sum_{i=0}^{N_{p}} c_{i} t^{i},
\label{tsmodel_linear} 
\end{equation} 
where $t$ denotes the observation time(s) referred to a modelling reference epoch $t_{\rm ref}$.
The amplitudes in each term of Eq.~\ref{tsmodel_linear} are fit by a linear model, whose results are re-expressed in the following form 
\begin{equation}
y = \sum_{n=1}^{N_{f}} \sum_{k=1}^{N_{h}(n)} 
    A_{n,k} \cos(2\pi k f_{n} t + \psi_{n,k}) +  \sum_{i=0}^{N_{p}} c_{i} t^{i}, 
\label{tsmodel} 
\end{equation} 
with amplitudes $A_{n,k}=\left(a_{n,k}^2+b_{n,k}^2\right)^{1/2}$ for the $n$th frequency $f_n$ and the $k$th harmonic, and phases $\psi_{n,k}=\mbox{atan2}(-a_{n,k},b_{n,k})$ which are subsequently mapped to the range from 0 to $2\pi$ (adding $2\pi$ to negative phase values).
The maximum gap in phase $\Delta \phi_{\max}$ of the light curve folded by the frequency $f_n$ is used to prevent over-modelling by limiting the addition of higher harmonics so that $N_h(n) < \pi/ \Delta \phi_{\max}$.
 
The first frequency determined by period search is used as input to the time series modelling to perform a combined polynomial and harmonic fit according to Eq.~(\ref{tsmodel_linear}). 
For this analysis, no trend was modelled and the polynomial component of the model includes only the mean magnitude. 
In addition to the first harmonic, increasingly higher harmonics are tested for significance, up to a maximum harmonic specified in the configuration (typically 15), unless reduced by the maximum phase-gap criterion. 
The calculated models are compared using one or more of several criteria such as an F-test, AIC or BIC to select the best model \citep[see e.g.][]{Hastie2009}. 
This process for determining which harmonics of a detected oscillation frequency are significant requires $N$ hypothesis tests, where $N>1$. To ensure that the 
significance level applied pertains to the family of $N$ hypothesis tests combined, the Bonferroni multiple comparison test is applied. 
Once the final best model is determined, non-linear modelling of the data is performed using the Levenberg-Marquardt non-linear fitting algorithm \citep{Levenberg_1944,Marquardt_1963}.
The freely varying parameters of the model now include the frequencies as well as the amplitudes and phases. 


The final output, in addition to the period and FAP, includes the coefficients of the polynomial and the Fourier series, an estimate of the precision of the fit (residual error) and a measure of the fit quality of the best-fitting model.
 \label{DataProcessing:Characterization}

\subsubsection{Classification} \label{sec:classification}
\label{DataProcessing:Classification}
The classification module uses automated supervised statistical classifiers to assign to each variable star a set of probabilities that it is a member of one of the pre-defined variability types. 

These probabilities are subsequently used by the Specific Object Studies (SOS) module (see Sect.~\ref{DataProcessing:SOS}) for an in-depth analysis.
The classification is done using three different classifiers: Gaussian Mixtures (GMs), Bayesian Networks (BNs), and Random Forests (RFs). These machine learning techniques have been used in the past to classify variables stars, and we therefore refer to the literature for a more  detailed description of the algorithms, e.g.~\citet{2011MNRAS.414.2602D}, \citet{2011MNRAS.418...96B}, and \citet{2011ApJ...733...10R}.

\begin{figure}
\centering
\includegraphics[angle=0,width=0.40\textwidth]{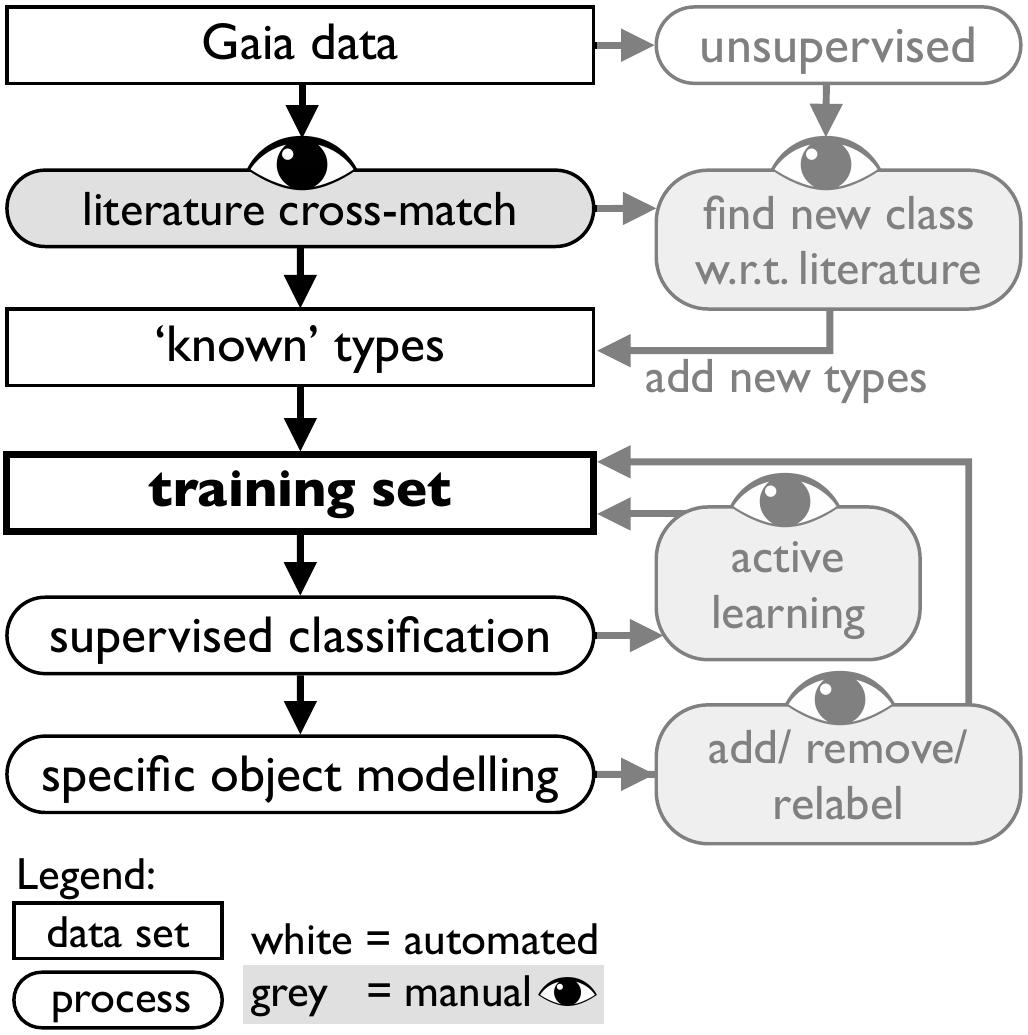}
\caption{The planned scheme to iteratively improve the training sets for the classifiers that classify 
all variable stars observed by {\it Gaia}. Elements drawn in grey on the right side have not been applied in this release, but are planned for future releases. See Section~\ref{sec:classification} for more details.
}
\label{fig:classificationTrainingsetImprovementPlan}
\end{figure}

Each of these classifiers requires a) a training set that shows bona fide examples of variable stars for each variability class, and b) a set of attributes that can be used to distinguish members of different variability classes.
The training sets will evolve from one data release to the next, as more bona fide {\it Gaia} examples of a variability class will be available throughout the mission.
The general iterative scheme we use to improve the training sets is show in Fig.~\ref{fig:classificationTrainingsetImprovementPlan}.
The initial training set is constructed by identifying those sources which are known to be variable in the literature and for which {\it Gaia} time series exist.
This training set is then used by supervised classifiers to classify all other variable {\it Gaia} sources.
The SOS module verifies the ones with a high classification probability, and the bona fide variables are added to the training set.
This way the training set will gradually grow in time.
Independently of the supervised classifiers, an unsupervised clustering algorithm will be developed to search for clusters of similar variables in the {\it Gaia} attribute space.
By comparing the detected clusters with the already known variability classes, this allows to detect new variability classes or new subdivisions of existing variability classes.
This information is then fed back to adapt the training set for the supervised classifiers. For the sake of brevity in the remainder of the article, we speak about ``classifiers'' when we refer to supervised classifiers, unless explicitly mentioned otherwise.

The three classifiers use different priors.
In a Bayesian framework, the posterior probability $P(C\ |\ A)$ that a variable with attribute set $A$ belongs to the variability class $C$ can be written as
\[
P(C\ |\ A) = \frac{P(A\ |\ C) \ P(C)}{P(A)},
\]
where $P(A\ |\ C)$ is the likelihood, $P(C)$ the prior probability of a star belonging to class $C$, and $P(A)$ the marginal likelihood (also referred to as model evidence).
Both the RF and BN algorithms (are designed to) use the relative fraction of variables in the training set as a prior $P(C)$ of each variability class.
For the GM classifier, we (can) use flat priors, as the relative fractions in the training set likely deviate from the true ones.

Each classifier assumes that any source to classify belongs to one of the variability classes present in the training set.
Since the membership probabilities must add up to one, this implies that high classification probabilities can be obtained even if the source belongs to a variability class not included in the training set, which in turn leads to an increased contamination.
The GM and BN classifiers mitigate this problem by filtering the input to the classifiers and removing outliers.
Outliers are defined as sources at large Mahalanobis distances \citep{Mahalanobis} from the centres of the Gaussian components that define the GM classifier.
That is, if $A$ is the attribute set of the source, $\boldsymbol{\mu}_{ij}$ is the mean vector of the $j$-th Gaussian component of variability class $i$, and $\mathbf{\Sigma}_{ij}$ is the corresponding covariance matrix, then the source is flagged as an outlier if
\[
\min_{ij}\sqrt{ \ (A - \boldsymbol{\mu}_{ij})^{T} \ \mathbf{\Sigma}_{ij}^{-1} \ (A - \boldsymbol{\mu}_{ij}) }\ > 3.8.
\]
The cut-off value of 3.8 corresponds to a $p$-value of 0.05, using the statistical result that the squared Mahalanobis distance has a $\chi^2$-distribution (with 7 degrees of freedom in this case, equal to the number of attributes used, see Sect.~\ref{sec:supervisedClassification}), assuming a multivariate normal distribution for the training-set attributes for each variability class.

\subsubsection{Specific Object Studies} \label{DataProcessing:SOS}

For certain types of variable objects, additional processing is performed based on their variability properties.
This is the case, for example, for Cepheids, RR~Lyrae stars, long period variables, stars showing solar-like rotational modulation, transient stars such as cataclysmic variables, B-type emission line stars or flare stars, as well as eclipsing binaries, microlensing events, exoplanetary transits, or Active Galactic Nuclei.
The relevant work package is named Specific Object Studies (SOS).

The purpose of SOS processing is twofold. Firstly it validates the classification and secondly it provides parameters specific to those types of variability.
The SOS modules, one per variability type considered, takes as input the results of either the Classification step described in the previous section, or of so-called extractors.
Extractors aim at extracting from the data variability types that would benefit from a specific detection algorithm.
Currently implemented extractors are dedicated to the selection of transients (using for example the Excess Abbe method described in \citealt{Mowlavi_2014}) and of microlensing events.
The detailed SOS schema for each SOS variability type will be described in papers at the time of publication of their data in {\it Gaia} data releases.
Here, we only briefly present the case of Cepheids and RR~Lyrae stars, the data of which is published in {\it Gaia} DR1.


The SOS data processing module for Cepheids and RR~Lyrae stars (hereafter referred to as the SOS~Cep\&RRL pipeline) performs a detailed analysis of all objects classified as such by the classification module.
It determines the period(s) of each Cepheid and RR~Lyrae star candidate using the Lomb-Scarge algorithm, models its light curve using a Fourier series, and refines the period using a non-linear fitting algorithm.
It then identifies its pulsation mode and classification type based on the period(s) and light curve properties.
The period-luminosity relations are used to identify Cepheids and  subdivide them into different types and pulsation modes, while the period-amplitude relation is used for RR~Lyrae stars.
In the process, the SOS~Cep\&RRL work package either confirms the initial classification provided by the Classification module, or changes it from an RR~Lyrae star to Cepheid (or vice-versa), or rejects the source as being either an RR~Lyrae star or a Cepheid.

Confirmed RR~Lyrae stars are classified into RRab, RRc and RRd types, and confirmed Cepheids into $\delta$~Cepheid, anomalous Cepheid and Type~II Cepheid types, the latter type being further subdivided into W~Virginis, BL~Herculis and RV~Tauri types.
Their basic properties are computed and saved for publication in {\it Gaia} data releases.

The SOS~Cep\&RRL pipeline is extensively presented in \citet{DPACP-13}, and is not repeated here.
The interested reader is referred to that paper for an in-depth description of the processing steps and procedures.

\subsubsection{Global Variability Studies} \label{sec:gvs-dataprocessing}

The Global Variability Studies work package aims at describing the
global properties of the variability zoo \citep{EyerMowlavi2008} as
seen with the eyes of {\it Gaia}. It encompasses two main sub-WPs: Quality
Assessment and Bias Estimation. The main aim of Quality Assessment is
to construct useful summaries of the distribution of variable sources
in the spaces of parameters that characterise the various variability
types and deploy automatic checks for consistency with respect to
extant reference surveys such as OGLE, EROS, {\it Hipparcos} as well as
upcoming surveys such as LSST \citep{ivezic2008lsst}, Pan-STARRS 
\citep[see e.g.][and references therein]{2013ApJS..205...20M}, 
TESS \citep{2015JATIS...1a4003R}.
Departures from the distribution of sources in reference data sets 
are expected in the
{\it Gaia} variability data due to the different selection effects that
result from the observational setups in the reference variability
surveys and {\it Gaia} itself. The Bias Estimation Work Package aims to
quantify these biases, especially those stemming from the different
time samplings and sensitivities.


The Quality Assessment sub-work package produces several diagnostics. First, 
it computes kernel density estimates of the
probability density functions (PDFs) that describe the distribution of
sources in the space of the parameters that characterise the various
variability types. 
These PDFs are computed from the output of each classifier with
several membership probability cuts. As we lower the membership
probability threshold, we expect the resulting set to be more
complete but also more contaminated, which is reflected in the
PDFs. We compute several two-dimensional PDFs instead of a single
high-dimensional one to facilitate the visualization, analysis and
comparison. The PDFs are estimated on a grid of points. The software
includes the possibility to (1) define irregular grids that
increase the density of grid points in the regions of steeper
gradients of the PDFs, (2) define different functional
kernels \citep[normal or Epanechnikov;][]{Epa69}, and (3) define adaptive
kernel sizes. All plots shown in Sect. \ref{sec:gvs-data} correspond to
orthogonal grids and PDFs estimated using constant normal kernels, in
planes containing the decadic logarithm of the period (expressed in
days) and one of the shape parameters $R_{21}$ (the amplitude ratio of
the second to first Fourier terms of the light curve decomposition)
and $\phi_{21}$ \citep[the phase difference between the first two terms of
the Fourier decomposition;][]{1981ApJ...248..291S}.

Individual PDFs only allow for a qualitative and subjective
evaluation. In order to make more quantitative comparisons, the
Quality Assessment software incorporates statistical tests of the
significance of the differences between PDFs. The first test proceeds
by dividing the samples produced by Characterisation and
Classification (and the reference survey sample) into several
subsamples that may be disjoint if the number of available sources
allows for it. In the case of small sample sizes, we sample with
replacement (and thus, the subsamples can have non empty
intersections). Replacement implies correlations between samples, an
underestimation of the variances and, as a consequence, a decrease in
the sensitivity of the tests. The {\it Gaia} and reference survey samples
are paired, a PDF is estimated for each subsample, and the difference
between the {\it Gaia} and reference survey PDFs is computed. For $k$
subsamples, this process results in $k$ differences between paired
PDFs. The $k$ PDF differences are used to estimate the standard
deviation at each grid point. Since the distance between grid points
is less than the typical kernel size, correlations appear amongst the
estimates of the PDF and its variances in neighbouring points.

Once the estimates of the variance at each grid point are available,
the difference between two PDFs ({\it Gaia} and the reference survey) is
measured in terms of the standard deviation. We will refer to these
differences as scaled differences. Thus, the resulting maps reflect
the statistical significance (not the magnitude) of the differences
between the two PDFs. 

The PDFs for the complete samples (that is, not for the subsamples)
are subsequently used to compute Kullback-Leibler \citep[KL;][]{KullbackLeibler} divergences to
measure the degree of departure of the different {\it Gaia} results (for
example, for samples from the various classifiers or combinations
thereof, and/or for different membership probability thresholds) from
the distribution of sources in the reference survey. These KL
divergences can be used to select the combination of classifier and
membership cuts that result in a PDF that is closer to the reference
survey if this consistency is assumed to be desirable, although this
approach always leads to neglecting new discoveries. In general, this
minimum divergence setup is used as a basis to study and validate the
departures from it. 

The Hotelling T$^2$ test \citep{hotelling1931} is also applied to the 
samples described above, in order to check for the
equality of the means of the distributions in the {\it Gaia} and reference
surveys only in cases where the distribution of sources in
parameter space is approximately multinormal. However, the T$^2$ test is a 
global summary that does not convey information on
the local differences between the two distributions.

The Quality Assessment sub-work package also includes modules for the detection of outliers based on the estimated PDFs, and for the comparison of
classical linear models constructed from {\it Gaia} samples with those in the literature. These modules are not described here.

\section{The data
\label{sec:theData}}
We introduce the data sets used to produce the variability results for the {\it Gaia} DR1 in Sect.~\ref{sec:dataUsedInGdr1}, as well as the  time and flux conversions used for the processing in Sect.~\ref{sec:time_and_conversions}.
The details of the applied observation filters are discussed in Sect.~\ref{sec:obsFiltering}. 

\subsection{Data used in the {\it Gaia} Data Release 1\label{sec:dataUsedInGdr1}}

The variability data processing results in this paper are derived from the two consecutive Data Reduction Cycles 0 and 1 of {\it Gaia} epoch photometry provided by CU5, which are abbreviated as `C0' and `C1' throughout this paper.
Our variability processing was first applied to the C0 dataset containing the photometric reduction of the first 3 months\footnote{
The 3 months of C0 data covers observations taken between July~25 and October~25 2014, during which several scanning laws were active (see Sect.~\ref{sec:epsl} and \ref{sec:nsl} for details): 28 days of EPSL, 1 month of transition NSL and 1 month of GAREQ NSL. 
%
}
of {\it Gaia} science data and finally to the C1 dataset containing an improved photometric reduction of almost 14 months\footnote{
The 14 months of C1 data covers observations taken between July~25 2014 and September~16 2015, i.e., the same start as C0 but with almost 1 year of GAREQ NSL. 
} of science data. 

\begin{figure}
\centering
\includegraphics[width=0.40\textwidth]{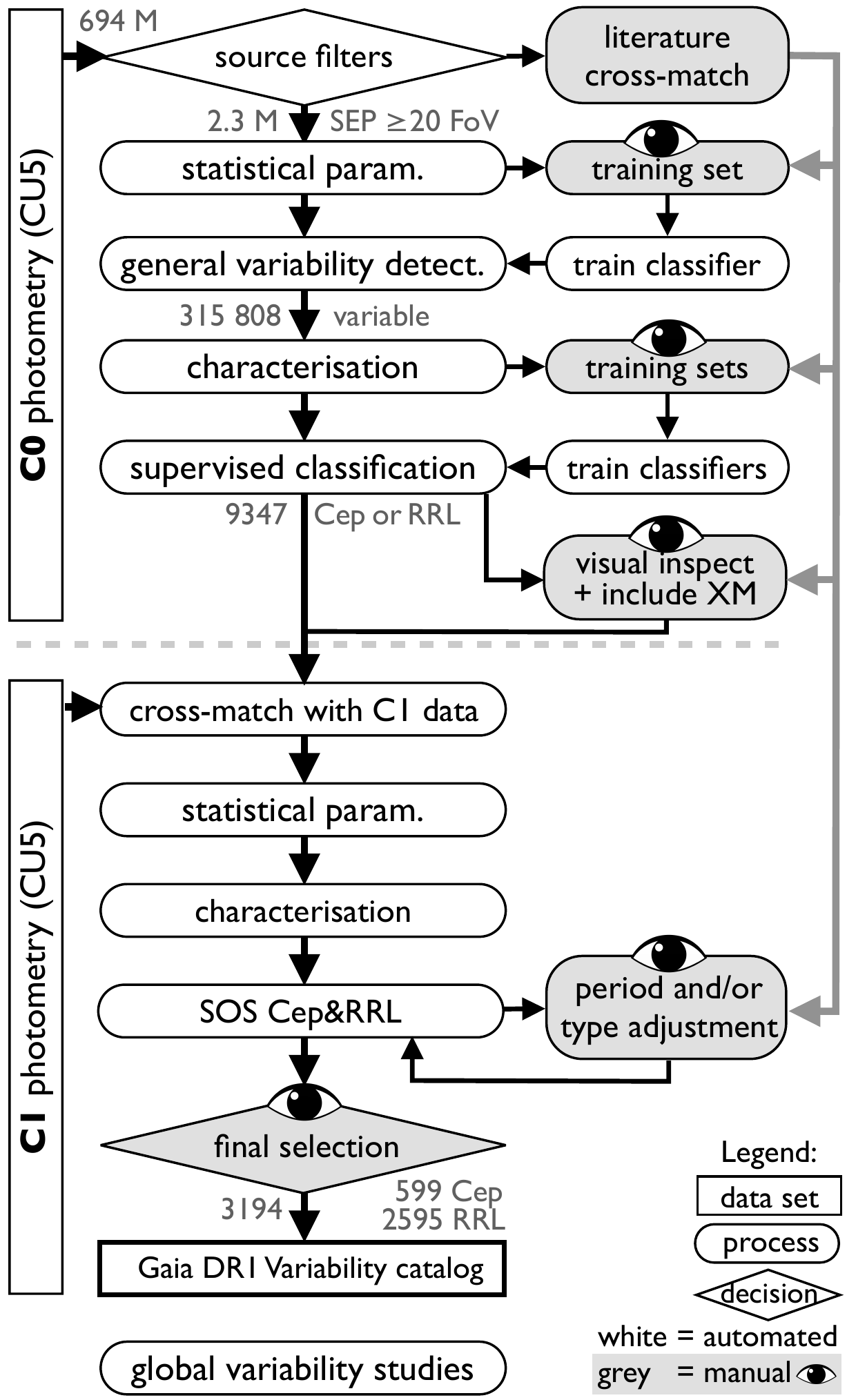}
\caption{{\it Gaia} DR1 processing diagram for the South Ecliptic Pole (SEP) data processing.
The processing was performed consecutively on two photometric datasets (see Sect.~\ref{sec:dataUsedInGdr1}), labelled `C0'  (top half of diagram), and `C1' (lower half of diagram) which are connected through a cross-match of identified sources.
The results of C0 and C1 data processing are presented in Sect.~\ref{sec:C0sepResults} and~\ref{sec:C1SepResults}, respectively.  
The grey numbers refer to the number of sources available at various stages of the processing.
The results of the global variability studies is presented in Sect.~\ref{sec:gvs-data}.
}
\label{fig:sepProcessing}
\end{figure}

The processing of these two datasets is presented in Sect.~\ref{sec:C0sepResults} and~\ref{sec:C1SepResults} and is summarised in Fig.~\ref{fig:sepProcessing}.
The use of two different datasets for the {\it Gaia} DR1 unfortunately made the processing more complicated, but it was a one-time exception due to time constraints without which the {\it Gaia} DR1 variable catalogue would not have been possible. 

Although various sources of input are planned to be used for the variability processing, as shown in Fig.~\ref{fig:generalProcessing}, the only calibrated data available in the C0 data were $G$-band photometric time series.
In addition, the average of the integrated Blue Photometer (BP) and Red Photometer (RP) time series were available, however not planned for publication.
Therefore, in this paper we present analyses based on the $G$-band photometry (averaged per FoV transit), with the additional use of the mean $G_{\rm BP}-G_{\rm RP}$ `colour' in classification, where it significantly improved the source identification in both crossmatch and variability-type classification.
This colour information is however not released in this \textit{Gaia} DR1.
 
To allow for a confident identification of sources, we only select sources with a minimum number of  20~FoV observations, which reduced the number of C0 photometry sources from 694~million to 3.7~million.

The threshold on the number of transits leaves enough data points per source for automated period search and allows for visual confirmations of sources identified by our pipeline.
In future data releases, these conditions will be relaxed as the quality and quantity of the data will increase and our automated procedures will improve in accuracy.
Note that the number of observations actually processed for each source was often reduced due to the removal of `outliers' and flux anomalies, as explained in Sects.~\ref{sec:obsFiltering} and~\ref{sec:C0resultStatistics_timesampling}, therefore occasionally resulting in time series with less that 20~FoV observations 
 (see Table~\ref{tbl:statistics} for more statistical details).

\begin{figure}
\centering
\includegraphics[angle=0,width=0.47\textwidth]{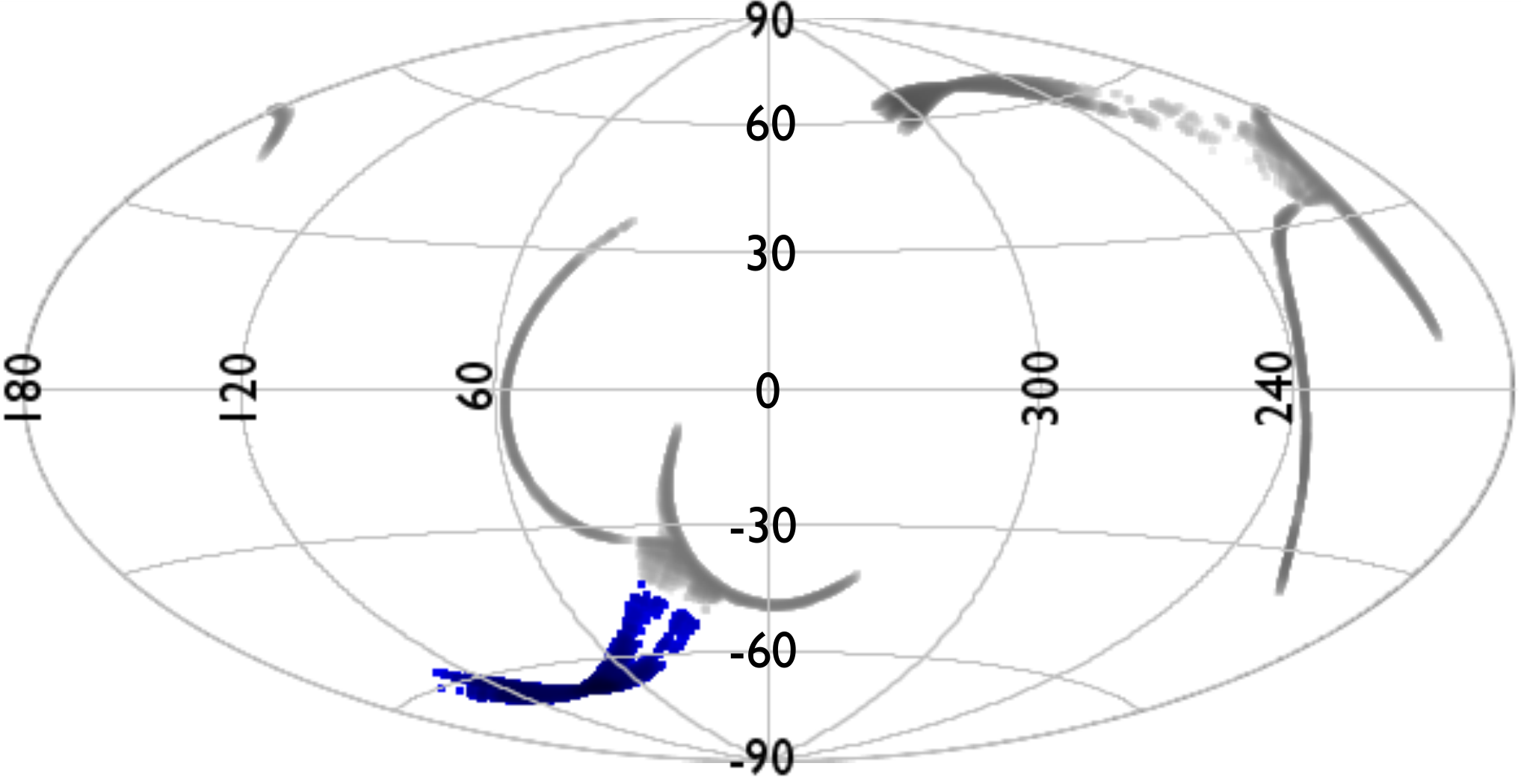}
\caption{Equatorial sky distribution of the 
3.7~million sources with $\geq$20~FoV observations in the $G$ band, that were available in the C0 data. Of these, 2.3~million sources in the `SEP' region (i.e., within 38 deg from the South Ecliptic Pole) have been processed, coloured in blue. The very particular sky coverage is due to the scanning laws that were active in the time interval of the examined data.  (C0 data) }
\label{fig:sepSrcSelection}
\end{figure}

Additionally, we only selected sources that are within~38 degrees from the South Ecliptic Pole, which hereafter is referred to as the `SEP' region. 
Figure~\ref{fig:sepSrcSelection} shows the sky distribution of these remaining 
2\,295\,539 
sources that were processed to search for Cepheid and RR~Lyrae stars.

In the following C1 dataset, 
only the source candidates identified in the C0 dataset were processed. All content in the {\it Gaia} DR1 is based on C1 data only, which is detailed in Sect.~\ref{sec:Gdr1varCatalog}.

\subsection{Time and flux conversions\label{sec:time_and_conversions}}
Published observation times are expressed in units of barycentric JD (in TCB) in days - 2\,455\,197.5, as detailed in the following. Observation times are first converted from On-board Mission Time (OBMT) into Julian date in TCB (Temps-Coordonn\'{e}e Barycentrique). Next we apply a correction for the light-travel time to the Solar system barycentre, resulting in Barycentric Julian Dates (BJD). Finally, an offset of 2\,455\,197.5 days (corresponding to a reference time $T_0$ at 2010-01-01T00:00:00) is applied to produce conveniently small numerical values. The time accuracy of the individual (currently not published) CCD observations is (much) below 1~ms, though the currently published per-FoV observation times are averaged over typically 9 CCD observations spread over about 44~sec.


In the variability pipeline both fluxes and magnitudes are used in different processing modules. CU5 provides calibrated 
photometry in flux. To convert to magnitude we use the $G$-band zero-point magnitude of 25.52477 in the Vega system  \citep{DPACP-12} published in the \textit{Gaia} DR1. 

\subsection{Observation filtering\label{sec:obsFiltering}} 
Even though extensive work has gone into the $G$-band photometry calibration, it is not yet optimal \citep[see][]{DPACP-11},
leading to some anomalies in the flux and uncertainty estimations. To deal with these anomalies, 
two consecutive \textit{operators} were put in place to filter out potential anomalous observations. Due to these
filters, we end up with some sources that have less than 20 observations per time-series, as detailed in Sect.~\ref{sec:C0resultStatistics}.

\begin{figure}[t]
\centering
\includegraphics[angle=0,width=\hsize]{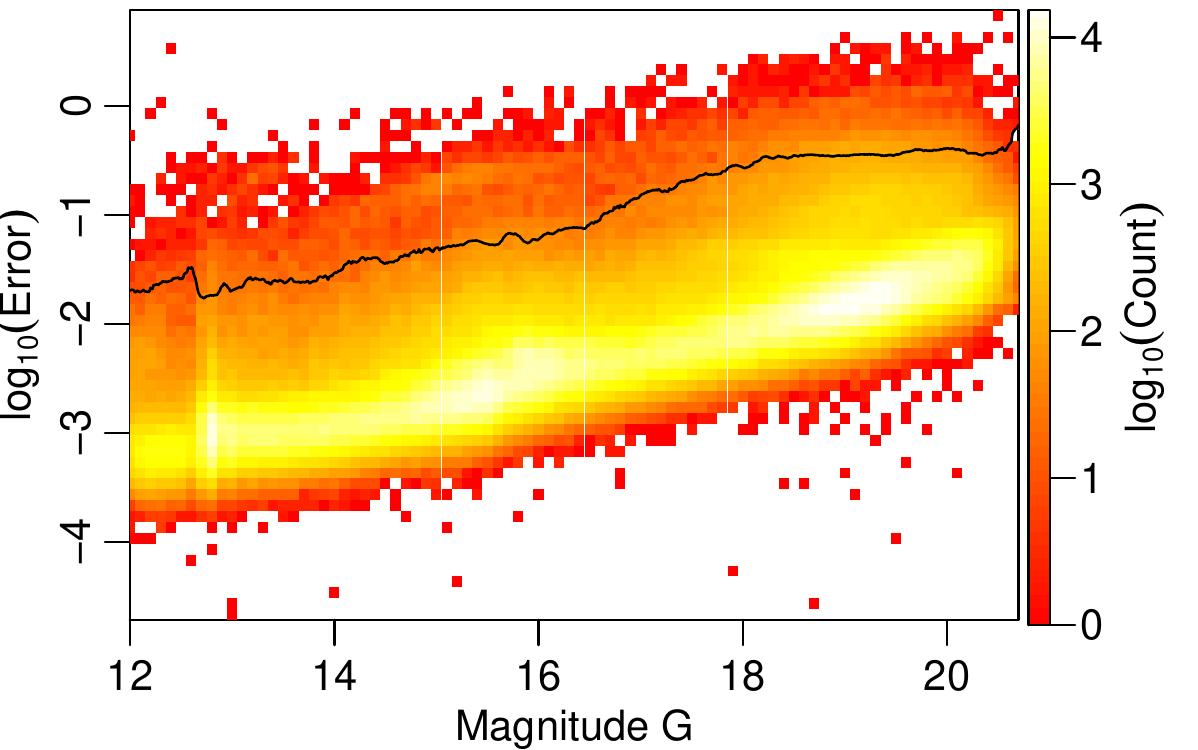}
\caption{Histogram of FoV $G$ magnitude versus $G$ errors of a C1 data sample including all Cepheid and RR~Lyrae candidates. The quantile selected threshold is superimposed as a black line, i.e., observations with errors above this threshold are excluded from processing. (C1 data)}
\label{fig:ExtremeErrorCleaning}
\end{figure}

The visualization of both C0 and C1 datasets revealed  mild as well as extreme FoV-transit $G$ magnitude outliers, usually associated with extremely large uncertainties, suggesting  possibly unreliable data points and the necessity hence of excluding them from the processing.
In order to define a cleaning operator based on the large error values, we determined a magnitude-dependent threshold.
In the C0 data, we determined this threshold based on a 2-dimensional histogram of  median errors vs. median magnitude per time series.
The analyses of the C0 data, which lead to the selection of candidate Cepheid and RR~Lyrae variables, used time series obtained by filtering out the observations with errors larger than these magnitude-dependent thresholds.
In the C1 data analyses, we used the magnitude-error pairs of a sample of 50K objects that included all the candidate Cepheid and RR~Lyrae stars.
To determine the optimal thresholds, we applied a sliding window technique to a sufficiently dense magnitude grid: for each magnitude $G$ on the grid $(10, 10.01, 10.02, \ldots, 20.7)$, we took all transits with magnitudes in the interval $[G-0.1,G+0.1]$, and computed a series of high quantiles $\{0.98, 0.981, \ldots, 0.999\}$ of all the uncertainties of these FoV transits.
A visual inspection of a sample of time series suggested the 99.5th percentile as the best local threshold at all magnitudes.
Figure \ref{fig:ExtremeErrorCleaning} shows the selected quantile superimposed on the 2D histogram of magnitude-error pairs. 
The processing of the C1 data included the removal of those $G$-band FoV transits with uncertainties larger than the value of the local interpolated threshold.

\section{C0: South Ecliptic Pole data processing
\label{sec:C0sepResults}}
The SEP (South Ecliptic Pole) data processing followed the processing outlined in Fig.~\ref{fig:sepProcessing}, which was split up in two successive data sets C0 and C1 described in Sect.~\ref{sec:theData}.
This section presents the properties of the time series and the classification results of the C0 data set, which is not published in {\it Gaia} DR1.

\subsection{Crossmatch with literature (C0)} \label{sect:crossmatch}
The {\it Gaia} data were crossmatched with a selection of surveys for use in both validation and classification training sets.
The crossmatch was performed before pipeline processing runs and thus it was allowed for a lower threshold of 10 (instead of 20) FoV transits per source in the {\it G} band, so that further selections could be applied to subsequent work packages as deemed appropriate. In the end, only GVD made use of sources with less than 20 (unfiltered) FoV observations per source.
Exceptionally, known planetary transits and quasars were crossmatched beyond the SEP region (defined in Sect.~\ref{sec:dataUsedInGdr1}) in order to increase the representation of these classes.

The positions of target objects of a given survey are used to search for {\it Gaia} neighbours within a radius of 5 arcsec (to account for different positional accuracies and limited proper motions).
If the number of objects to crossmatch with the {\it Gaia} sources is less than about a thousand, a positional-based match (typically the nearest neighbour) is confirmed by visual inspection, 
otherwise the selection of crossmatched objects takes into account positional, photometric, and time-series (if available) attributes by means of dedicated Random Forest classifiers, as described in \citet{ADASS_crossmatch}.

\begin{figure*}
\centering
\includegraphics[angle=0,width=0.85\textwidth]{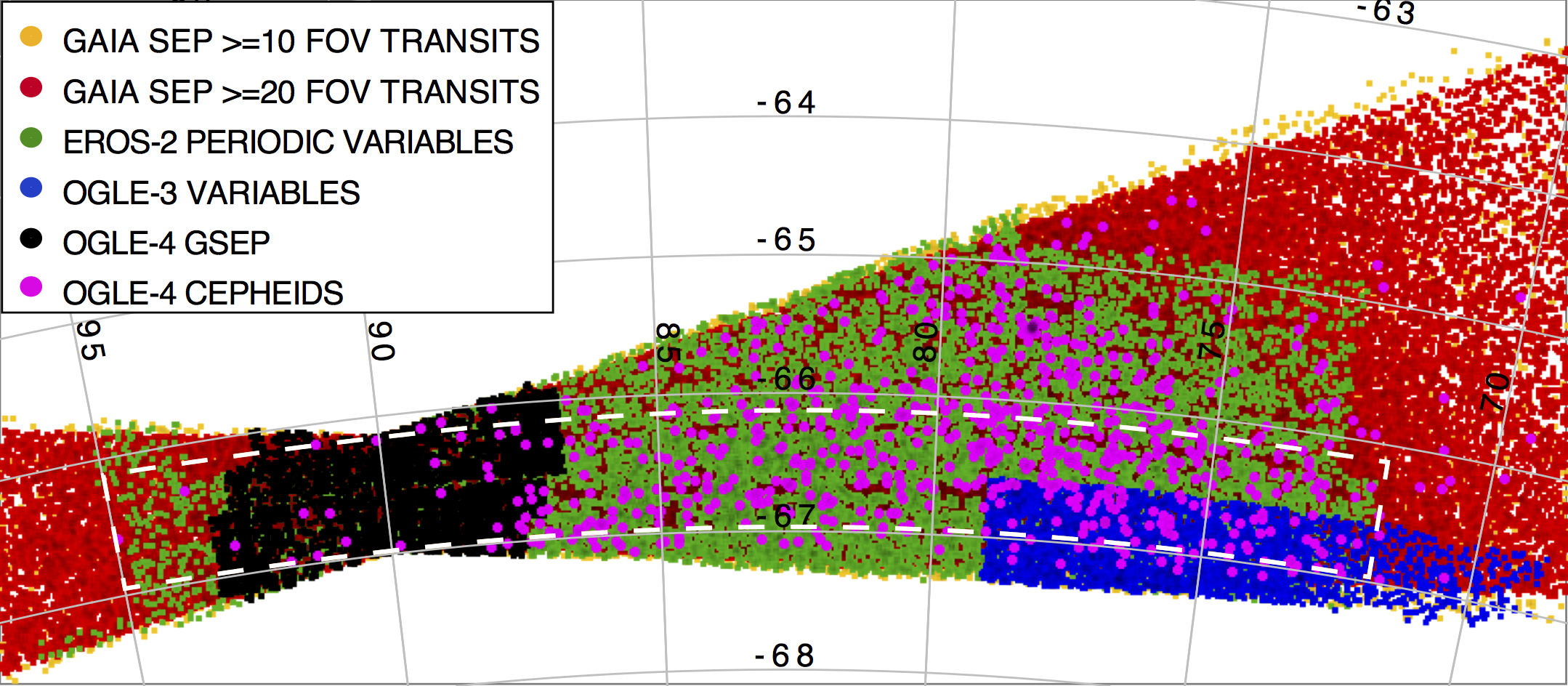}
\caption{Sky map of crossmatched catalogues near the South Ecliptic Pole in Equatorial coordinates. The dashed white line outlines the test region ($72.00\degr < \hbox{RA}<95.00\degr$  and  $-66.93\degr <\hbox{Dec}< -66.15\degr$) used for the sky completeness estimates of Table~\ref{tab:completenessOverview}, as described in Sect.~\ref{sec:completeness}. (C0 data)}
\label{fig:xm_skymap}
\end{figure*}

The use of different surveys with their own peculiarities provides objects from a variety of types, coverage in different regions of the sky, measurements in different photometric filters, with diverse sensitivity, resolution, etc.  
For the first {\it Gaia} release, the OGLE-IV {\it Gaia} South Ecliptic Pole (GSEP) is particularly useful as it combines the highest source number density (from the nearby Large Magellanic Cloud) with the greatest number of observations per source (as per the EPSL scanning law during the first 28 days, see Sect.~\ref{sec:epsl}).
Other surveys covering the region of the Large Magellanic Cloud near the South Ecliptic Pole are also considered, in addition to a few others that cover different regions of the sky:
\begin{enumerate}
\item The OGLE-IV GSEP variable stars \citep{2012AcA....62..219S}.\footnote{\url{ftp://ftp.astrouw.edu.pl/ogle/ogle4/GSEP/var_stars/}}
\item The OGLE-IV GSEP constant star candidates from objects with the smallest variations in both {\it V} and {\it I} bands.\footnote{\url{ftp://ftp.astrouw.edu.pl/ogle/ogle4/GSEP/maps/}}
\item The OGLE-IV Cepheids \citep{2015AcA....65..233S,2015AcA....65..297S}.\footnote{\url{ftp://ftp.astrouw.edu.pl/ogle/ogle4/OCVS/lmc/}}
\item The OGLE-III variable stars \citep{2008AcA....58...69U}.\footnote{\url{ftp://ftp.astrouw.edu.pl/ogle/ogle3/OIII-CVS/}}
\item The periodic variable stars extracted from the EROS-II survey by \citet{2014A&A...566A..43K}.
\item High-confidence (99\%) SDSS photometric quasar candidates which also show radio and/or X-ray association, extracted from the Half Million Quasar catalogue \citep{2015PASA...32...10F}.
\item Confirmed planetary transits listed by Southworth (Aug.\ 2015)\footnote{See the list of `basic observable properties' at \url{http://www.astro.keele.ac.uk/jkt/tepcat/observables.html}}.
\end{enumerate} 
The numbers of crossmatched sources per survey are listed in Table~\ref{tab:xm_all} and the sky coverage of the OGLE and EROS matches near the South Ecliptic Pole is depicted in Fig.~\ref{fig:xm_skymap}.

\begin{table}
 \centering
  \caption{The number of crossmatch sources as a function of catalogue, for {\it Gaia} sources with at least 10 FoV transits. (C0 data)}
  \vspace{2mm}
\label{tab:xm_all} 
  \begin{tabular}{lr}
  \hline
  \noalign{\smallskip}
Catalogues from Literature &  Crossmatched\\
  &  Sources \\
\hline
\noalign{\smallskip}
EROS-II Periodic Variables & 13\,296 \\
OGLE-III Variable Stars & 3501\\
OGLE-IV Cepheids & 748\\
OGLE-IV GSEP Constant Stars & 4311\\
OGLE-IV GSEP Variable Stars & 2064\\
SDSS Photometric Quasars & 632\\
Planetary Transits & 1\\
\hline
\end{tabular} 
\end{table}

\subsection{Statistics (C0)} \label{sec:C0resultStatistics}
We present here some important statistical properties of the $G$-band of all  2\,295\,539 sources considered in the C0 photometry relevant to the characterisation 
and subsequent classification of variable objects. Statistical parameters are computed as the first step in the CU7 general pipeline without any prior information on variability.

\subsubsection{Characteristics of the time sampling} \label{sec:C0resultStatistics_timesampling}
\begin{figure*}
\centering
\includegraphics[width=\textwidth]{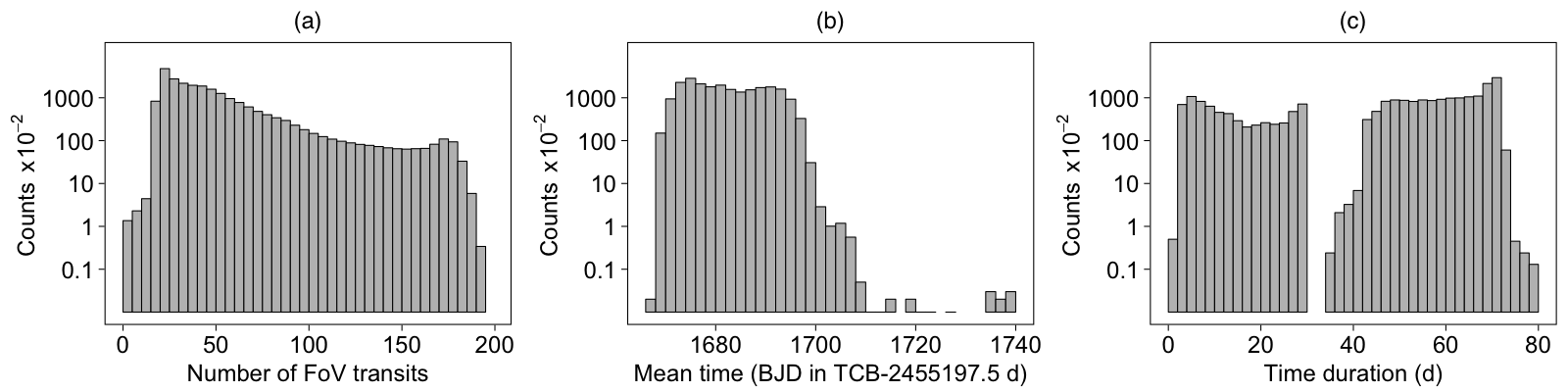}
   \caption{Characteristics of the time sampling of the $\approx$ 2.3 million SEP sources (C0 data).
   (a)~Distribution of the number of FoV transits, (b)~Distribution of the mean observation time, (c)~Distribution of the time series duration}
  \label{fig:stats_sampling_C0}
\end{figure*}

\begin{figure}
\centering
\includegraphics[width=0.45\textwidth]{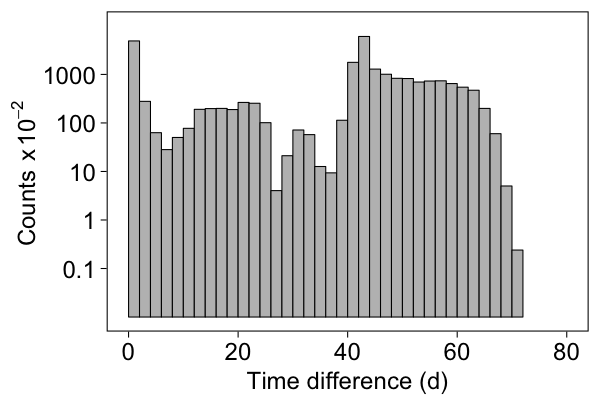}
   \caption{Distribution of the maximum  time difference between two successive FoV transits in the 2\,295\,539 sources in the C0 dataset. }
  \label{fig:stats_sampling_maxdeltatime_C0}
\end{figure}

Figure~\ref{fig:stats_sampling_C0} presents distributions and plots of the time sampling characteristics of the $\approx$2.3 million SEP C0 sources.
As described in Sect. \ref{sec:obsFiltering}, due to some anomalies in the flux and uncertainty estimations, 
$\approx$3.5$\%$ of the sources in this dataset had $<$20~FoV transits following application of the {\it operators} (see Sect. \ref{sec:obsFiltering}).  This can 
be seen clearly in Fig.~\ref{fig:stats_sampling_C0}(a). The average number of  FoV transits per source is 50 but can be up to almost 
200 due to the dense sampling during the 28 days of EPSL.  The total duration of these time series extends up to almost 80 days  as shown in Fig.~\ref{fig:stats_sampling_C0}(b). 
The dense sampling and 80 day duration are  favourable for the detection and characterisation of many variability types. 
The distribution of the mean observation time in Fig.~\ref{fig:stats_sampling_C0}(c) 
shows that, for the majority of sources, most FoV transits occur during the ecliptic pole scanning time interval. 
Figure~\ref{fig:stats_sampling_maxdeltatime_C0} shows the distribution of the maximum difference between two successive observations in the C0 time series.
It can be seen that in some cases there are large gaps in time of up to 71 days.  
Such large gaps can have an adverse effect on the detection of variability and search for periodicity. 

\subsubsection{Characteristics of the $G$-band magnitude}
\begin{figure*}
\centering
\includegraphics[width=\textwidth]{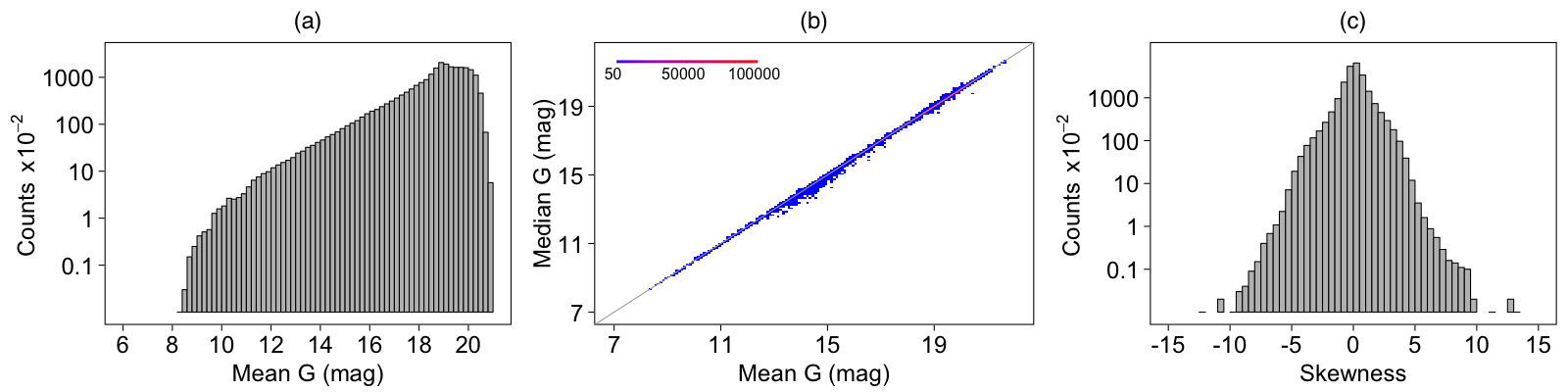}
   \caption{Histograms showing characteristics of the $G$-band magnitude of the $\approx$2.3 million SEP sources (C0 data). 
   (a)~Distribution of the mean magnitude, (b)~The median vs the mean magnitude, (c)~Distribution of the skewness of the magnitude distribution.}
  \label{fig:stats_moments_C0}
\end{figure*}

Figure~\ref{fig:stats_moments_C0} shows characteristics of the magnitude distribution of the $\approx$2.3 million SEP sources  in the C0 dataset.  The median 
magnitude spans a range from $\approx$7 to 21~mag. A break in the completeness rate of detected objects is visible at $G$$\approx$19~mag, followed by a sharp decline after $G$$\approx$20~mag.
Figure~\ref{fig:stats_moments_C0}(b) shows a strong agreement between the mean and median values, indicating that the filtering operators described in Sect. \ref{sec:obsFiltering}
have worked well in removing anomalies in the flux and uncertainty estimations. Figure~\ref{fig:stats_moments_C0}(c) illustrates the distribution of the $G$-band time-series skewness.

\subsubsection{Characteristics of the $G$-band magnitude uncertainties}
\begin{figure*}
\centering
\includegraphics[width=\textwidth]{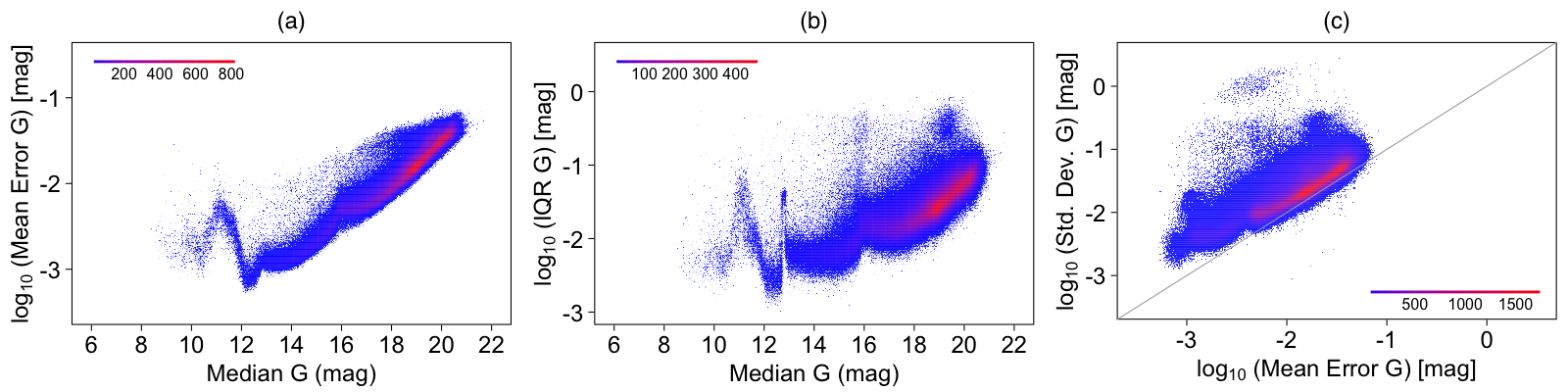}
   \caption{Density plots showing characteristics of the $G$-band magnitude uncertainties of the $\approx$ 2.3 million SEP sources (C0 data). 
   (a)~Distribution of the mean error as a function of the median magnitude, (b)~The IQR as a function of the median magnitude, 
   (c)~Standard deviation vs the mean error.}
  \label{fig:stats_uncertainties_C0}
\end{figure*}

The distribution of mean FoV photometric uncertainties as a function of median magnitude per object is shown in Fig.~\ref{fig:stats_uncertainties_C0}(a).
The large abrupt increase in the mean FoV uncertainties per time series at the bright end ($G$$\approx$11~mag) is due to inaccurate onboard estimates of the brightness which prevent TDI gates from reducing the integration time of bright objects precisely, resulting in saturation.

\subsection{Variability detection (C0)} \label{sec:GVD_run}
The GVD package was executed on C0 data to determine the variable objects to subsequently characterise and classify.
A set of 10 GVD attributes were selected according to their usefulness in the training of a Random Forest (RF) classifier:
 the reduced $\chi^2$ in $G$ band,
 the Stetson variability index in the single $G$ band and in the $(G,G_\mathrm{RP})$ band pair \citep{1996PASP..108..851S},
 the greatest absolute deviation from the median in $G$ band normalised by the uncertainty of the corresponding extremal measurement,
 the percentile-based skewness (defined in Sect.~\ref{DataProcessing:Classification}),
 the standardised kurtosis cumulant, the difference between the $G$-band interquartile range (IQR) of a source and its most common value (as a function of magnitude),
 the Spearman correlation in the $(G,G_\mathrm{BP})$ and $(G,G_\mathrm{RP})$ band pairs,
 and the number of FoV transits per source.

\begin{figure}
\centering
\includegraphics[angle=0,width=0.25\textwidth]{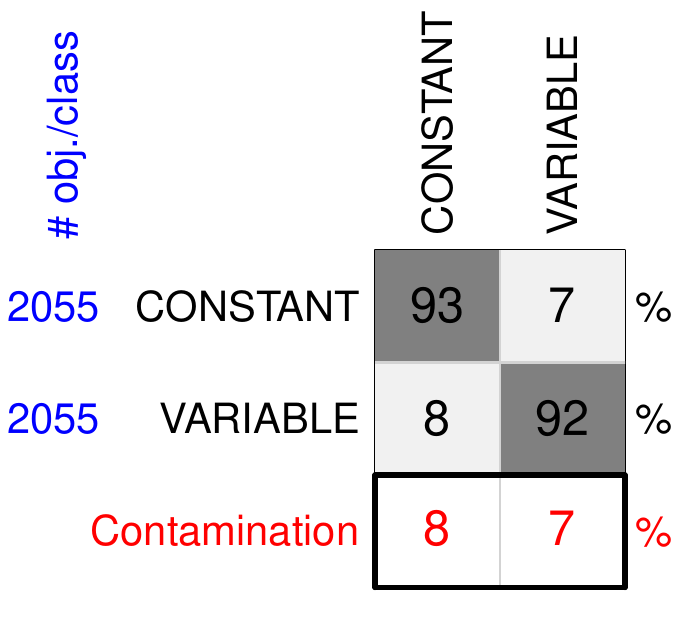}
\caption{Confusion matrix of the variability detection Random Forest classifier. (C0 data)}
\label{fig:GVD_confMatrix}
\end{figure}

The training set comprised an equal number of variable and non-variable objects:
2055 {\it Gaia} sources crossmatched with the OGLE-IV {\it Gaia} South Ecliptic Pole (GSEP) variable stars \citep{2012AcA....62..219S} and the same number of constant star candidates inferred from the least varying objects in the OGLE-IV GSEP maps (see Sect.~\ref{sect:crossmatch}).
The RF classifier was trained with 500 trees and the confusion matrix of the training set (estimated from the `out-of-bag' objects of RF) shows a completeness rate of 92-93\% (Fig.~\ref{fig:GVD_confMatrix}).
This GVD-classifier was applied to the 
2\,295\,539 
SEP sources and 
315\,808  
of them (13.8\%) were classified as variable in the C0 data (as annotated in Fig.~\ref{fig:sepProcessing}).

\subsection{Variability characterisation (C0)} \label{sec:C0Characterization-results}
The Characterisation pipeline, described in Sect.~\ref{DataProcessing:Characterization} and comprising both period search and time series modelling, was run on the 315\,808 sources detected as variable by GVD in the C0 data (see Sect.~\ref{sec:GVD_run}).
The $G$-band time series of each source was first independently scanned for periodicity, the most significant period retained, and the data fit with a Fourier series model using this period. 
No de-trend was applied prior to running period search for this first analysis, as some initial indications of systematic trends in the EPSL were not confirmed by or not propagated into the NSL data.
We are aware that this carries some risk of masking some periodicity as well as of introducing some artificial variability, but we are confident that these cases are rare and will be filtered out by the classification process.

For the period analysis, the Least Squares period search method \citep[see][]{2009A&A...496..577Z} was chosen since it is one of the best methods generally across all variability types and it has a good performance for Cepheid and RR~Lyrae variables.
A minimum of 9 FoV transits was required, as studies showed that a reliable estimate of the period is almost impossible with fewer observations.
This cut further reduced the number of sources processed by characterisation to 315\,649.
The non-optimal estimates of the errors on magnitudes, as reported in Sect.~\ref{DataProcessing:Statistics}, together with the fact that extreme outliers and observations with excessively large uncertainties had been removed, resulted in a rather flat error distribution and consequently did not provide any impetus for a weighted period search.
Adaptive sampling of the frequency interval was used whereby the start frequency and the frequency step were determined from the total duration of the observations of each individual time-series, as described in  Sect.~\ref{DataProcessing:Characterization}. The highest frequency searched was 6d$^{-1}$ and the processing was limited to mono-periodic models, because of the limited time span and/or limited number of observations.
\begin{figure*}
\centering
 \includegraphics[width=\textwidth]{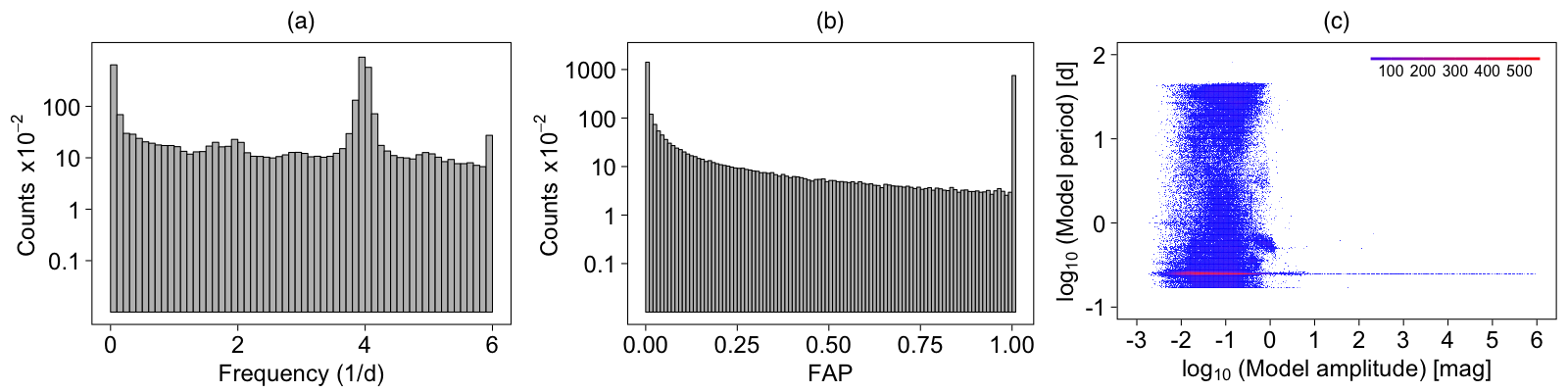}
   \caption{Distributions of frequencies and corresponding False Alarm Probability (FAP), together with a plot of period versus the peak-to-peak model amplitude from Characterisation using the Least Squares period search method \citep[see][]{2009A&A...496..577Z} 
   for all 315\,649 variable sources with >8 FoV transits (C0 data).}
    \label{fig:char_hist_frequency_all_recovery_C0}
\end{figure*}

During EPSL time interval, the sampling was, for many sources, very regular and often uninterrupted. The time sampling is related to the rotation period (6h) of the spacecraft and as a consequence, we notice parasite frequencies at 4d$^{-1}$ and/or at the harmonics and subharmonics of this frequency.  It should also be remarked that variables with periods very close to 6h that have only FoV transits from the EPSL time interval are likely to be sampled at very similar phases in their cycle 
and might not be recognised as  variable.  
The relevance of parasite frequencies will however gradually reduce as more NSL observations enter the data sets. Figures \ref{fig:char_hist_frequency_all_recovery_C0}(a) and (b)  show distributions of the period recovery with the Least Squares methods and the corresponding False Alarm Probability (FAP) for all 315\,649 variable sources with $>$8 FoV transits in the C0 dataset.  A peak at the 4d$^{-1}$ parasite frequency can be clearly seen.

The most significant period found by the period search process is used as input to the modelling module, which automatically determines the best model of the data and improves the overall fit. 
Sources with a frequency resulting  in a  maximum gap in phase $\Delta \phi_{\max}>0.5$ are not modelled beyond a constant term, further reducing the total number of variables with a significant frequency to $255\,706$. 
Attributes such as, but not limited to, the final period and the peak-to-peak amplitude of the model are then used by the classification module.  Figure \ref{fig:char_hist_frequency_all_recovery_C0}(c) shows
a log-log plot of the model period as a function of the model peak-to-peak amplitude for the $255\,706$ variables sources with a significant period.  The sharp cut-offs in period represent the minimum and maximum search periods of $\approx$0.17 and 40 days respectively. The peak-to-peak amplitude of the vast majority of all variables falls in the range from 0.005 to 2~mag. A strong line with a high density of sources is clearly seen at the 4d$^{-1}$ parasite frequency, extending from 
very small to extremely large amplitudes. Some of these are sources with a true frequency around 4d$^{-1}$, however the majority are not. The very large amplitudes are clearly not physical 
and are an artefact of the automated modelling system that tries to model data with a 4d$^{-1}$  parasite frequency.  All sources with a peak-to-peak amplitude larger than 8~mag have data only in the 28~day EPSL  time interval. As more NSL observations are included in these time series, such high-amplitude artefacts will naturally disappear.

\subsection{Supervised classification of variable stars (C0)} \label{sec:supervisedClassification}

For this particular data release, we aim to classify only two types of variable stars: RR~Lyrae stars and Cepheids.
To do so, we use the 3 different classifiers we mentioned in Sect.~\ref{DataProcessing:Classification}, which we configured in such a way that they are complementary rather than supplementary.
This is reflected in the choice of training sets and classification attributes.
The RF classifier was configured to optimise completeness, at the expense of more contamination.
Here, we define completeness as $TP / (TP + FN)$ and the contamination as $FP / (TP + FP)$, where $TP$ and $FP$ are the number of true and false positives respectively.
The other two classifiers, GM and BN, were optimised towards low contamination, at the expense of a decreased completeness.

\subsubsection{The training sets}\label{sec:classificationTrainingsets}

The GM and BN classifiers were trained with a training set containing {\it Gaia} $G$-band time series of 524 RR~Lyrae stars, 602 Cepheids, 10 $\delta$ Scuti stars, and 177 eclipsing binaries.
All training time series contain at least 20 {\it Gaia} time points in the $G$ band, have a {\it Gaia}-derived pulsation period that deviates no more than 5\% in phase with respect to the literature value, and were each visually inspected for their quality.
We included more classes in the training set besides the target classes RR~Lyrae and Cepheids in order to help avoid members of these other classes to be erroneously classified as RR~Lyrae or Cepheid.
All of the training sources were previously known variables in the OGLE-IV catalogue \citep{2012AcA....62..219S,2015AcA....65..233S,2015AcA....65..297S}.

\begin{figure*}
\centering
\includegraphics[angle=0,width=0.9\textwidth]{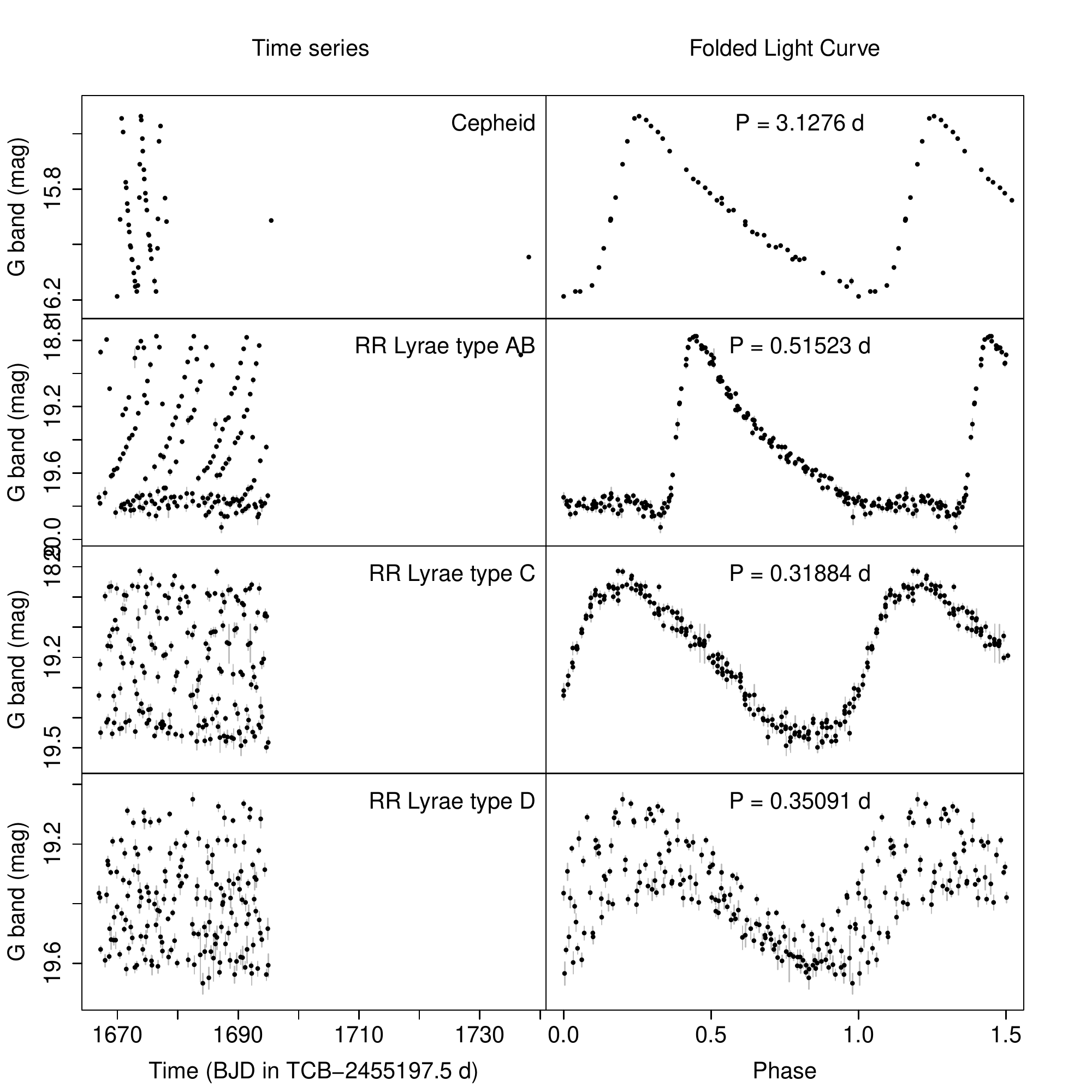}
\caption{Time series and folded light curves for a sample of periodic variable stars, with (dominant) period $P$ in days computed by the pipeline. Represented classes include Cepheid and RR~Lyrae (types AB, C, and D). (C0 data)}
\label{fig:lightcurve_1}
\end{figure*}

\begin{figure*}
\centering
\includegraphics[angle=0,width=0.9\textwidth]{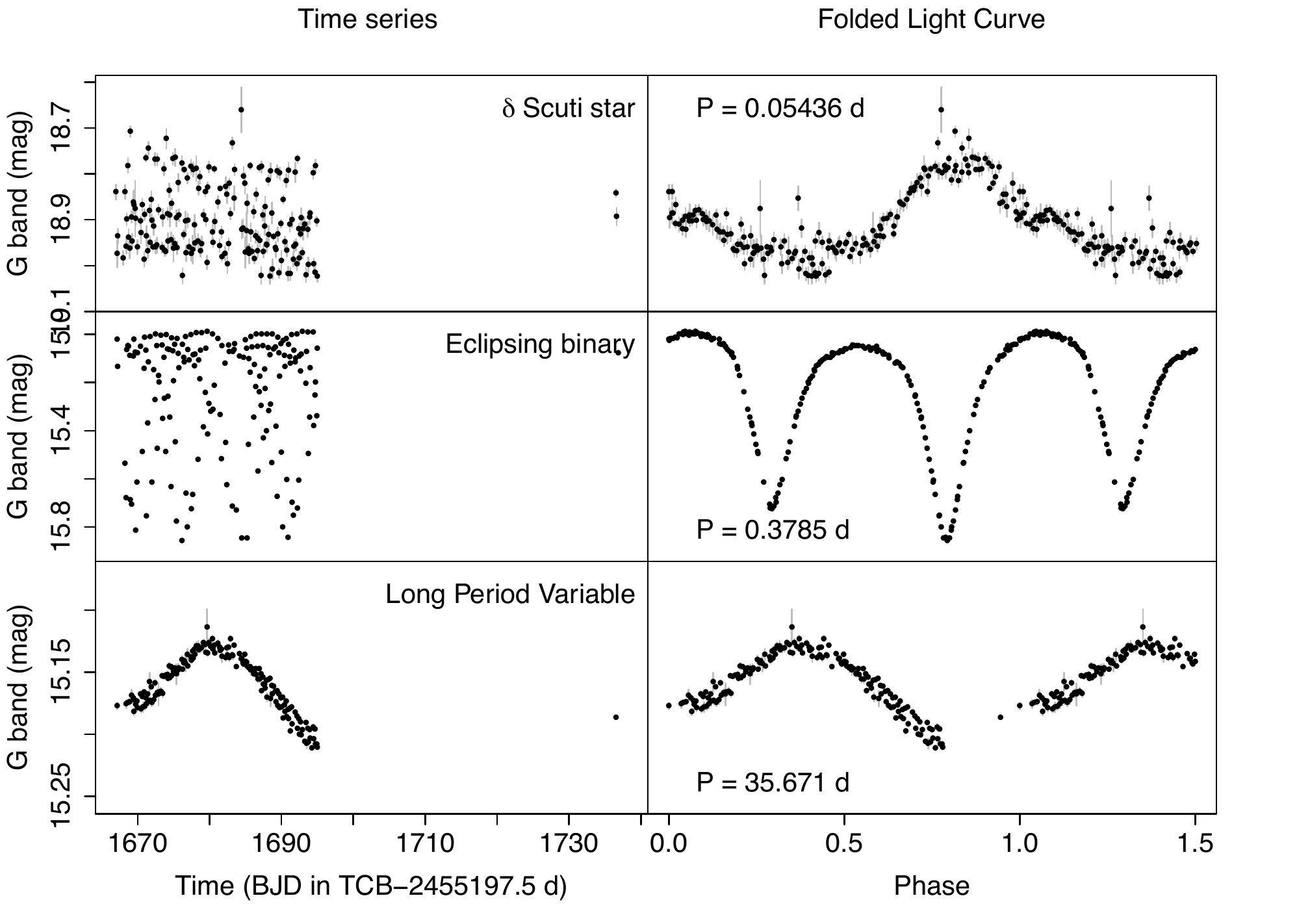}
\caption{
Same as Fig~\ref{fig:lightcurve_1}, for the classes of $\delta$~Scuti, eclipsing binary, and long period variables. The period used in the folded light curve of the $\delta$~Scuti star was set from literature \citep{2012AcA....62..219S} as it was outside the range optimised for RR~Lyrae and Cepheids (which are the only variability types published in this data release). The periods of eclipsing binaries were doubled, as the method used in the pipeline was known to recover half of the correct period for most of these stars \citep{2011MNRAS.414.2602D}. (C0 data) }
\label{fig:lightcurve_2}
\end{figure*}

The RF classifier was trained with a training set containing {\it Gaia} $G$-band time series of 2562 RR~Lyrae variables + Cepheids combined in one group (to enhance the completeness of both targeted classes and not let the classifier penalise misclassifications between them), 71 $\delta$ Scuti stars, 844 eclipsing binaries, 12 ellipsoidal binaries, 1711 LPV variables, and 117 quasars, all containing at least 20 points.
All of the training sources were previously known variables in the OGLE-IV, OGLE-III \citep{2008AcA....58...69U}, EROS-II \citep{2014A&A...566A..43K}, and Half-Million Quasar \citep{2015PASA...32...10F} catalogues.
However, contrary to the GM and BN classifiers, no requirement was imposed on whether the {\it Gaia} time series of the target shows the correct periodicity (the time sampling produced by the {\it Gaia} Scanning Law may be insufficient for an accurate determination of the frequency).
The expected advantage is that the RF classifier learns that members of each variability type can be correctly characterised by the {\it Gaia} pipeline but also may suffer from processing artifacts such as aliasing or unexpected noise characteristics.
Recognizing the true but unclear cases helps towards completeness.
On the other hand, a significant amount of unclear cases in the training set makes the distinction between the different variability classes more fuzzy, working against a low contamination.
In addition to the variable stars, the training set contains a subset of 436 constant stars, selected from the crossmatch results with match probability $\geq$99\% (see Sect.~\ref{sect:crossmatch}).
Sample light curves of periodic variable stars are illustrated in Figs.~\ref{fig:lightcurve_1} and \ref{fig:lightcurve_2}.


\subsubsection{The classification attributes}\label{sec:C0clasAttributes}

The GM and BN classifiers use the same top 7 classification attributes that were found the most effective by \citet{2011MNRAS.414.2602D}:
\begin{enumerate}
\item $\logten P$: where $P$ is the period found with the unweighted least-squares method \citep{2009A&A...496..577Z}.
\item $G_\mathrm{BP}-G_\mathrm{RP}$: the (possibly reddened) median {\it Gaia} colour index.
\item $\phi_{21} \equiv \psi_{1,2} - 2 \psi_{1,1}$ where the phases $\psi_{n,k}$ are defined in Eq.~(\ref{tsmodel}).
\item $k_3$: a robust measure of the skewness, computed as $(P_{95}+P_{5}-2\,P_{50}) / (P_{95}-P_{5})$ where $P_n$ is the $n$-th percentile.
\item $\sigma_{\rm res}/\sigma_{\rm raw}$: the ratio of the median of the residuals and the median absolute deviation of the raw time series.
\item $\logten a$: where $a$ is the peak-to-peak amplitude of the best-fit model to the light curve.
\item $\logten(1+R_{21})$, with $R_{21}=a_2/a_1$, where $a_2$ and $a_1$ are the amplitudes of the 2nd and 1st harmonics of the first frequency, respectively.
\end{enumerate}
Unless specified otherwise, all attributes were computed using \textit{G}-band time series.

The RF classifier, using more variability classes in its training set, requires more attributes to distinguish them. Moreover, unlike for the
GM and BN classifiers, the attributes $\logten a$ and $\logten(1+a_2/a_1)$ do not appear among the best ones as not relevant to 
constant stars and (non-periodic) quasars. Instead, the standard deviation $\sigma_{\rm raw}$ of the raw time series was found to be a more
suitable proxy. In addition to attributes 1--4 listed above, the RF classifier, also uses:
\begin{enumerate}
\setcounter{enumi}{4}
\item $\sigma_{\rm raw}$: the unweighted standard deviation of the raw time series.
\item $J_S$: the single-band Stetson variability index \citep{1996PASP..108..851S} pairing observations within 0.1 days.
\item $\logten\chi^2_{\rm QSO}$: the logarithm of the reduced $\chi^2$ with respect to a parametrized quasar variance model (see \citet{2011AJ....141...93B}), using parameter values corresponding to the SDSS \textit{g} band at magnitude 19.
\item $P_{\rm non-QSO}$: a quantity proportional to a symmetric Beta distribution for the null-hypothesis distribution of $\chi^2_{QSO}$ given the data, proportional to $P(\chi^2_{QSO}|x,\mbox{not quasar})$ in \citet{2011AJ....141...93B}.
\item $\logten(\chi^2_{\rm false}/\nu)$: the logarithm of a potential false alarm defined in \citet{2011AJ....141...93B}.
\item $\sigma_{\rm p2p}$: median of the absolute values of the differences between successive magnitudes, normalised by the median absolute deviation around the median \citep[see][]{2011MNRAS.414.2602D}.
\item $\mathcal{A}$: the Abbe value \citep{vonNeumann1941, vonNeumann1942}.
\end{enumerate}

Figures~\ref{fig:trainingsetRF} and \ref{fig:trainingsetGMBN} in the
Appendix show the distribution of these attributes for the training
sets described above.

\subsubsection{Setup and assessment of the classifiers}

\begin{figure}
\centering
\includegraphics[angle=0,width=0.48\textwidth]{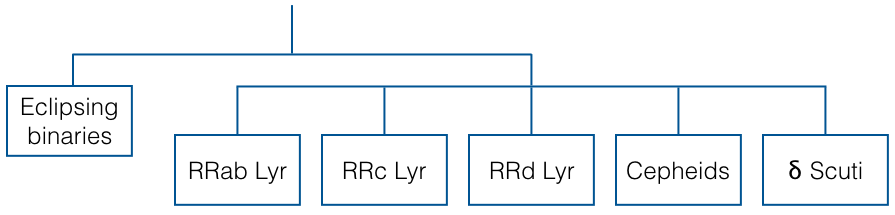}
\caption{The multi-stage classification tree used with Bayesian Networks in each node.}
\label{fig:multistagetree}
\end{figure}

For the BN classifier, a multi-stage tree was set up. Starting from the complete set of all variability classes, each node in the tree consists a classifier that separates branches of one or more variability classes.
These separations proceed down to the lowest level nodes that branch into a single variability class.
Using the chain rule of conditional probabilities, the final probability of a source belonging to a particular variability class is then the product of all conditional probabilities computed in each node that leads to this variability class.
The advantage of this multi-stage approach is that one can tailor the attributes in each node depending on the variability classes that are split off, eliminating uninformative attributes that may cause confusion.
This avoids overfitting and degraded classification performance. The multi-stage tree used to produce the results described later in this section is shown in Fig.~\ref{fig:multistagetree}, where a BN classifier is used in each node.
It turns out that in this case, because of the small number of variability classes and attributes, only the eclipsing binaries are worth splitting off in a separate branch.
We tested whether the GM classifier would also benefit from this approach, but this turned out not to be the case: the tree did not significantly outperform a single-stage classifier fitting a Gaussian mixture to each variability class.

\begin{figure}
\centering
\includegraphics[angle=0,width=0.4\textwidth]{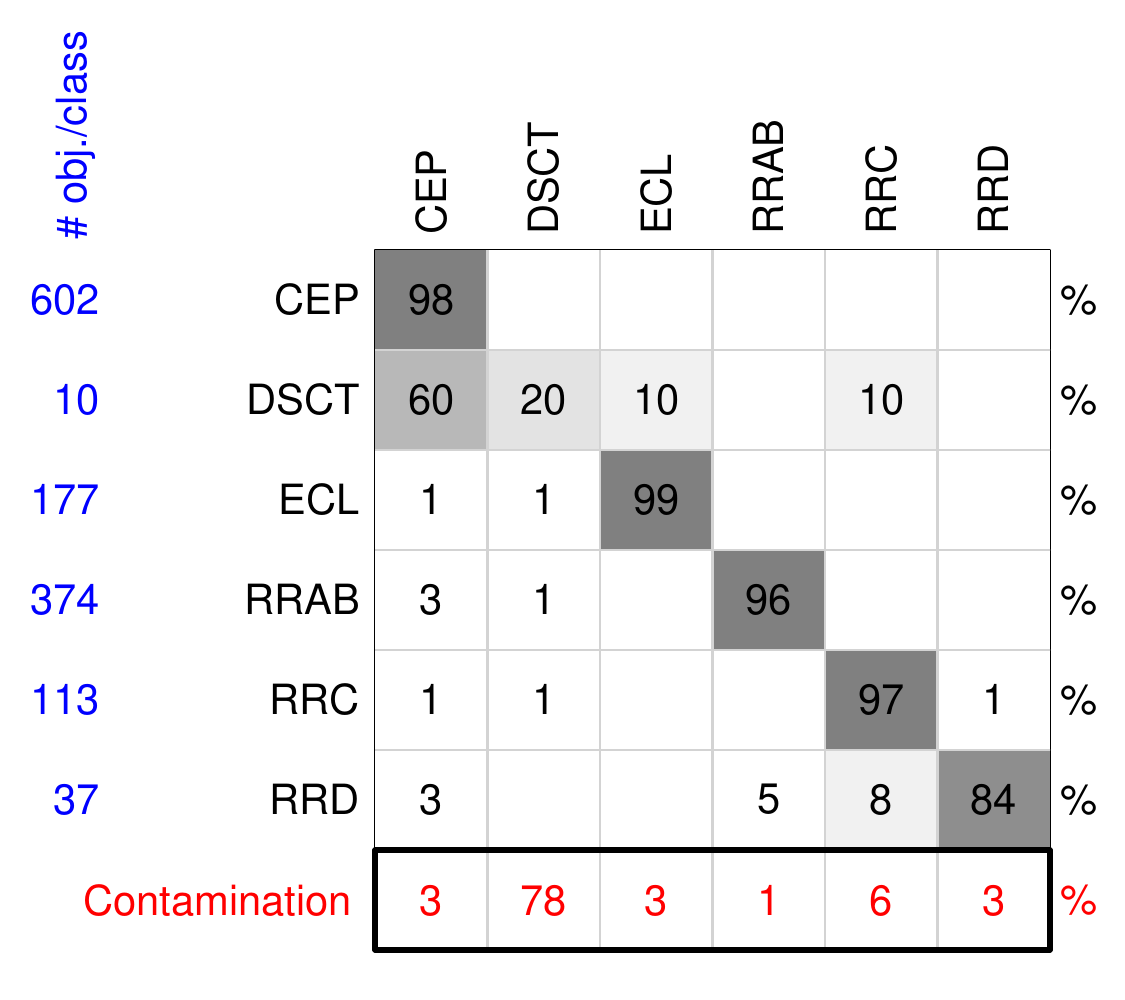}\\
(a) Gaussian Mixture
\includegraphics[angle=0,width=0.4\textwidth]{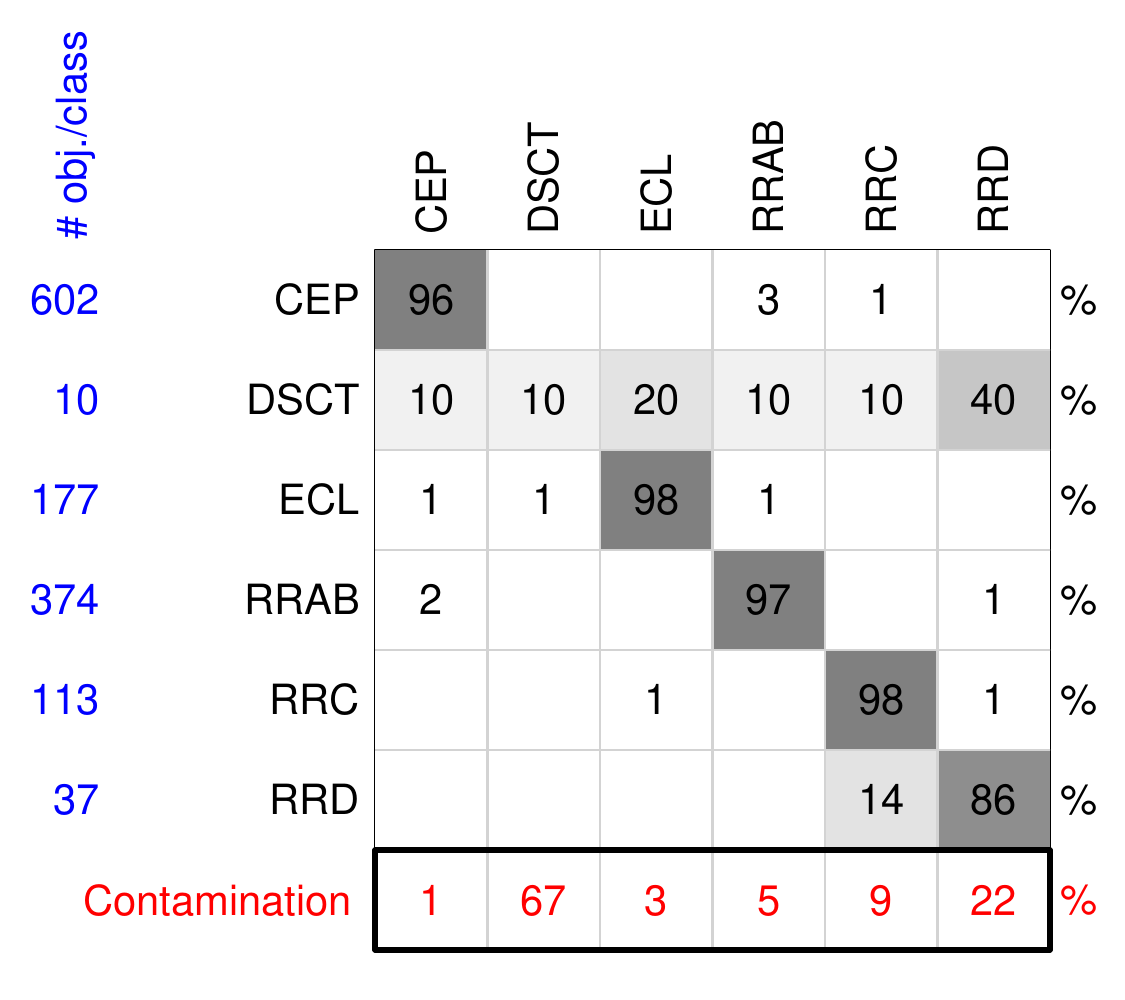}\\
(b) Bayesian Network
\includegraphics[angle=0,width=0.4\textwidth]{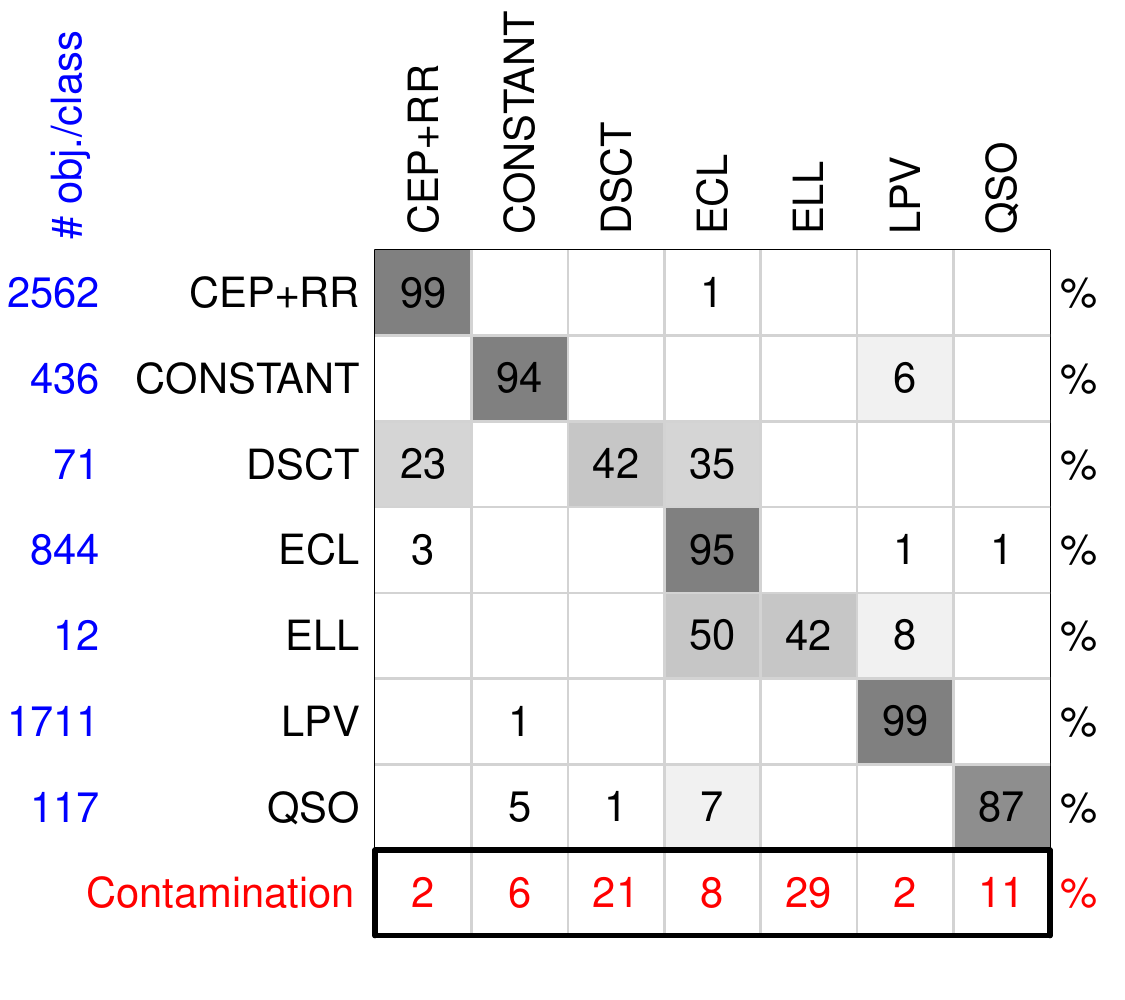}\\
(c) Random Forest
\caption{Confusion matrices of the training sets of the three classifiers (by 10-fold cross-validations and with no missing attributes) on C0 data: (a) Gaussian Mixture, (b) Bayesian Network, and (c) Random Forest. Class labels per row and column denote trained and classified variability types, respectively. For example, the GM classifier classified 3\% of the RRab training examples erroneously as Cepheids. Note that we rounded to the nearest integer percentage, which sometimes results in percentages not adding up exactly to 100\%. }
\label{fig:confMatrix}
\end{figure}

We assessed each of our 3 classifiers using 10-fold cross-validation.
The resulting confusion matrices are listed in Fig.~\ref{fig:confMatrix}, which shows that the classification performance for the target classes of RR~Lyrae variables and Cepheid variables is excellent.

\subsubsection{Classification of the South Ecliptic Pole variables} \label{sec:clasSepVariables}

\begin{figure}[h!]
\centering
\includegraphics[angle=0,width=0.47\textwidth]{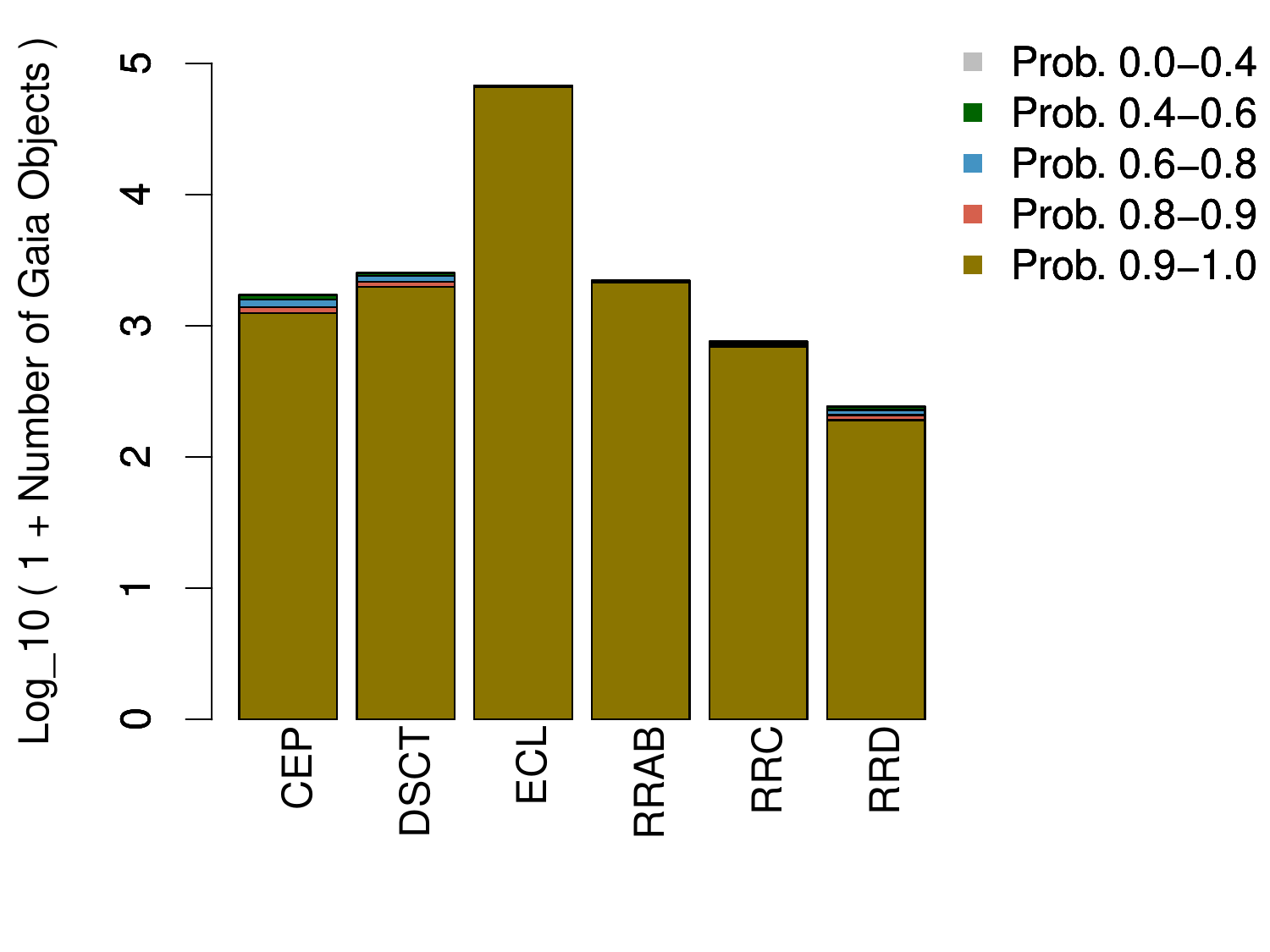}\\
(a) Gaussian Mixture \\
\includegraphics[angle=0,width=0.47\textwidth]{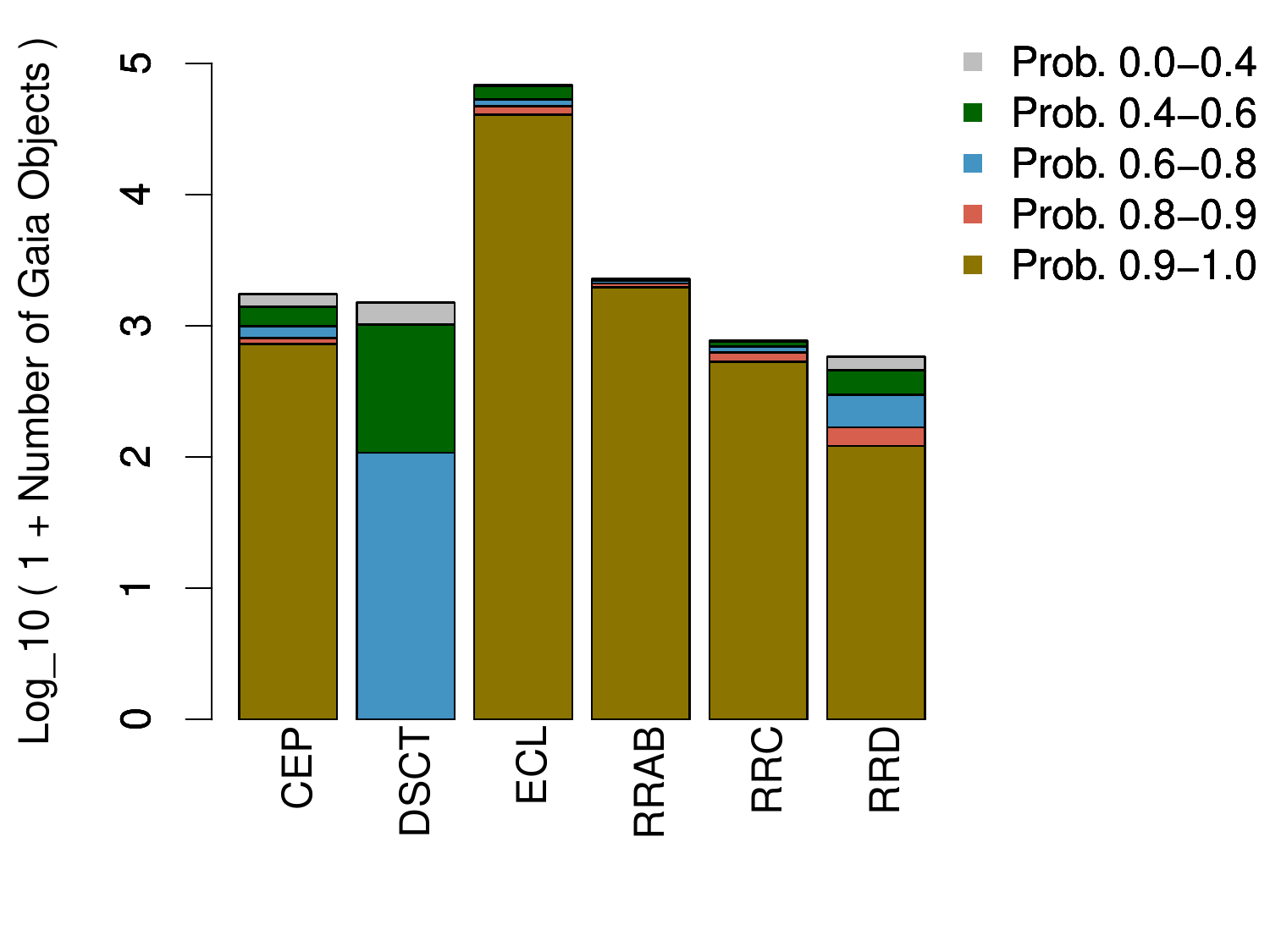}\\
(b) Bayesian Network \\
\includegraphics[angle=0,width=0.47\textwidth]{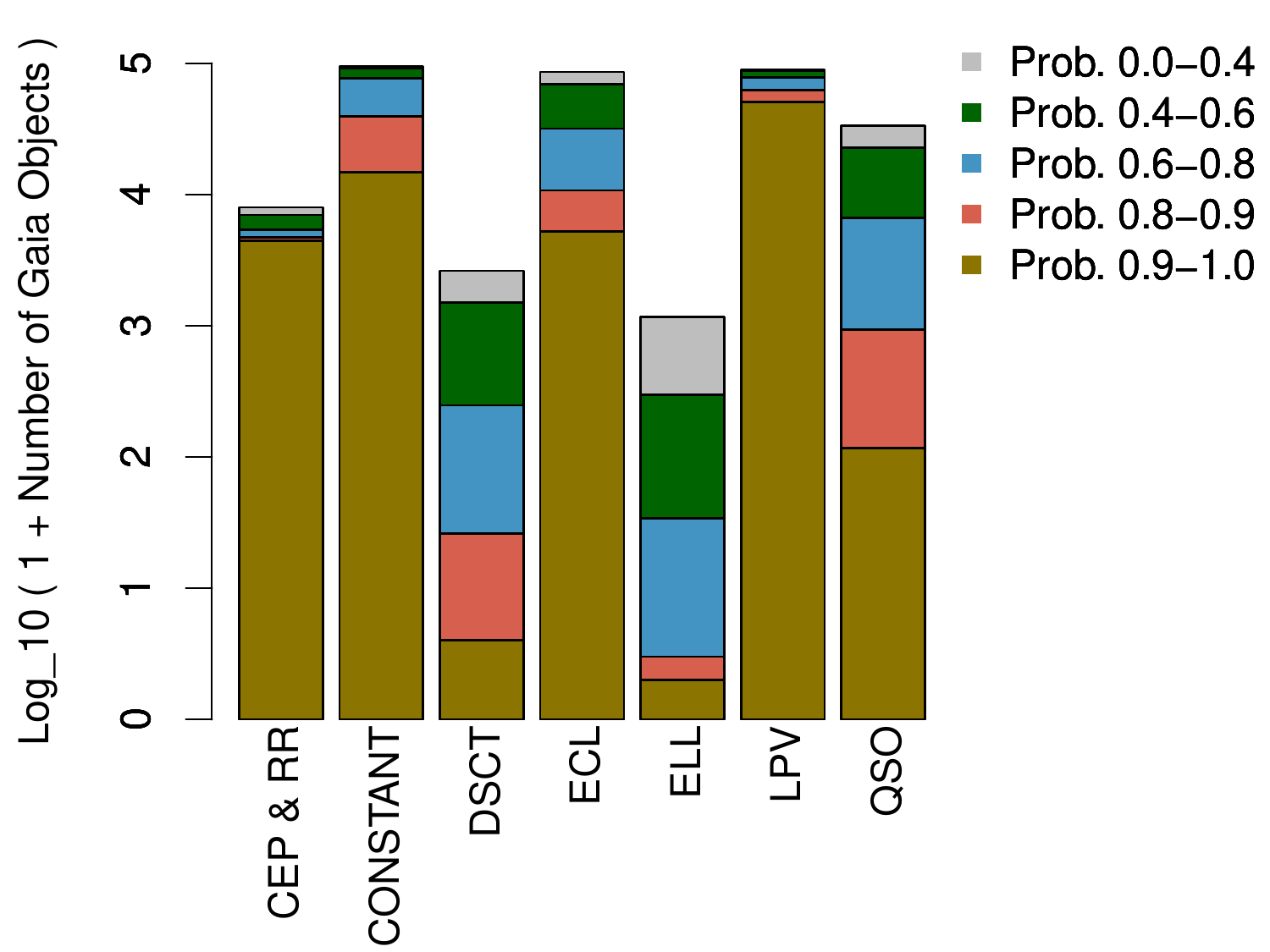}\\
(c) Random Forest
\caption{Supervised classification results per class of the three classifiers on C0 data: (a) Gaussian Mixture, (b) Bayesian Network, and (c) Random Forest, binned per classification probability as indicated in the colour-coded legend.}
\label{fig:class_res}
\end{figure}

We applied the classifiers described in the previous section (one random forest classifier, one single-stage Gaussian mixture classifier, and one multi-stage Bayesian network classifier) to the $315\,808$ variable stars in the SEP region detected by {\it Gaia}.
For each classifier and for each variable star this led to a probability vector that the star is a member of one of the predefined variability classes.
Figure~\ref{fig:class_res} shows for each classifier the resulting histogram of the number of variables per variability class. 
The different colours correspond to different probability bins.
The GM classifier considered most of the variables not a member of one of the included variability classes, and flagged them as outliers.
The remaining stars were classified with a high degree of confidence (with probabilities greater than 0.9).
The same is true for the multi-stage BN classifier, which also ignored the outliers flagged by the GM classifier, but produced lower probabilities for the $\delta$ Scuti variables.
The RF classifier did not include any outlier detection mechanism, and therefore assigned a probability vector to all variables.
Thus, a high probability from the RF classifier does not imply proximity to training set examples in the attribute space.
As expected for a classifier that focuses on completeness rather than a low contamination, many of the derived probabilities are lower, leading to a somewhat fuzzier overall classification.
The confidence in the RR~Lyrae and Cepheid variables is however still very strong.

\begin{figure}[h!] 
\centering
\includegraphics[angle=0,width=0.47\textwidth]{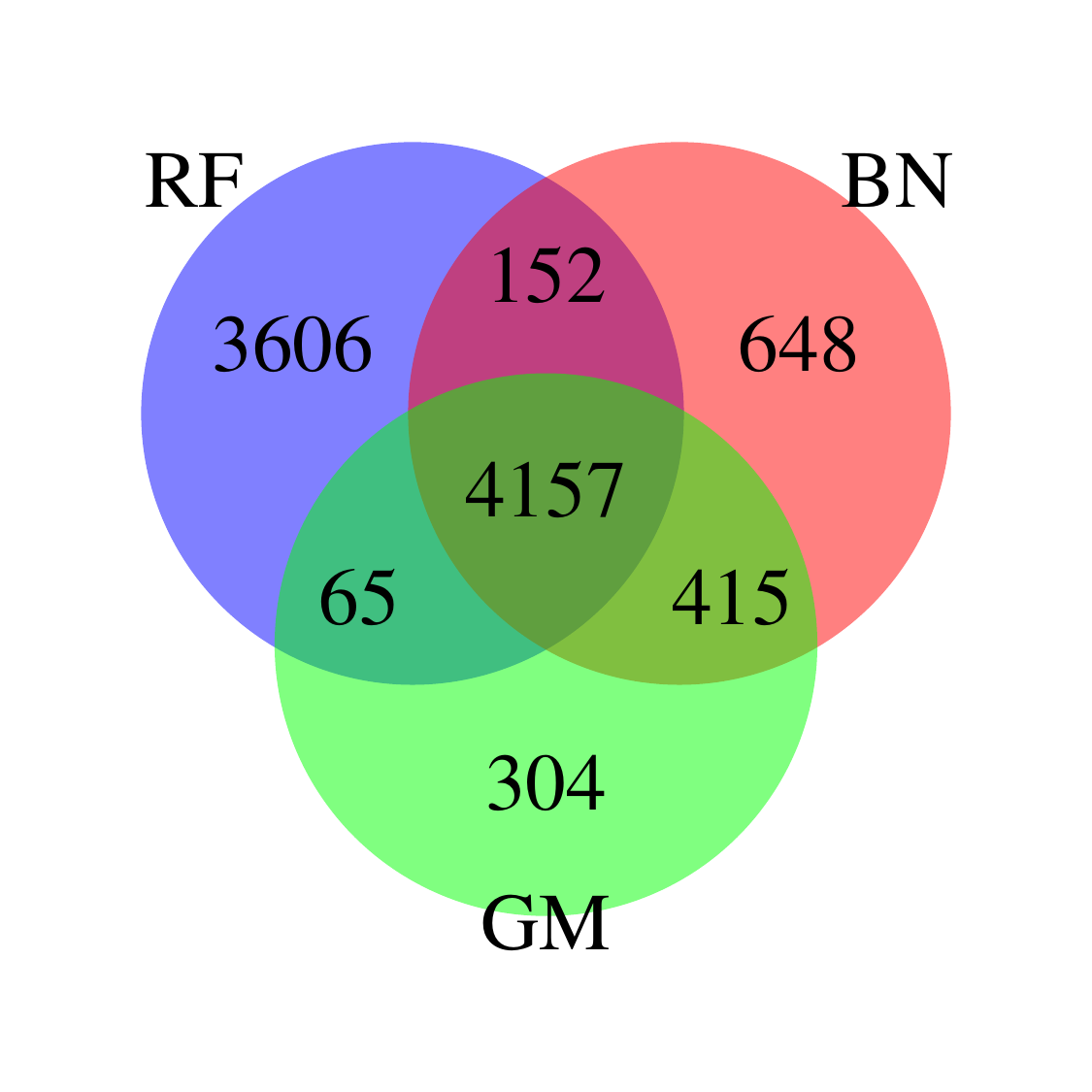}
\caption{Venn diagram of the 9347 sources classified as Cepheid or RR~Lyrae in the SEP region by the three classifiers used, the numbers correspond to those shown in Fig.~\ref{fig:class_res}, and are further subdivided in probability ranges in Table~\ref{tab:KLdistances}. (C0 data)}
\label{fig:sepVennClassifierCepAndRrlNoGDR1}
\end{figure}

To determine to what extent the different classifiers agree on whether a variable star is an RR~Lyrae star or a Cepheid, we first need to set a probability threshold for membership in these variability classes.
Given that the classifiers use different training sets, and/or different setups, this threshold may not be the same for each classifier. 
Focusing on RR~Lyrae variables and Cepheids, the comparison between the different classifiers is summarized in a Venn diagram in Fig.~\ref{fig:sepVennClassifierCepAndRrlNoGDR1}.
This diagram shows the benefit of using more than one classifier to get a global view.
For about half of the total number of classified RR~Lyrae and Cepheid variables, there are at least two classifiers that agree.
Most of the results not shared by other classifiers are due to RF, which focuses on completeness rather than low contamination, and therefore includes many erroneous classifications.
But even if we focus on the stricter GM and BN classifiers, they agree only up to 80\% of their combined total number of RR~Lyrae stars and Cepheids.

\begin{figure}
\centering
\includegraphics[angle=0,width=0.47\textwidth]{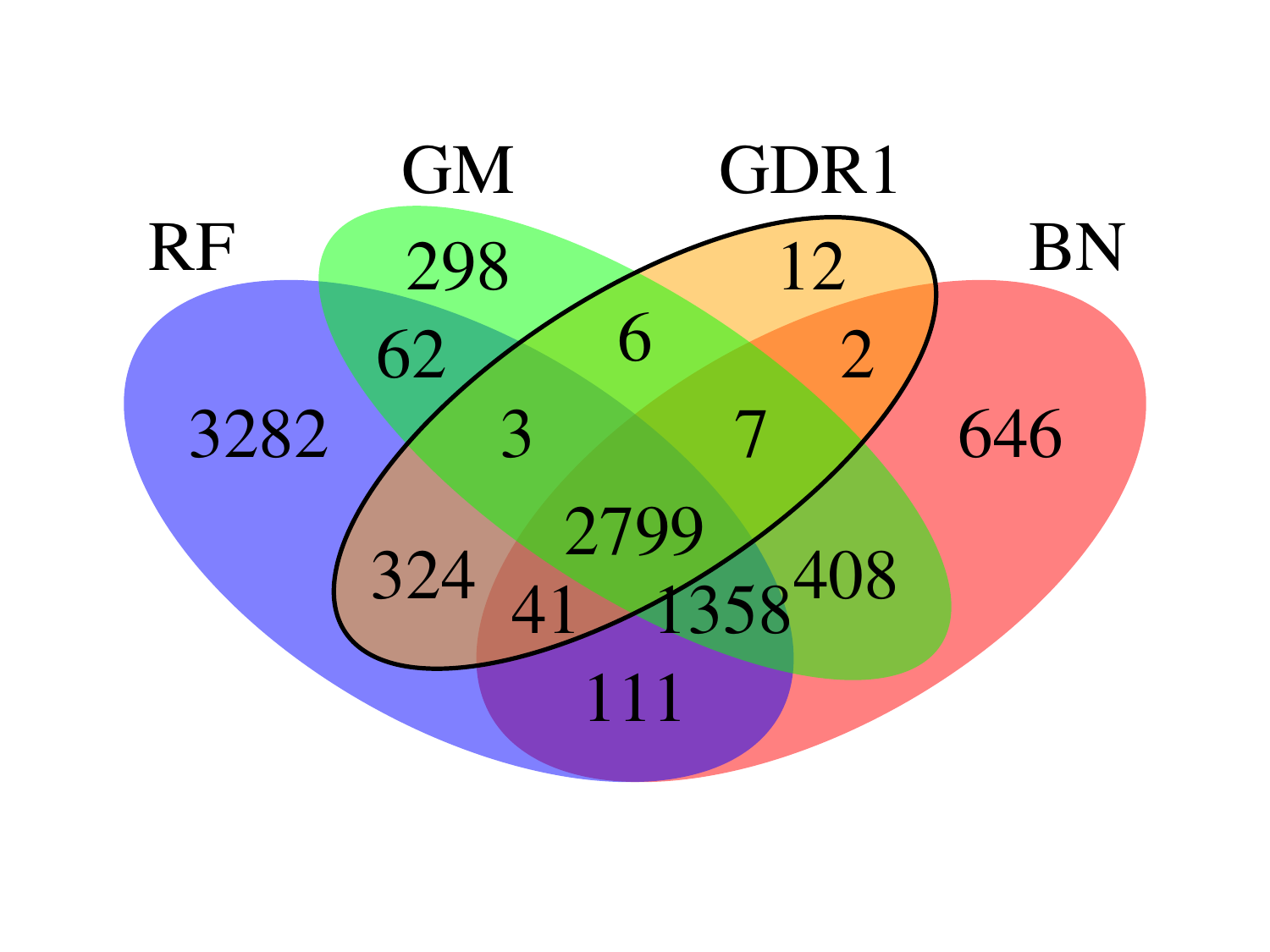}
\caption{Venn diagram of the number of sources classified as Cepheid and RR~Lyrae in the SEP region (as in Fig.~\ref{fig:sepVennClassifierCepAndRrlNoGDR1}) intersecting with the 3194 published {\it Gaia} DR1 sources 
 (see Sect.~\ref{sec:clasSepVariables} for details). 
(C0 data)}
\label{fig:sepVennClassifierCepAndRrl}
\end{figure}

The Venn diagram in Fig.~\ref{fig:sepVennClassifierCepAndRrl} shows how many variables that each classifier labeled as RR~Lyrae or Cepheid ended up in the final {\it Gaia} DR1 variability catalogue.
Note that this catalogue is the end product of the SOS package that reviewed the output of the classifiers, and imposed strict quality control criteria. 
Almost 57\% 
of the variables classified as RR~Lyrae or Cepheid by the GM classifier ended up in the catalogue, 
53\%  
for the BN classifier, and
40\% 
 for the RF classifier. On the other hand, if we would have used only the GM or BN classifiers, the catalogue would be 
 12\% 
 resp.~11\% 
 smaller, while using only the RF classifier would have decreased the size of the {\it Gaia} DR1 catalogue by 
 1\%. 
As shown in Fig.~\ref{fig:sepVennClassifierCepAndRrl}, twelve of the variable stars in this catalogue are not classified by any of the classifiers, as the automated characterisation (period search and/or harmonic fitting) was off target, leading to classification attributes that are too far from the training examples.\footnote{Seven of these objects were added a posteriori from a crossmatch with the OGLE-IV and EROS-II catalogues, and other five ones were picked up from visual inspection of the classification outliers. In fact, as part of a larger inspection of the classification results to be used for SOS~CEP\&RRL, an enlarged set of $19\,923$ Cepheid and RR~Lyrae candidates was created, in which $10\,576$ outliers from Bayesian Network were included on top of the 9\,347 non-outlier candidates of all three classifiers \citep[for details, see][]{DPACP-13}. }

\section{C1: South Ecliptic Pole data processing
\label{sec:C1SepResults}}
The SEP (South Ecliptic Pole) data processing followed the processing outlined in Fig.~\ref{fig:sepProcessing}, which was split up in two successive data sets C0 and C1 described in Sect.~\ref{sec:theData}.
This section presents the results of the processing of the C1 data set selected by \citet{DPACP-13}, which consists of the  3\,194 Cepheids and RR~Lyrae stars published in the {\it Gaia}~DR1.

\subsection{Identification of C0 Cepheid and RR~Lyrae candidates in C1 data} \label{sec:C0toC1crossMatch}
The C0 and C1 datasets result from different photometric calibration runs of CU5.
The difference is not limited to the time span and number of time series, but also to changes of source identifiers, at least for the objects not included in the Initial Gaia Source List \citep[which matches {\it Gaia} sources with known objects from a compilation of catalogues, see][]{2014A&A...570A..87S}.
Thus, the sources identified in the C0 data as Cepheid and RR~Lyrae candidates were crossmatched with the C1 counterparts with positional and photometric information, according to the method described in \citet{ADASS_crossmatch}.
About 8\% of C0 stars could not be found in the C1 data, possibly as a result of stricter quality criteria at earlier stages.

\subsection{Statistics (C1)}  \label{sec:C1resultStatistics}
\begin{figure*}
\centering
\includegraphics[width=\textwidth]{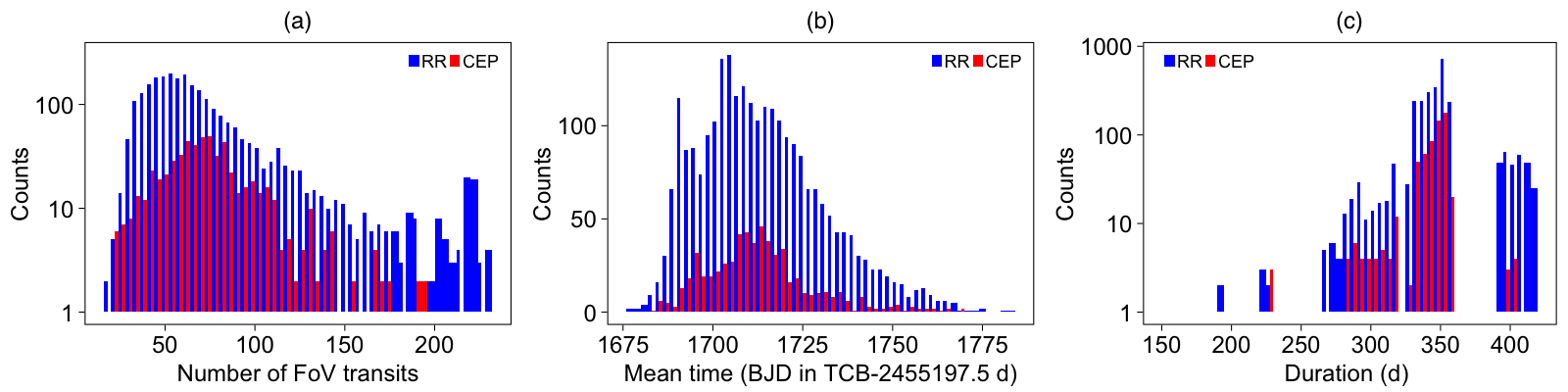}
   \caption{Histograms showing characteristics of the time sampling for the selected 599 Cepheid and 2595 RR~Lyrae candidates (C1 data). 
   (a)~Distribution of the number of FoV transits, (b)~Distribution of the mean observation time, (c)~Distribution of the time series duration.
   }
  \label{fig:stats_sampling}
\end{figure*}

\begin{figure}
\centering
\includegraphics[width=0.40\textwidth]{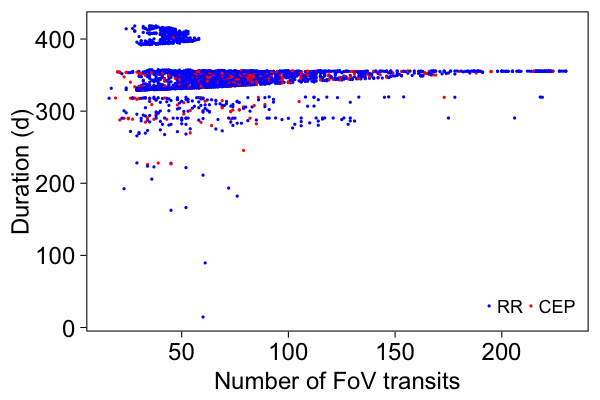}
   \caption{Time series duration as a function of the number of FoV transits. (C1 data)}
  \label{fig:stats_sampling_durvsnfov}
\end{figure}

Figure~\ref{fig:stats_sampling} presents distributions of the time sampling  of the final Cepheid and RR~Lyrae variables. 
From the initial cut of sources with $\ge$20 measurements (Sect.~\ref{sec:dataUsedInGdr1}) and the subsequent application of filters (Sect.~\ref{sec:obsFiltering}), some of the time series remain with less than 20 observations.
A minimum of 16 and a maximum of 230 observations is seen for all final Cepheid and RR~Lyrae variables, as shown in Figs.~\ref{fig:stats_sampling}(a) and~\ref{fig:stats_sampling_durvsnfov}. 
The median number of observations of 72--75 is similar to what is globally expected after the full 5 years of the {\it Gaia} mission.
Such a high number of observations is due to the EPSL sampling during the first 28 days, which was key to the early processing 
and detection of variable stars. 

Figure \ref{fig:stats_sampling_durvsnfov} shows a plot of the time series duration as a function of the number of FoV transits for the selected 599 Cepheid and 2595 RR~Lyrae candidates (C1 data), from which we see that the vast majority of Cepheid and RR~Lyrae candidates contain observations covering a time duration of over 300 days.  

\begin{figure}
\centering
\includegraphics[width=0.47\textwidth]{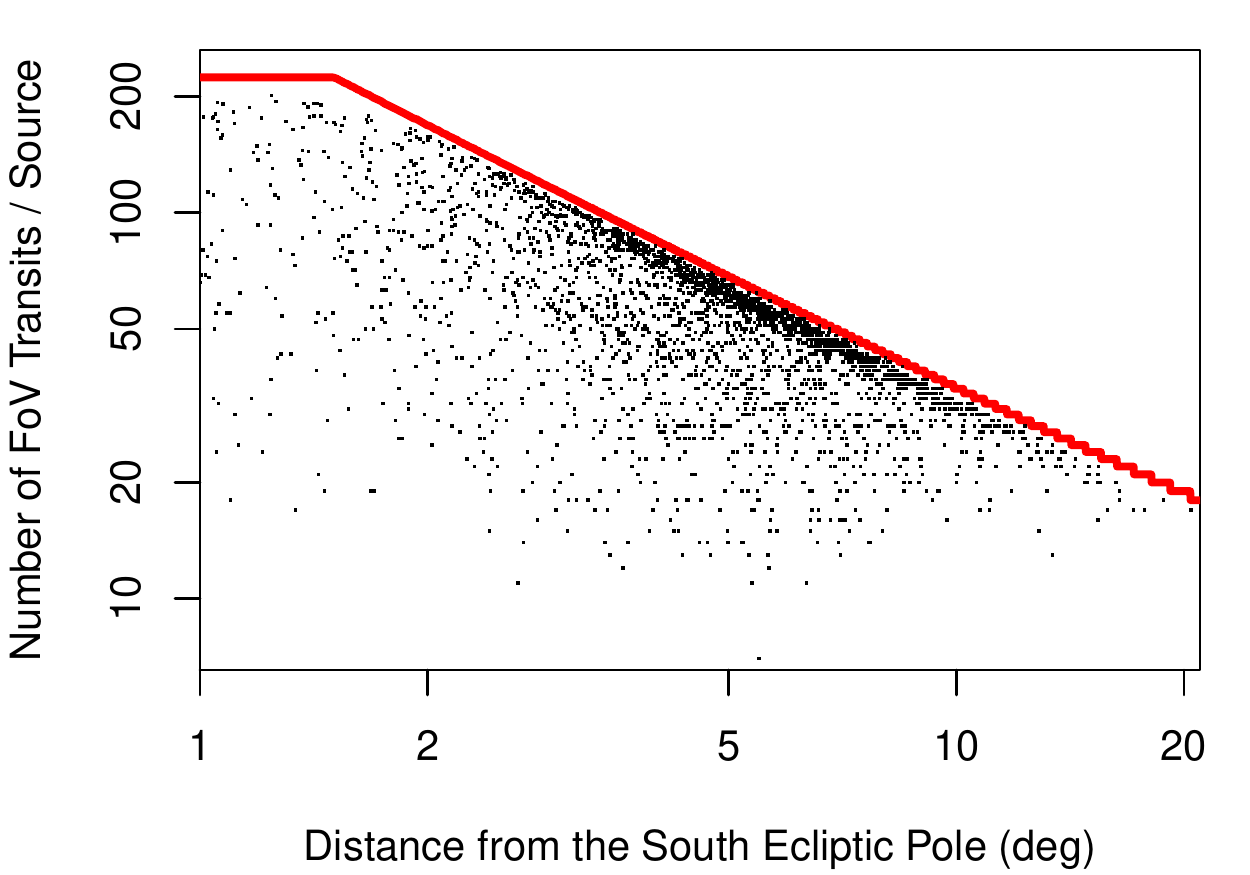}
\caption{Number of FoV $G$-band transits per source as a function of angular distance from the SEP, shown for the published time series and limited to the 28 days of EPSL (until Aug.~22, 2014). The threshold highlighted in red defines the maximum number of observations possible according to the scanning law. (C1 data) }
\label{fig:epsl_numobs}
\end{figure}

The number of processed FoV transits per object was lower than expected from simple scanning law considerations because of several reasons:
 time gaps for decontamination campaigns,\footnote{There are two decontamination campaigns within the published data time range, on Sept.~23-24, 2014 and June~3-4, 2015.}
 FoV refocusing,\footnote{Refocusing of FoV-2 on Oct.~24, 2014 and of FoV-1 on Aug.~3, 2015.} transit rejections by the photometric processing,
 incomplete transit information (photometry or astrometry), outlier filtering,
 and other causes (see Sect.~\ref{sec:obsFiltering}). 
This incompleteness is illustrated in Fig.~\ref{fig:epsl_numobs} by the number of FoV $G$-band observations per source with respect to the expectations from ideal observations (with no data loss) as a function of distance from the SEP (using only time series of the {\it Gaia} DR1 variable stars and limited to the EPSL phase).
However, it is reassuring that the sampling of most sources is rather close to optimal.

\begin{figure*}
\centering
\includegraphics[width=\textwidth]{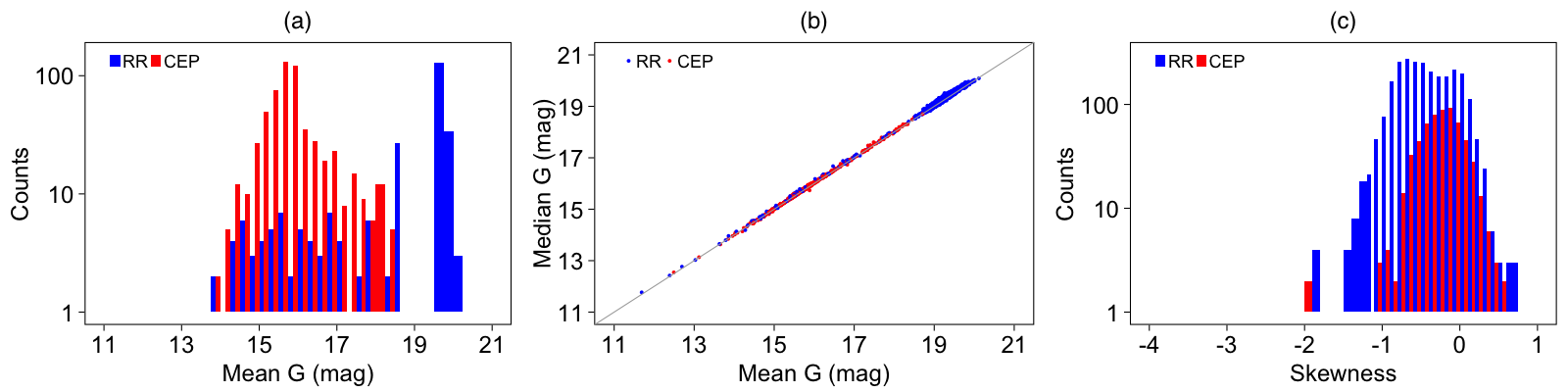}
   \caption{Diagrams showing characteristics of the $G$ magnitude for the 599 Cepheid and 2595 RR~Lyrae variables (C1 data). 
   (a)~Distribution of the mean magnitude, (b)~The median vs the mean magnitude, (c)~Distribution of the skewness of magnitudes.}
  \label{fig:stats_moments}
\end{figure*}

\begin{figure*}
\centering
\includegraphics[width=\textwidth]{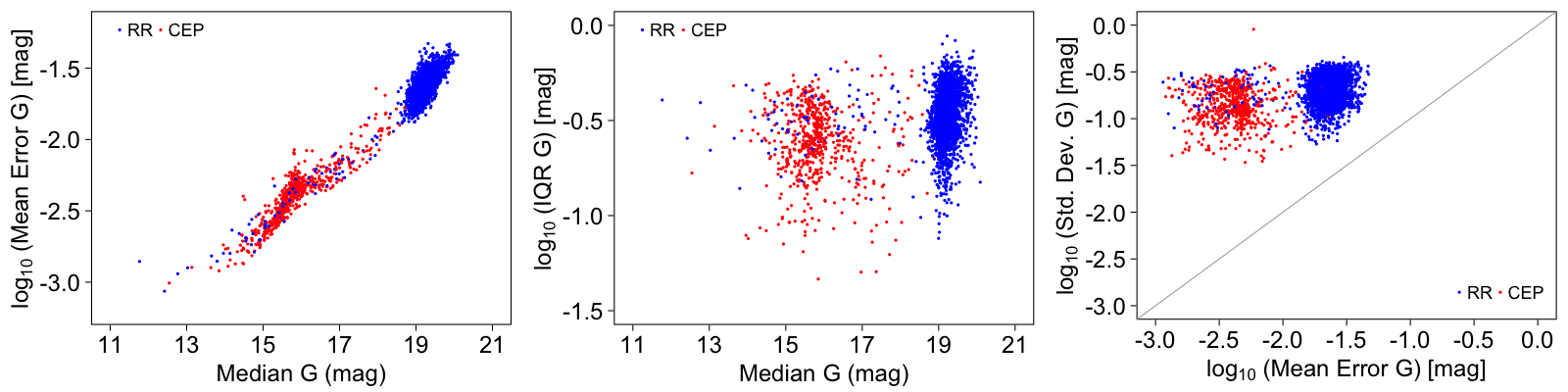}
   \caption{Diagrams showing characteristics of the uncertainty on the magnitude for 599 Cepheid and 2595 RR~Lyrae variables (C1 data). 
   (a)~The mean error as a function of the median magnitude,
   (b)~The IQR as a function of the median magnitude, 
   (c)~The standard deviation as a function of the mean error.}
  \label{fig:stats_uncertainties}
\end{figure*}

Figure~\ref{fig:stats_moments} shows characteristics of the magnitude distribution of the Cepheid and RR~Lyrae  variables. 
The mean magnitude of the Cepheids is centred around 16 whereas the majority of the RR~Lyrae variables are fainter.
Figure~\ref{fig:stats_moments}(b) shows the median magnitude as a function of the mean and it provides strong  verification that the cleaning operators, described in Sect.~\ref{sec:obsFiltering}, worked well and that the time series do not contain any remaining extreme, unphysical outliers. 
The mean of the magnitude skewness distribution in Fig.~\ref{fig:stats_moments}(c) is confirmed to be negative, as expected from the light-curve shape of most Cepheids and RR~Lyrae stars.
%
Distributions of the mean uncertainties and of the time-series IQR with respect to the median magnitude are presented in Fig.~\ref{fig:stats_uncertainties}, which also shows
that the mean uncertainty of the Cepheid and RR~Lyrae variables is clearly smaller than the time-series standard deviation.
A summary of the range for all the relevant statistical parameters published in the {\it Gaia} DR1 for the Cepheid and RR~Lyrae variables is given in Table~\ref{tbl:statistics}. 
%
\begin{table*}
\centering
\vspace*{1cm}
\caption{ Statistical parameters of $G$-band FoV transits per source, published in the {\it Gaia} DR1 of the final list of all 599 Cepheid and 2595 RR~Lyrae variables published in the {\it Gaia} DR1. (C1 data)} 
\vspace{2mm}
\begin{tabular}{lrrrrrr}
\toprule
Parameter &\multicolumn{2}{c@{}}{1st percentile} &  \multicolumn{2}{c@{}}{99th percentile} & \multicolumn{2}{c@{}}{Median}   \\
\cmidrule(l){2-7}
 & Cepheid & RR~Lyrae & Cepheid & RR~Lyrae & Cepheid & RR~Lyrae \\ 
\midrule
  Number of FoV transits & 16.00 & 19.00 & 230.00 & 224.00 & 71.97 & 75.57 \\ 
  Time duration (d) & 14.51 & 226.04 & 418.20 & 405.95 & 348.92 & 343.21 \\ 
  Mean observation time (d) & 1677.41 & 1682.95 & 1782.13 & 1768.61 & 1713.99 & 1712.80 \\ 
  Min. magnitude & 14.83 & 13.87 & 19.39 & 18.02 & 18.72 & 15.44 \\ 
  Max. magnitude & 15.50 & 14.27 & 20.22 & 18.63 & 19.47 & 15.96 \\ 
  Mean magnitude & 15.30 & 14.03 & 19.79 & 18.27 & 19.16 & 15.73 \\ 
  Median magnitude & 15.50 & 14.27 & 20.22 & 18.63 & 19.47 & 15.96 \\ 
  Magnitude range & 0.30 & 0.14 & 1.32 & 1.13 & 0.74 & 0.44 \\ 
  Standard deviation (mag) & 0.08 & 0.04 & 0.37 & 0.32 & 0.20 & 0.14 \\ 
  Skewness of magnitudes & -1.25 & -0.99 & 0.34 & 0.78 & -0.44 & -0.25 \\ 
  Kurtosis of magnitudes & -1.61 & -1.61 & 1.72 & 1.13 & -0.96 & -1.24 \\ 
  MAD (mag) & 0.10 & 0.05 & 0.39 & 0.35 & 0.22 & 0.17 \\ 
  Abbe & 0.74 & 0.07 & 1.39 & 1.18 & 1.08 & 0.26 \\ 
  IQR (mag) & 0.14 & 0.07 & 0.63 & 0.55 & 0.34 & 0.25 \\ 
\bottomrule
\end{tabular}
\label{tbl:statistics}
\end{table*}

\subsection{Variability characterisation (C1)} \label{sec:C1Characterization-results}
The sources classified as Cepheids or RR~Lyrae in C0 (see Sect.~\ref{sec:supervisedClassification}) were identified in C1 data and processed by the Characterisation pipeline, as described in Sect.~\ref{sec:C0resultStatistics} but with a few modifications.
Here the upper search frequency was fixed to 3.9d$^{-1}$ rather than 6d$^{-1}$ as used in C0 processing, in order to avoid the parasite frequency at 4.0d$^{-1}$ caused by the spacecraft rotation and because there was no need to scan shorter periods, as the present study focused on Cepheid and RR~Lyrae variables.
As for C0 data, the processing and subsequent modelling of C1 photometry was limited to mono-periodic models. 

\begin{figure}[h]
\centering
 \includegraphics[width=0.45\textwidth]{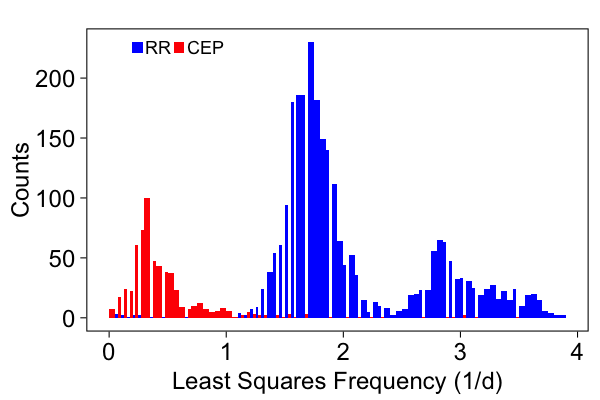}
   \caption{Histogram of the frequencies recovered using the Least Squares period search method \citep[see][]{2009A&A...496..577Z} for all 3194 processed sources. (C1 data)}
    \label{fig:char_hist_frequency_all_recovery}
\end{figure}

Figure~\ref{fig:char_hist_frequency_all_recovery} shows a histogram of the recovered frequencies for all 3194 processed Cepheid and RR~Lyrae variables.
The distribution of Cepheid variables is centred around a period of $\approx$3~days.
The distribution of RR~Lyrae variables presents 2 peaks, one around a period of half a day and the other around 1/3 of a day, comprising RRab and RRc subtype variables, respectively. 

\begin{figure}
\centering
 \includegraphics[width=0.45\textwidth]{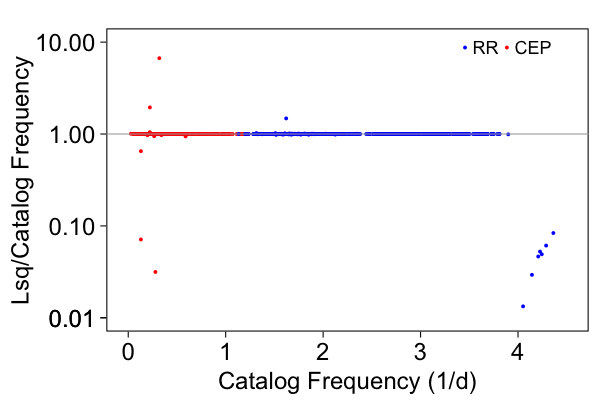}
    \caption{Scatter plot of the ratio of the recovered frequencies using the Least Squares period search method \citep[see][]{2009A&A...496..577Z} to the corresponding OGLE-IV GSEP frequency as a function of the OGLE-IV frequency for 2044 validation sources. (C1 data)}
      \label{fig:char_scatter_frequency_recovery}
\end{figure}

\begin{figure}
\centering
 \includegraphics[width=0.45\textwidth]{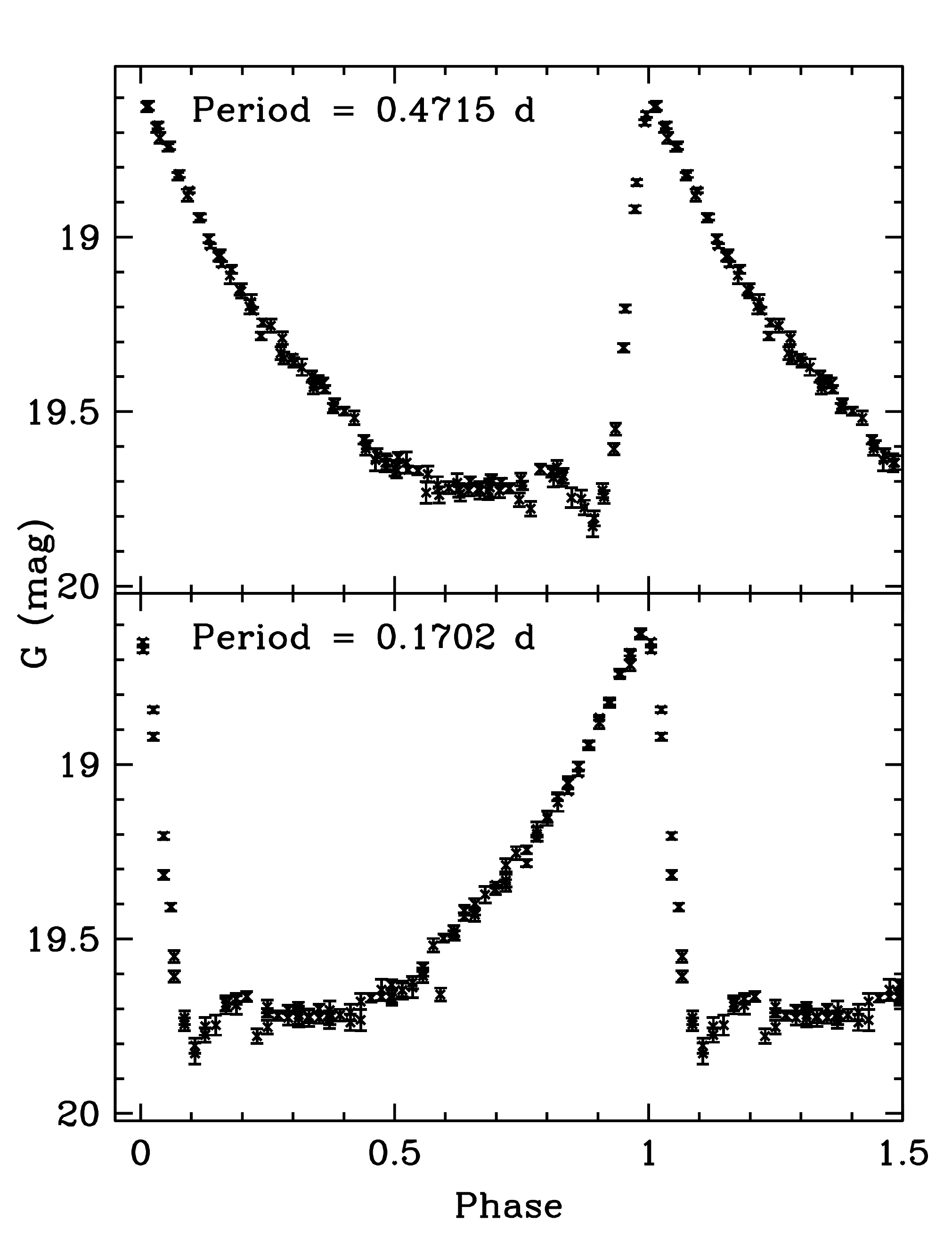}
   \caption{The top panel shows the classical light curve of an RR~Lyrae star of period $P=0.475$~d, with a typical rapid increase in brightness in brightness and a slower decline. The lower panel shows the alias where the arrow of time seems reversed. The alias originates from the fact that the sampling is very regular, so high peaks are present in the spectral window. The peak which provides a confusion results from the convolution of the spectral window with the true signal into negative frequencies and is then mirrored in the positive side, in other words a negative frequency with a positive arrow of time can be seen as a positive frequency with the reverse arrow of time. Three points were removed to make this aliasing effect more conspicuous. (C1 data)}
    \label{fig:AliasedRRL}
\end{figure}

The performance of the Characterisation system was studied using 2044 sources from the OGLE-IV GSEP crossmatch set of variable data (Sect.~\ref{sect:crossmatch}). 
Figure~\ref{fig:char_scatter_frequency_recovery} presents a scatter plot of the ratio of the recovered frequencies (using the Least Squares period search method) to the corresponding OGLE-IV frequency as a function of the OGLE-IV frequency for the 2044 validation sources. 
Because of the selection of Cepheid and RR~Lyrae variables, the {\it Gaia} period deviates from the OGLE-IV values only for a very few variables.
A total of 1940 validation sources have a correctly recovered period (532 Cepheid and 1408 RR~Lyrae variables).
Deviations could in almost all cases be explained by the low number of {\it Gaia} data or by the odd distribution of data from the EPSL combined with the NSL. 
If only the EPSL data were considered, the sampling was sometimes so regular that the alias frequency at 4d$^{-1}-f$ was found instead of the catalogue frequency $f$, resulting in rather surprising light curves. 
These ambiguities disappeared as soon as NSL data were added. In a few cases (in the lower-right hand side of Fig.~\ref{fig:char_scatter_frequency_recovery}), the alias frequency $f-4$d$^{-1}$ was found only because the frequency search was limited to frequencies below 3.9d$^{-1}$. 
The elaborated analysis in the SOS work package could easily remedy these cases. 
An example of alias (wagon-wheel/stroboscopic effect) is illustrated in Fig.~\ref{fig:AliasedRRL}, for the case of an RR~Lyrae star.

\begin{figure}
\centering
  \includegraphics[width=0.45\textwidth]{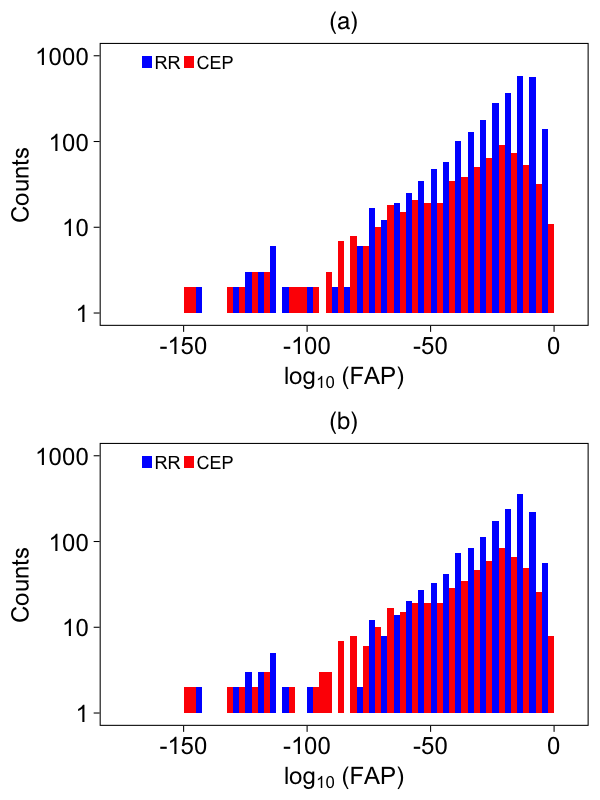}
   \caption{Histogram of the Baluev FAP for Cepheid and RR~Lyrae variables. (a)~All 599 Cepheid and 2595 RR~Lyrae variables.
   (b)~Only those 532 Cepheid and 1408 RR~Lyrae variables which are crossmatched and whose periods agree with the ones in the literature. (C1 data)}
    \label{fig:char_hist_fap}
\end{figure}

Figure \ref{fig:char_hist_fap} shows the distribution of the FAP \citep{Baluev2008} for (a)~all the processed Cepheid and RR~Lyrae variables and (b)~those which are crossmatched and whose periods agree with the ones in the literature. The large majority of the Cepheid and RR~Lyrae stars have a low to very low FAP, only a few have a FAP above 0.01: 14 in the full dataset of 3194 sources and only 2 in the dataset of validation sources with a correctly recovered period. 

\begin{figure}
\centering
\includegraphics[width=0.45\textwidth]{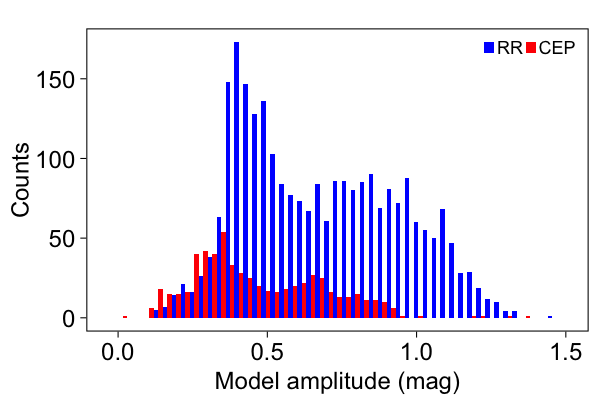}
\caption{Histogram of the peak-to-peak amplitude of the model in the $G$ band. (C1 data)}
\label{fig:char_model_p2p_amplitude}
\end{figure}

\begin{figure}
\centering
\includegraphics[width=0.45\textwidth]{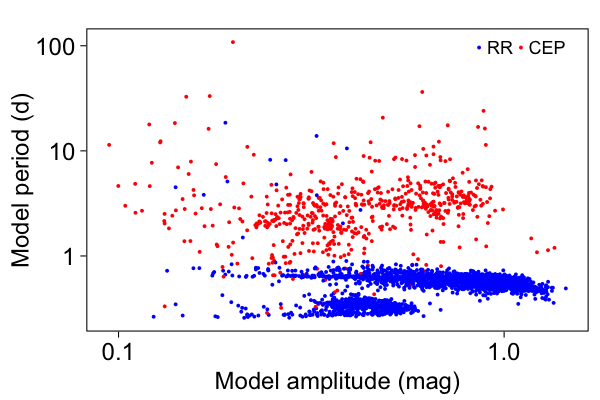}
\caption{Plot of the log of the characterisation model period vs the log of the peak-to-peak amplitude of the model. (C1 data)}
\label{fig:char_logP_p2p_amplitude}
\end{figure}

The most significant period found by the period search process is used as input to the modelling module, which automatically determines the best model to the data and improves the overall fit. 
Attributes such as, but not limited to, the final period and the peak-to-peak amplitude of the model are then used by the classification module.  
Figures~\ref{fig:char_model_p2p_amplitude} and 
\ref{fig:char_logP_p2p_amplitude} show the distribution of the final model peak-to-peak $G$-band amplitude and a log-log plot of the model period as a function of the model amplitude. The latter demonstrates 
how these attributes can be used to separate the Cepheid and  RR~Lyrae variables. 

\subsection{Specific Object Studies Cep\&RRL (C1)} \label{Results:SOS}
The processing of C1 data by the SOS~Cep\&RRL pipeline confirmed, characterised, and classified 2595~RR~Lyrae stars and 599~Cepheids.
The RR~Lyrae stars contain 1910 RRab and 685 RRc type stars.
The Cepheids, on the other hand, include 558 $\delta$~Cepheids, 16 anomalous Cepheids and 25 type~II Cepheids, of which 11 could be classified as BL~Her, 10 as W~Vir, and 2 as RV~Tau.
Double mode stars are not considered in {\it Gaia}~DR1.
The full results are presented in \citet{DPACP-13}, to which the reader is referred for more details.

\subsection{Data Release 1 final selection (C1)} \label{sec:Gdr1Selection}
In this first {\it Gaia} data release, several filters were applied to the data resulting from SOS Cep\&RRL. We rejected some sources due to incomplete light-curve sampling or too many outlying values and/or errors, leading to unreliable source parameter estimates. 
Other objects were rejected due to low signal-to-noise levels or not convincing classifications (e.g., from possibly misclassified eclipsing binaries or $\delta$~Scuti stars). 
While a relevant fraction of excluded sources includes bona-fide RR~Lyrae stars and Cepheids, our exclusion list aims at a high reliability of the published results.
Furthermore, during quality assessment, we decided to postpone all 239 identified double/multi-mode sources to a future data release. 
Additionally, CU9 \citep{DPACP-16} filtered out 169 more sources processed by CU7, due to strict data quality cuts on data released by all CUs contributing to the {\it Gaia} DR1 (CU3, CU5 and CU7), resulting in the 3194 Cepheids and RR~Lyrae stars published in the {\it Gaia} DR1.

\section{Sky completeness estimate
\label{sec:completeness}}
In this section we estimate the 
sky completeness of the identified Cepheid and RR~Lyrae stars in the {\it Gaia} data, by measuring the rate of success in the recovery of known Cepheids and RR~Lyrae stars. 
Note that this {\it sky completeness} is different from the completeness computed for our supervised classifiers (Sect.~\ref{sec:supervisedClassification}), which only measures the recovery rate with respect to the training set.
To measure the sky completeness for {\it Gaia}, one needs to examine a sky region which 
is sufficiently well covered by the {\it Gaia} sampling
and is included in our selection of processed sources. For this release,
this means that all test positions should have received $\geq$20 FoV transits and be located within our defined SEP region ($<38\degr $ from the SEP). The chosen test region with sufficient coverage is defined by  $72.00\degr< \hbox{RA}<95.00\degr$ and ${-66.93}\degr <\hbox{Dec}< {-66.15}\degr$, as outlined in Fig.~\ref{fig:xm_skymap}. 

The reference known stars are taken from the OGLE-IV GSEP RR~Lyrae and OGLE-IV Cepheid catalogues. 
The overlap of these catalogues with our test region includes 231 and 389 stars, respectively. 
With a maximum $V$-band mean magnitude of 19.2 and 19.7, respectively, all sources in these samples should be detectable in the {\it Gaia} data. 
The fact that the test region extends beyond the coverage of the respective catalogues is not an issue because we are only interested in the recovery fraction of the stars selected from these catalogues. 
Assuming the detection efficiency of {\it Gaia} is independent of the location in a sufficiently covered area
\footnote{To test this assumption, the completeness analysis was repeated with a test region restricted to $80.00\degr < \hbox{RA}<90.00 \degr $,  which reduced the overlap with the reference star catalogues roughly by half, resulting in a reference star sample of 131 RR~Lyrae and 170 Cepheid, i.e., about 57\%  and 44\% of the original test samples, respectively. The resulting completeness percentages change by less than 2\% throughout Table~\ref{tab:completenessOverview}, indicating that the assumption is valid within these limits.}, the exact size and position of the test region within it are not critical, although the test area should of course include a sufficient number of reference sources. 

The first row in Table~\ref{tab:completenessOverview} shows the number of reference stars in the test region. 
In the following steps, we compute the numbers of recovered {\it Gaia} objects and express percentages with respect to these OGLE-IV star counts.

The second row in Table~\ref{tab:completenessOverview} lists the number of sources in the test region with at least 20 FoV observations found with the crossmatch procedure described in Sect.~\ref{sect:crossmatch} and illustrated in Fig.~\ref{fig:xm_skymap}.
It shows that almost all OGLE-IV Cepheids in the test region, and about 93\% of the OGLE-IV GSEP RR~Lyrae stars, are identified in the {\it Gaia} data. 
This loss of sources is unfortunate but is expected to decrease in the coming years, as at the moment the {\it Gaia} astrometric and photometric calibration and observation rejection criteria are not yet finalised.
Given that in the test region only these crossmatched sources were subsequently processed, this row indicates the maximum number of these two variable types that can potentially be recovered by the following stages.

The third row in Table~\ref{tab:completenessOverview} shows the number of sources classified as either Cepheid or RR~Lyrae by one of the classifiers.
All sources in the test region are correctly identified by the various classifiers, except for one RR~Lyrae star misclassified for reasons explained in Section~\ref{sec:clasSepVariables}.
Note that most of the test region sources were part of the training sets, and therefore this should not be interpreted as a measure of completeness of predicted Cepheid and RR~Lyrae types by our classifiers
(see Sect.~\ref{sec:supervisedClassification} for a 10-fold cross-validation of the training sets).

The fourth row in Table~\ref{tab:completenessOverview} demonstrates the unfortunate loss of about 10\% of the sources in our test region due to the incomplete match between dataset C0 and C1, as described in  Sect.~\ref{sec:C0toC1crossMatch}.

Of all Cepheids and RR~Lyrae sources identified by the SOS Cep\&RRL module, more than 400 sources were finally excluded due to several criteria described in Sect.~\ref{sec:Gdr1Selection}.
This results in sky completeness (recovery) rates of 58\% for RR~Lyrae stars and 67\% for Cepheid stars of the corresponding OGLE-IV catalogues, as shown in the last row of Table~\ref{tab:completenessOverview}.
In particular, the exclusion of the 239 double mode sources identified in the C1 data, accounts for a loss of 8\% of the RR~Lyrae and 5\% of the Cepheid stars in the test region, which were still present in the `SOS Cep\&RRL + cleaning' results (the fifth row in the table).

The completeness estimates found by \citet{Udalski16} are consistent with our results. The low completeness should however not be a reason for concern as it is a natural consequence of the various selections explained above, the most severe of which will significantly decrease in future data releases.

\begin{table}
\begin{footnotesize}
\setlength\tabcolsep{5pt}
 \centering
  \caption{Sky completeness for the sources inside the test region outlined in Fig.~\ref{fig:xm_skymap}. See Sect.~\ref{sec:completeness} for details.
  }
  \vspace{2mm}
\label{tab:completenessOverview} 
  \begin{tabularx}{\columnwidth}{Xrr} 
  \hline
\noalign{\smallskip}
  Set & OGLE-IV-GSEP& OGLE-IV \\
		  & RR~Lyrae  &  Cepheid   \\
\hline
\noalign{\smallskip}
  Reference stars in test region  & 231 (100\%) & 389 (100\%) \\
   - C0:  Number of FoV transits $\geq 20$  & 214 (~~93\%) & 384 (~~99\%)  \\
   -- C0: Class: Cep\&RRL (in 9347) & 213 (~~92\%)   & 384 (~~99\%)  \\
    -- C0 to C1 matching &190 (~~82\%)  &  347 (~~89\%) \\
    --- C1: SOS Cep\&RRL + cleaning   &154 (~~66\%)  &  282 (~~72\%) \\
    ---- C1: Only single mode $\rightarrow$ DR1 &135 (~~58\%)  &  262 (~~67\%) \\
 \hline
\end{tabularx} 
\end{footnotesize}
\end{table}

\section{Global Variability Studies
\label{sec:gvs-data}}

In Sect. \ref{sec:gvs-dataprocessing}, we described the key components of the Quality Assessment module.
Here we summarise and describe the main results of this analysis for the RR~Lyrae and Cepheid candidate sources.

Given the spatial coverage of the EPSL data, we have only considered the OGLE-IV 
catalog of Cepheid stars and the OGLE-III catalog of RR~Lyrae stars as reference catalogs for validating our results based on \textit{Gaia}.
In the following, we compare the lists of Cepheids and RR~Lyrae stars generated by the three automated classifiers discussed in Sect. \ref{DataProcessing:Classification} (C0 data) with the final lists produced by the SOS module (C1 data).
We only take into consideration {\it Gaia} sources classified as Cepheids or RR~Lyrae stars by the automated classifiers with significant first harmonics (see Sect. \ref{DataProcessing:Characterization}), so that the $R_{21}=a_2/a_1$ and $\phi_{21}$ light-curve shape parameters can be calculated. 
Sources where only the fundamental frequency is detected and no harmonics thereof, are assigned null phase differences and amplitude ratios.

We compare the samples in the 3D space of periods, $R_{21}$ and $\phi_{21}$.
While no difference is expected between the periods inferred from time series in the $G$ band and those available in the reference catalogues, the amplitude 
ratio $R_{21}$ and the phase difference $\phi_{21}$ may show different distributions as a result of the different bands used in the two surveys.
In order to check for the presence of such systematic differences and correct for them if needed, we compared the values of these light-curve shape parameters 
in the OGLE-IV catalogue of Cepheid variables and OGLE-III catalogue of RR~Lyrae stars with our own estimates obtained with the {\it Gaia} CU7 processing 
pipeline.
Simple frequentist tests showed no statistical significance for neither the
zero-point offset nor first order terms of any kind in the case of the $R_{21}$
ratio. However, in the case of $\phi_{21}$ we find
statistical evidence of a zero-point offset ($p$-value
smaller than 0.005) in both the Cepheid and RR~Lyrae samples. 
We compared the two sets of differences (for Cepheid and RR~Lyrae
variable stars) and discarded equality of the zero-point offsets, 
again for a $p$-value of 0.005. 

We have adopted a value for the $\phi_{21}$ offsets that minimizes 
the Kullback-Leibler (KL) divergence or, loosely speaking, the 
difference between the PDFs derived from OGLE and {\it Gaia} samples
of each type: 0.24~rad for Cepheid stars and 0.28~rad 
for RR~Lyrae variables. These offsets are consistent (within the 
confidence intervals) with the estimates that would be derived from 
least-squares fits but have the advantage that they result in PDF 
subtractions free from systematics.

The values of the computed KL divergence (including the
offset correction described above) between the reference survey (OGLE)
and the {\it Gaia} samples are summarised in Table \ref{tab:KLdistances} for
the three different classifiers described in
Sect.~\ref{DataProcessing:Classification}, as a function of the
membership probability threshold used to define the samples. 
Throughout all this Section, the results of the RF classifier
refer to the sample of Cepheids plus RR~Lyrae candidates together
since these were not split into separate classes by the classifier.

Table \ref{tab:KLdistances} shows a relatively stable behaviour of the
KL divergences per classifier and across different membership
probability thresholds. This stability is exceptional in the case of
the GM classifier, due to a very stringent removal of
outliers prior to the classification stage and to the distribution of
GM membership probabilities, both described in
Sect. \ref{sec:clasSepVariables}. This produces GM samples that
are almost the same for the several membership probability thresholds
tested. The Cepheid sample size only increases by 4\% (or 3\% for
RR~Lyrae) when we lower
the membership probability threshold from 0.9 to 0.5. 
For comparison, the corresponding values for the BN
classifier are 17\% and 14\% for Cepheid and RR~Lyrae stars
respectively, and 10\% for the RF classifier (for the combined class
of Cepheids and RR~Lyrae stars).



\begin{table*}
 \centering
  \caption{Kullback-Leibler distances between the reference OGLE data
    sets of Cepheids (OGLE IV) and RR~Lyrae (OGLE III) and the data
    sets produced after thresholding the membership probabilities $p$
    yielded by the three classifiers described in
    Sect. \ref{DataProcessing:Classification}. The resulting sample
    size for each threshold is shown in parenthesis. (C0 data)}
  \vspace{2mm}
\label{tab:KLdistances} 
  \begin{tabular}{llcccccc}
  \hline
Classifier & Class & $p>0.5$ &$p>0.6$ &$p>0.7$ &$p>0.8$ &$p>0.9$  \\
\hline
 RF  & RR~Lyrae \& Cepheids  &0.37 (4182) &0.35 (4050) &0.34 (3959) &0.34 (3901) &0.33 (3789)  \\
 BN  & RR~Lyrae \& Cepheids  &0.34 (3552) &0.35 (3471) &0.36 (3397) &0.37 (3293) &0.40 (3088)  \\
 GM  & RR~Lyrae \& Cepheids  &0.34 (3512) &0.34 (3497) &0.34 (3485) &0.34 (3459) &0.34 (3401)  \\
 BN  & Cepheids              &0.63 (~~769)  &0.62 (~~731)  &0.60 (~~704)  &0.59 (~~679)  &0.59 (~~657)  \\
 GM  & Cepheids              &0.57 (~~735)  &0.56 (~~730)  &0.56 (~~727)  &0.56 (~~724)  &0.56 (~~707)   \\
 BN  & RR~Lyrae              &0.29 (2783) &0.30 (2740) &0.30 (2693) &0.32 (2614) &0.36 (2431)  \\
 GM  & RR~Lyrae              &0.28 (2777) &0.28 (2767) &0.28 (2758) &0.28 (2735) &0.28 (2694)  \\
\hline
\end{tabular} 
\end{table*}

\begin{figure*}
\centering
\includegraphics[width=\textwidth]{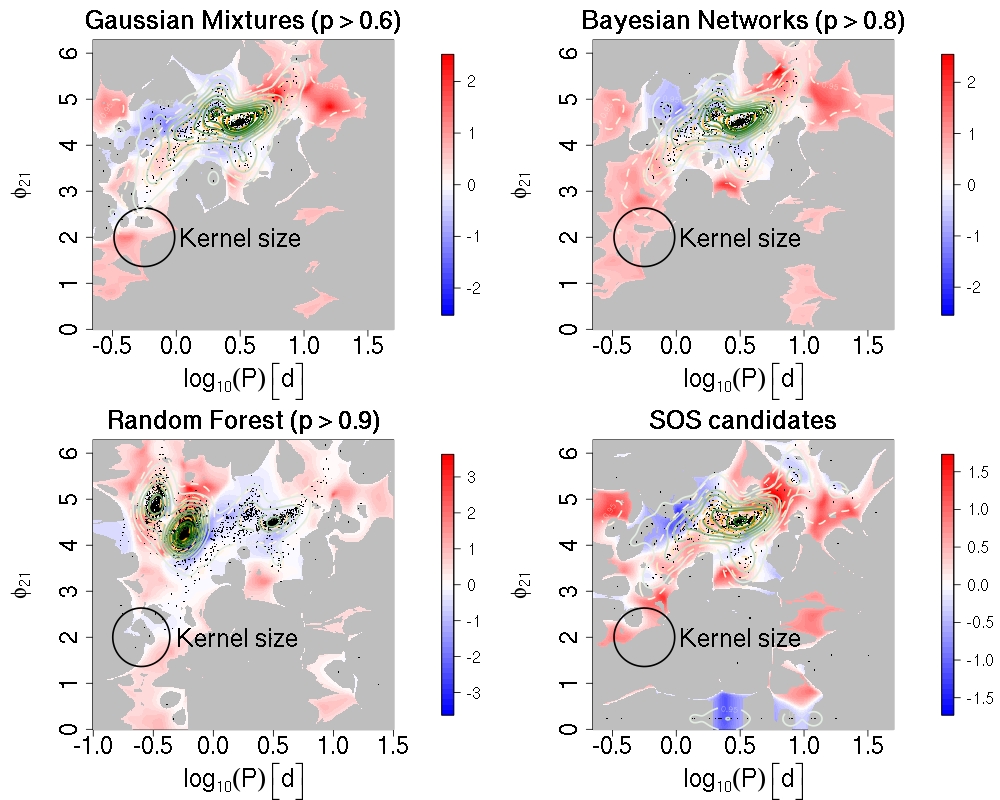}
\caption{Maps of the scaled differences between the kernel density
  estimates of the PDFs that describe the distribution of Cepheids in
  the $\log(P)-\phi_{21}$ space. The sub-panels correspond to the
  Gaussian Mixture classifier with a membership probability threshold
  $p > 0.6$ (left); Boosted Bayesian Networks with $p > 0.8$ (centre-left); the Random Forest classifier with $p > 0.9$ for the PDF of
  Cepheids and RR~Lyrae combined (centre-right, see
  Sect.~\ref{DataProcessing:Classification}); and the SOS sample. (C0 data)}
\label{fig:pdfsubs-ceps-phi21}
\end{figure*}

Figures~\ref{fig:pdfsubs-ceps-phi21} and
\ref{fig:pdfsubs-ceps-r21}--\ref{fig:pdfsubs-rr-phi21} show maps of
the scaled differences between the PDF estimates for the OGLE and {\it Gaia}
samples of Cepheids and RR~Lyrae stars. We show, from left to right,
the results obtained for the  probability thresholds that minimize 
the KL divergence of the GM, BN and RF classifiers with respect to 
the reference OGLE sample, and for the full 
SOS sample. The scaled differences
correspond to the subtraction of the two PDFs divided by the standard
deviation of the estimate as inferred from subsampling (see
Sect. \ref{sec:gvs-dataprocessing}). The colour scale ranges from red
for a {\it Gaia} density deficit with respect to the OGLE reference PDF, to
blue for the opposite situation ({\it Gaia} excess with respect to
OGLE). The grey zones correspond to grid values that did not pass the
normality hypothesis test. We superimpose contour lines of the two
PDFs with a colour scale (orange for OGLE, green for {\it Gaia}) that fades
to white as the probability enclosed by the contour approaches
1. Labels interrupting the contour lines show the probability mass
enclosed by the contour. Finally, we overplot the {\it Gaia} samples with
black circles. Double-mode pulsators of the two variability types 
have been removed from the OGLE reference sample used to validate the SOS candidates 
\citep{DPACP-13} for the sake of consistency. The kernel 
sizes are shown as ellipses that represent the square root of the 
covariance matrices.

Figures \ref{fig:gvs-scatter-cep} and \ref{fig:gvs-scatter-rr} show scatter 
plots of the samples used to estimate the PDFs. Black dots correspond to the 
OGLE reference catalogue and transparent red circles represent the different 
{\it Gaia} samples (the minimum KL divergence GM, BN and RF samples, and the 
final SOS candidates). Again, we have removed double-mode pulsators from the 
OGLE reference sample represented in the rightmost panels for consistency with 
the SOS selection procedure.

Figure \ref{fig:gvs-scatter-rr} in particular shows a prominent lack of 
RR~Lyrae candidates around $\log(P)\approx-0.3$ days (frequencies around 
2 cycles day$^{-1}$ or 12 hours). This effect is related to the data processing 
described in Sect.~\ref{sec:C0Characterization-results}, and will disappear 
in future data releases as the time span increases and the sampling cadence corresponds 
to the nominal scaning law.
Except for this effect, all other systematic differences apparent in Figs. 
\ref{fig:pdfsubs-ceps-phi21} and \ref{fig:pdfsubs-ceps-r21}--\ref{fig:gvs-scatter-rr} 
merely reflect biases introduced in the training sets due to the relative lack 
of examples observed by {\it Gaia} (it should be borne in mind that it is not 
possible to use the classification attributes from other surveys to classify 
{\it Gaia} sources; the colour index or the amplitude are obvious examples 
of attributes that are strongly band dependent).


\section{Conclusions
\label{sec:conclusions}}
We presented the general approach and methods to systematically and automatically analyse the variable phenomena observed by the {\it Gaia} mission.
The analysis was applied for the {\it Gaia} data from the first 14 months of the spacecraft's operations on a subset centred around the South Ecliptic Pole.
This analysis led to the classification into variability types of this region and a further detailed analysis of 3194 variable stars, i.e. 2595 (of which 443 new) RR~Lyrae stars and 599 (of which 43 new) Cepheids presented in a separate article \citep{DPACP-13}.
A comparison with literature showed that the classifiers are performing very well, given the initial {\it Gaia} data.
Their performance will improve when other {\it Gaia} data types (astrometry, spectrophotometry, RV) will be added to the analysis pipeline and when the training sets are improved.

The goal of this analysis was mostly to present a showcase of time-series photometry of the {\it Gaia} mission. The multi-epoch $G$-magnitudes are of very good quality. Still, the quality is expected to significantly improve in the coming years.

Although we processed several millions of stars in an automated way, the detailed inspection and visual examination of subsamples of the data remain necessary to better understand peculiar features of the data.

In future releases, the number of investigated sources will gradually scale up to 1.5 billions and, most importantly, we will introduce other sources of information which render {\it Gaia} unique:
 the integrated $G_{\mathrm{BP}}$ and $G_{\mathrm{RP}}$ magnitudes,
 the BP and RP spectrophotometric data,
 the per-CCD photometry,
 the radial velocities and the RVS spectra (all these quantities can be used as mean or as time series depending on the brightness of the object and on the variability type),
besides the contribution of astrometry, which is a fundamental pillar for stellar astronomy.
On the side of the variability types, we have started to focus on some regular and large amplitudes variables, i.e. Cepheids and RR~Lyrae stars, and in the future releases, we will address variability classes of stars which are not strictly periodic and/or have lower variation amplitudes.
In conclusion, we can assert that {\it Gaia} will provide an absolutely unique variability data set of a space-quality homogeneous all-sky nature.

\begin{acknowledgements}
We would like to express our thanks to Dr Marc Audard for a detailed reading of the manuscript.

This work has made use of data from the ESA space mission {\it Gaia}, processed by the {\it Gaia} Data Processing and Analysis Consortium (DPAC).
Funding for the DPAC has been provided by national institutions, some of which participate in the {\it Gaia} Multilateral Agreement, 
which include, 
for Switzerland, the Swiss State Secretariat for Education, Research and Innovation through the ESA Prodex program, the ``Mesures d’accompagnement”, the ``Activit\'{e}s Nationales Compl\'{e}mentaires”, the Swiss National Science Foundation, and the Early Postdoc.Mobility fellowship;
for Belgium, the BELgian federal Science Policy Office (BELSPO) through PRODEX grants;
for Italy, Istituto Nazionale di Astrofisica (INAF) and the Agenzia Spaziale Italiana (ASI) through grants I/037/08/0,  I/058/10/0,  2014-025-R.0, and 2014-025-R.1.2015 to INAF (PI M.G. Lattanzi);
for France, the Centre National d'Etudes Spatiales (CNES).\\
We gratefully acknowledge Mark Taylor for creating the astronomy-oriented data handling and visualization software TOPCAT
\citep{2005ASPC..347...29T}.
\end{acknowledgements}

\bibliographystyle{aa}
\raggedbottom
\bibliography{local}

\clearpage
\onecolumn

\begin{appendix}

\section{The \textit{\textbf{Gaia}} time sampling \label{sec:Timesampling}}
In this section we summarise the key properties of the time sampling up to the 5-year nominal operations of {\it Gaia}. 
Section~\ref{sect:ccd} discusses per-CCD $G$-band photometry, while subsequent sections will focus on the statistics of field-of-view transits as available for the $G$-band photometry as well as BP and RP photometers. The number of transits of the RVS instrument is scaled by about 4/7 with respect to G, BP and RP due to the smaller across-scan coverage of only 4 CCDs versus 7 for the others.

\subsection{Per-CCD photometry} \label{sect:ccd}
When an object crosses the focal plane \citep[see fig.~4 in][]{DPACP-1}, it is first detected by one of the CCDs of the Sky-Mapper, and then typically by 9 other astrometric CCDs in the $G$-band \citep{DPACP-1}. 
Yet, during a focal plane crossing, this number of detections can be smaller than the 9 theoretically achievable for various reasons.
For example, when a star image crosses the Field of View (FoV) in the 4th row of CCDs, the CCD in the 9th strip is used only for on-board metrology (the wave-front sensor measurement), leaving only 8 CCDs with $G$-band measurements in this row.
Also, if an object crosses the field near the upper or lower edge of the focal plane, it may exit the focal plane before reaching the last CCD, since there is a small across-scan motion of the image due to the spin axis precession (this happens very rarely for stars, but is not exceptional for solar system objects because they often have a discernible motion on the CCDs due to their proximity).
Furthermore, there might be CCDs with perturbed measurements, causing a reduced quality and possible rejection.
The crossing of the focal plane is referred to as `FoV transit', independent of the number of CCD measurements that was obtained or kept.

The integration time per CCD is about 4.4~seconds in TDI (Time-Delayed Integration) mode. However, for bright sources ($G$\nolinebreak<12.5~mag),  gates are activated reducing the integration time \citep{DPACP-1} to stay within the full well capacity. 
Due to the gaps between CCDs, the actual interval  between CCD measurements is about 4.85~seconds.
In conclusion, {\it Gaia} probes the whole sky in an unprecedented way with a high cadence rate (from about 5 to 44~seconds).

\subsection{Ecliptic Pole Scanning Law (EPSL)}\label{sec:epsl}



During the first 28 days of operations  (from July 25 till Aug. 22, 2014), {\it Gaia} rotated without spin precession.
The spin axis laid constantly in the ecliptic plane at 45$\degr$ with respect to the {\it Gaia}-Sun direction (in the trailing direction), causing the scan plane to rotate by almost 1$\degr$ per day around the ecliptic poles as a result of the annual motion around the Sun. During each revolution of 6 hours, the North and South Ecliptic Pole areas were scanned and stars at high ecliptic latitude were observed up to 8 times per day (4 revolutions, 2 FoVs), i.e. up to 224 FoV transits in 28 days. 

\subsection{Nominal Scanning Law (NSL)}\label{sec:nsl}
The NSL is designed for {\it Gaia}'s operation during the nominal mission to scan the full sky and provide optimal constraints for the astrometric solutions.
In particular, the spin axis of the spacecraft has a fixed angle of 45$\degr$ with respect to the {\it Gaia}-Sun axis, and precesses around it with a period of 63 days.

The normalised distribution of intervals between successive FoV transits is displayed in Fig.~\ref{fig:SuccessiveGaps}.
The time interval between successive transits $i, i+1$ is defined as $t_{i+1} -t_i$ and only the intervals larger than 2 days are shown in this figure (see Sect.~\ref{sect:sequences} for shorter intervals).
We remark that the most common time interval is between 20 and 40 days.
For details on recurrence of specific  time intervals (of any pair of transits) see Sect.~\ref{sec:recurrenceTimeIntervals}.

Compared to the EPSL, the NSL sampling is sparser and more randomized in time, which makes NSL times series less prone to aliasing.
The highest peaks in the spectral window of NSL are significantly lower than the spectral window of EPSL, see Sect.~\ref{Sec:ProperOfNmesandSpectWindow}.
However, in terms of number of observations, the 28 days of EPSL for sources close to the ecliptic poles is representative of the 5-year NSL.

\paragraph{Transition NSL.}
At the end of the EPSL phase, the precession motion of the spin axis was set in, which marked the start of the Nominal Scanning Law, albeit in a transitional form for about one month so that the orientation of the spin axis should reach smoothly a transition point where the final parameters of the NSL could be reached saving as much consumables as possible for the attitude control. This scanning law was used from Aug. 22 till Sept. 23, 2015.

\paragraph{GAREQ NLS.}
On September 25, 2015 the satellite was manoeuvred 
to the GAREQ NSL, which
is a nominal scanning law (meaning  aspect angle of  45$\degr$, revolution of 6h, precession of 63 days) but with a particular choice of the two free initial conditions for the precession and revolution, in order to favour the observation of bright stars in the vicinity of Jupiter, when the planet is predicted to be in the FoV. This optimisation is designed for the {\it Gaia} Relativistic Experiment on Quadrupole light deflection (GAREQ) by Jupiter. This scanning law is intended to be used till the end of the nominal 5~yr mission.

\subsection{Expectations from simulations of the 5 year mission}
The results are based on two extensive simulations of the FoV transits using 
100\,000 and 786\,432 directions uniformly distributed over the whole celestial sphere.
The resulting cadence properties are not exactly those that are seen in the {\it Gaia} catalogue, because obviously the stars are not uniformly distributed across the sky and missed observations are expected due to various reasons, such as decontamination periods, data loss \citep{DPACP-1}, insufficient information for calibration, etc. 
However, removing a few data points does not significantly change how the time sampling varies with the ecliptic latitude and how often we can expect to have continuous sequences of observations.

\subsubsection{Consecutive Field-of-Views} \label{sect:sequences}
The {\it Gaia} spacecraft spins with a 6-hour period, within which observations are recorded from both FoVs when sources are detected by the on-board system.
The angular separation between the two FoVs is 106.5$\degr$, leading ideally to sequences of FoV transits separated by 1h~46min (rotation of 106.5$\degr$ from  Preceding FoV to Following FoV)  and 4h~14min (rotation of $253.5\degr$ from Following FoV to Preceding FoV) and possibly repeating over the next few revolutions. In general, a source is seen in a consecutive pair of FoVs and, due to the precession motion of the spin axis, is no longer visible in the next satellite revolution. But longer sequences of repeated transits always occur in the observation history of a star, providing a nice time coverage of the light curves over a day or so. 

The occurrence rate of each possible sequence can only be established with a dedicated simulation. Using $100\,000$ uniformly distributed directions across the sky and running the nominal scanning law over 5 years, we found the respective frequencies for each kind of possible sequence. A continuous sequence of transits is defined as a repeated succession of  FoV transits with no time gap larger than 2 revolutions (12h) between two successive transits. When a larger gap occurs, it is always of several days or weeks and this long gap partitions nicely the set of crossings into groups of short time sequences of different lengths which may recur after a longer interval. A typical sequence is denoted by a string, for example, by PFP for a succession of PFoV, FFoV, PFoV before a long gap appears. The respective probabilities (namely occurrence rates derived from the simulation) are given in Table~\ref{tbl:sequences}, averaged over the sky. In fact, we found that the dependence on ecliptic latitude is very small, unlike the case of the number of FoV crossings, which is highly dependent on the distance to the ecliptic plane.

The results are shown as conditional probabilities for two broad categories: P-sequences (resp. F-sequences), when the first transit in the sequence occurs in the PFoV (resp. FFoV). The absolute frequencies of each group are respectively 70\% and 30\%, as related to the proportion of the basic angle  ($106.5 \degr$) and its complement  ($253.5 \degr$) and do not change with ecliptic latitude. This can be understood with a simple argument:  if the occurrence of a first transit after a time gap takes place in the FFoV, this means that the star has entered the observability area above or below the scanning plane in the small arc between the two FoVs, while it must be in the large arc for a first PFoV transit. The probability for either instance is proportional to the size of the arcs. 

In Table~\ref{tbl:sequences}, we list the conditional probability of occurrence of every significant sequence within each of the two categories.When the first transit is in the PFoV, we have in general a new transit taking place in the FFoV, and then longer sequences are less frequent. When a sequence begins with the FFoV, a second transit in the PFoV  happens only in one case out of three. The last lines (``others") in Table~\ref{tbl:sequences} aggregate the much rarer longer sequences, which can be very long with 10 or 20 consecutive transits recorded over a few days. Although these are rare, almost every star is observed in a long sequence (at least) once during the mission lifetime, which is a nice feature for the time sampling of the light curves.

\begin{table}[ht]
        \caption{Conditional occurrence rate distribution of the different sequences of successive transits without large time gaps. Only sequences with occurrence rate $ \geq 0.5 \% $  are listed. Note that sequence PFPFP is below this threshold while the  PFPFPF is more frequent and listed.  }
        \centering          
        \label{tbl:sequences}
        \begin{tabular}{lccclcc}
	\hline\hline
        \noalign{\smallskip}
       \multicolumn{3}{c}{ P-sequences} && \multicolumn{3}{c}{ F-sequences} \\
        Sequence && Occurrence   &&  Sequence && Occurrence   \\
                  &&  rate (\%)       &&          &&   rate (\%) \\
         \hline         
        \noalign{\smallskip}
P      &&  22 && F      && 48  \\[3pt]
PF     &&  63 && FP     && 33 \\
PFP    &&   4 && FPF    &&  8 \\
PFPF   &&   6 && FPFP   &&  5 \\
PFPFPF &&   1 && FPFPF  &&  1 \\
others &&   4 && FPFPFP &&  2 \\
       &&     && others &&  4 \\
        \hline
        \end{tabular}
\end{table}

\subsubsection{Properties of the number of transits and of the spectral window} 
\label{Sec:ProperOfNmesandSpectWindow}
The number of FoV transits as a function of the ecliptic latitude  $\beta$ was simulated for $786\,432$ sources regularly distributed over the sky\footnote{Using the HEALPix sky segmentation scheme \citep[\url{http://healpix.sourceforge.net}]{2005ApJ...622..759G} .} for a 5-year mission (28 day EPSL, 1 month transition NSL and 4.83 year of GAREQ NSL), as shown in Fig.~\ref{fig:betanmes}.
The effect of the ecliptic latitude is conspicuous: the zone near the ecliptic is under-observed and the peaks at $|\beta|\sim 45 \degr$ are linked to the Solar aspect angle.
The maximum number of FoV transits of about 310 the at $|\beta|\sim 90 \degr$ is due to the EPSL. 

\begin{figure*}[t]
\centering
\includegraphics[width=\textwidth]{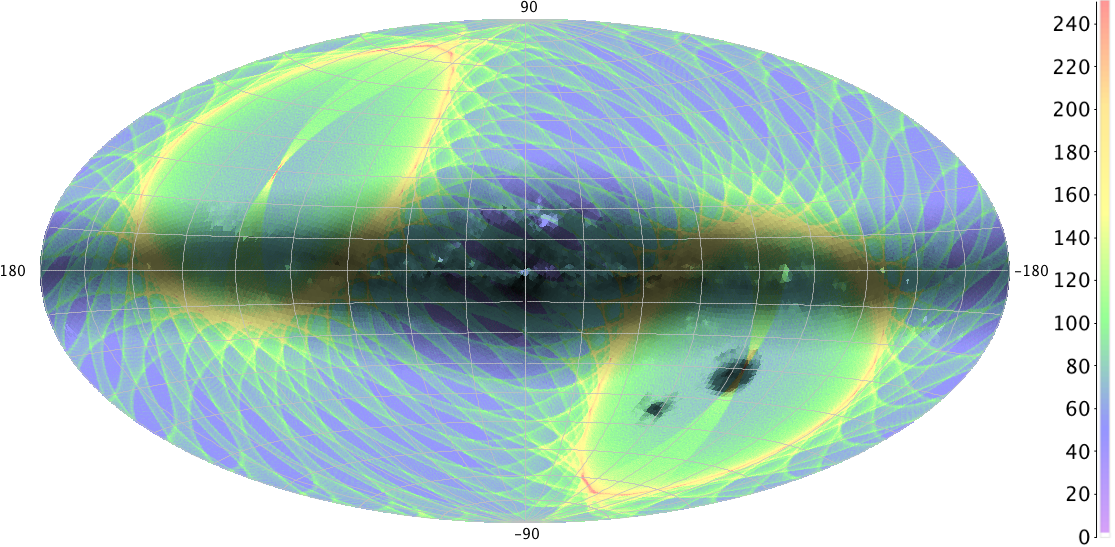}
\caption{Galactic Aitoff projection of the simulated number of FoV transits for 5~yr of {\it Gaia} scanning law blended with the \cite{2005ESASP.576..163D} Galactic cumulative stellar density up to G=20 (darker is more dense). 
Clearly visible are the LMC and SMC near the South Ecliptic Pole, the former partially overlapping with the EPSL footprint. The FoV transit count is clipped at 250 but reach up to 310 transits at the ecliptic poles. Galactic map origin $l=b=0$ is at the centre and $l$ is increasing to the left. Transits are computed for one time series per 0.05~deg$^2$. 
Figure~\ref{fig:betanmes} shows these data as a function of ecliptic latitude.}
\label{fig:galaxyNslBlend}
\end{figure*}

\begin{figure}[b]
\centering
\begin{minipage}[t]{.45\textwidth}
\centering
\includegraphics[width=0.95\linewidth]{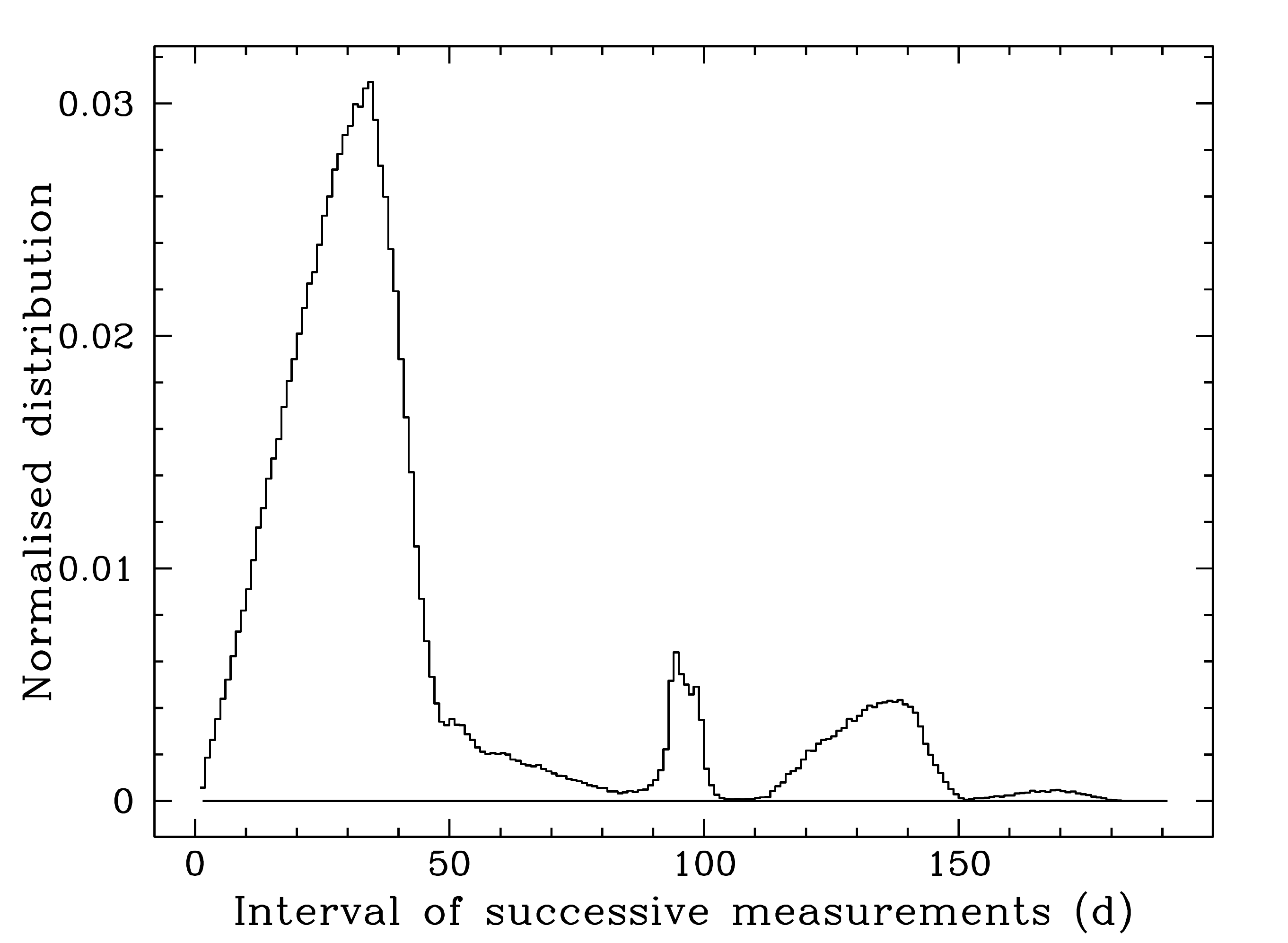}
\captionof{figure}{Normalised histogram of the intervals ($> 2$ days) between successive FoV transits of the same source (in days) over 5 years based on simulated {\it Gaia} sources randomly distributed over the sky.}
\label{fig:SuccessiveGaps}
\end{minipage}%
\begin{minipage}{.05\textwidth}
\hspace{0.1\linewidth}
\end{minipage}%
\begin{minipage}[t]{.45\textwidth}
\centering
\includegraphics[width=0.95\linewidth]{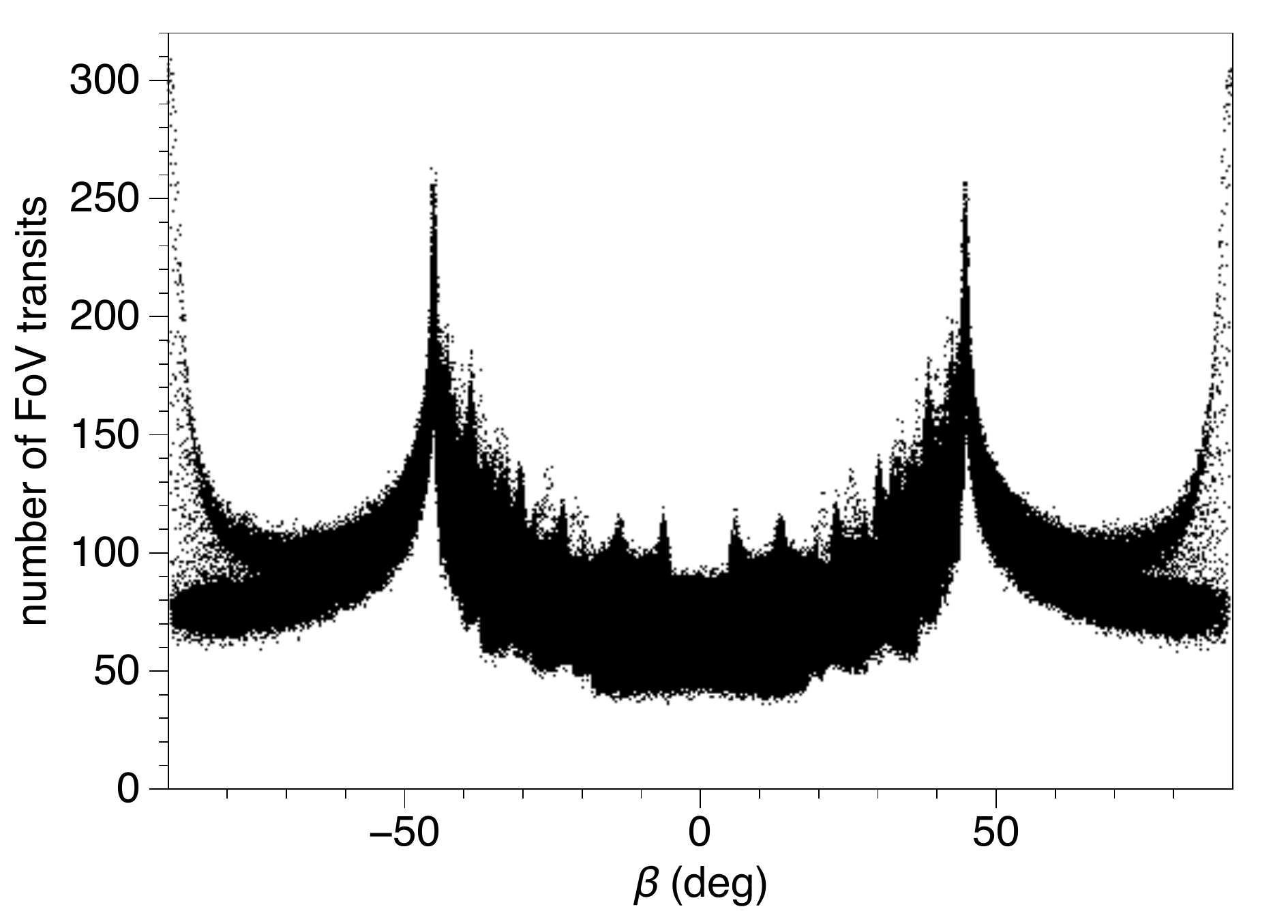} 
\captionof{figure}{Number of FoV transits as a function of the ecliptic latitude $\beta$ for a simulated 5-year mission. The high peaks of about 260 at $ \lvert \beta \lvert \sim  45 \degr$ are coming from the Nominal Scanning Law. There are also more discrete but higher peaks at the ecliptic poles ($ \lvert \beta \lvert \sim  90 \degr$) up to 310, 
caused by the Ecliptic Pole Scanning Law (EPSL). 
The minimum is 36 and the average 86, not taking into account any observation dead-time. See Fig.~\ref{fig:galaxyNslBlend} for a Galactic sky projection of these data.}
\label{fig:betanmes}
\end{minipage}%
\end{figure}

Figure~\ref{fig:galaxyNslBlend} shows the same simulated data projected across the sky, together with a density model of the Galaxy. 
The pattern of the Nominal Scanning Law (NSL) is roughly symmetric in the ecliptic plane.
Ecliptic latitudes $ \lvert \beta \lvert<  45 ^\circ$ are rather not homogeneous and under-sampled, causing `spoke' structures crossing the Galactic centre. Around $\lvert \beta \lvert \sim 45^\circ$ the sky is densly sampled (more observations per interval of time), while  for $ \lvert \beta \lvert >  45 ^\circ$ the number of transits is roughly constant, except for the extensively observed area around the poles during the 28d of EPSL.

\begin{figure}
\centering
\begin{minipage}[t]{.45\textwidth}
\centering
\includegraphics[width=0.95\linewidth]{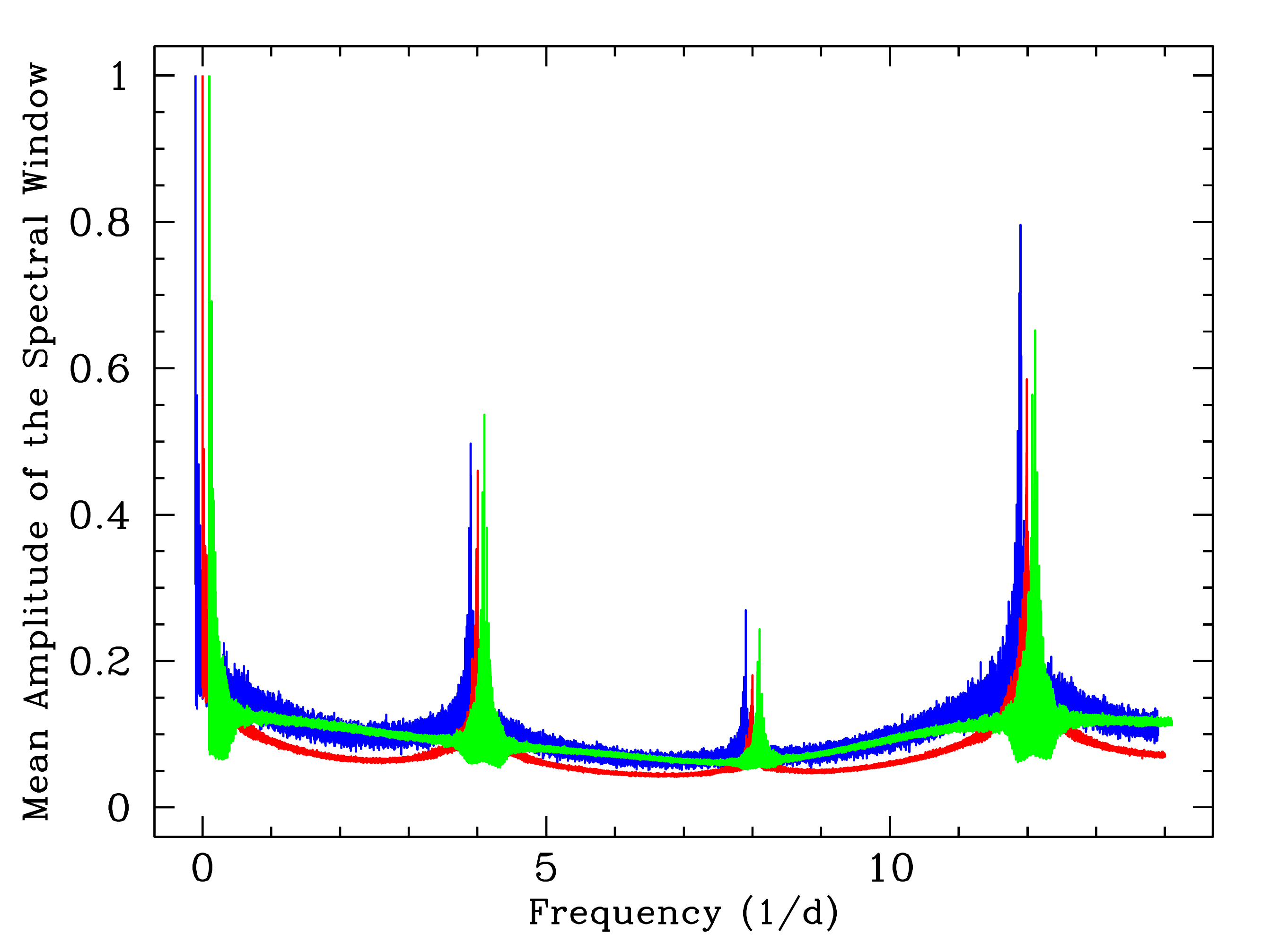}
\captionof{figure}{Mean amplitudes for the NSL of the spectral window near the regions of ecliptic latitude $\beta =0^{\circ}$ (ecliptic equator) in blue, $\beta=45^{\circ}$ in red and $\beta=90^{\circ}$ (ecliptic poles) in green. Each region covers 4000 squared degree. The blue and green spectral windows are shifted in frequencies by -0.1 and 0.1~d$^{-1}$ respectively to ease the comparison.}
\label{fig:mean_amplitude_spectral_window}
\end{minipage}%
\begin{minipage}{.05\textwidth}
\hspace{0.1\linewidth}
\end{minipage}%
\begin{minipage}[t]{.45\textwidth}
\centering
\includegraphics[width=0.95\linewidth]{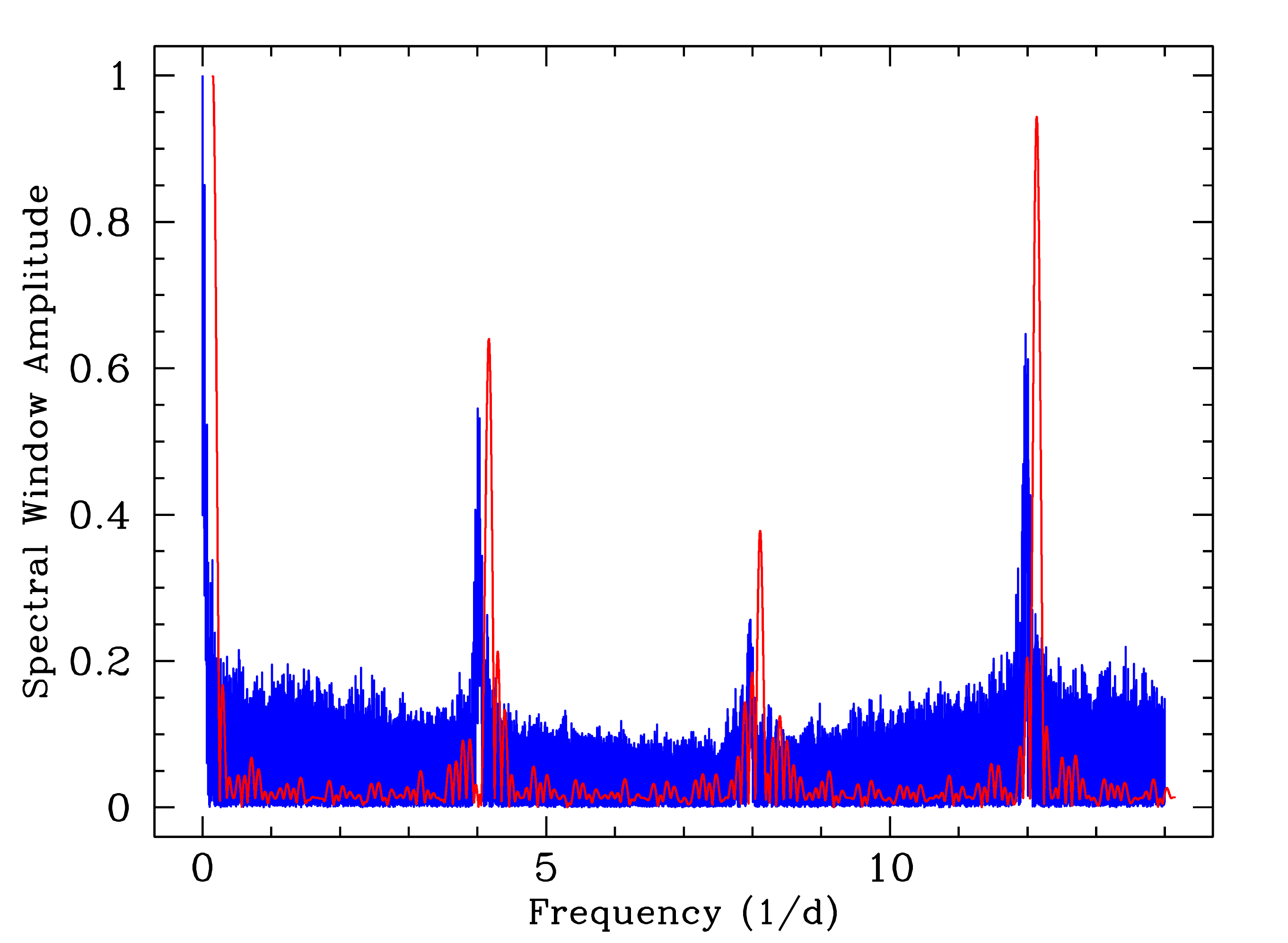}
\captionof{figure}{Amplitudes of the spectral window for a star with similar number of observations in EPSL and NSL: in red the Spectral window with only EPSL data, and in blue EPSL$+$NSL data. The amplitudes of the highest peaks are reduced significantly. The two curves are shifted by 0.15 cycle/d to better compare them.}
\label{fig:epsl_nsl_spectral_window}
\end{minipage}%
\end{figure}

The properties of the spectral window depend on the considered direction.
In Fig.~\ref{fig:mean_amplitude_spectral_window}, we present the average amplitudes of the spectral windows near $\beta =0^{\circ}$ (ecliptic equator), $\beta=45^{\circ}$, and $\beta=90^{\circ}$ (ecliptic poles) for regions covering 4000 square degrees each.
To ease the comparison, the blue and green spectral windows are shifted by -0.1 and 0.1 cycle per day, respectively.
We remark that the high peaks in Fig.~\ref{fig:mean_amplitude_spectral_window} are near the multiples of the rotation frequency of the spacecraft. 
The comparison of spectral windows from ground-based large scale surveys, as in presented in fig.~1 of \cite{2009sf2a.conf...45E}, with the ones presented here reveals that \textit{Gaia} is an outstanding survey to analyse periodic objects, because the spacecraft will be less prone to aliasing problems.
The rate of correct detection of periodic signals is very high as shown by \cite{2005MNRAS.361.1136E} (though that study is based on an old scanning law, the main results hold).

In Fig.~\ref{fig:epsl_nsl_spectral_window}, we show how the high amplitude peaks in the spectral window of EPSL sampling reduce when NSL observations are added.
As in Fig.~\ref{fig:mean_amplitude_spectral_window}, the spectral window is shifted for the EPSL (red line) by 0.15 cycle per day to ease the comparison.

\subsubsection{Recurrence of time intervals} \label{sec:recurrenceTimeIntervals}
\begin{figure}
\centering
\includegraphics[width=0.47\textwidth]{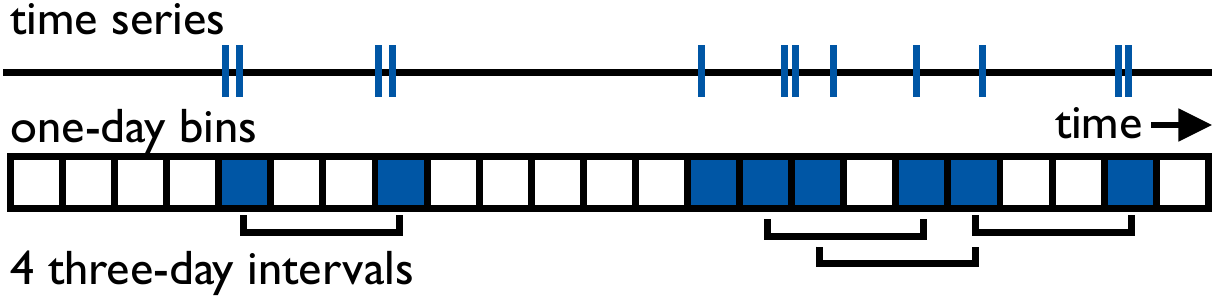}
\caption{
The adopted counting of recurrence of time intervals in a time series. The example shows 4 distinct three-day intervals, 3 four-day intervals and 1 five-day interval. When interested in a three-to-five day interval range this results in a total of 8~counts. This counting is applied to the 
sky maps in Fig.~\ref{fig:reoccurrenceTimeIntervals}.
}
\label{fig:intervalExplanation}
\end{figure}

Another interesting property of the scanning law is the recurrence of specific time intervals in the time series.
This is relevant for the sampling and detection of variable phenomena, especially if they are transient or stochastic in nature. 
Here we will focus on quantifying the recurrence of a specific time interval in a time series on time-scales longer than 1~day.
First we aggregate the time series in one-day bins (acting as a crude low-pass filter on the typical short sequences of consecutive FoV transits described in Sect.~\ref{sect:sequences}).
Next, we count (with this one-day granularity) the number of times that two bins with observations are separated by a given interval, as exemplified in Fig.~\ref{fig:intervalExplanation}.
We illustrate the recurrence counts of time interval ranges between 2 and 135 days in Fig.~\ref{fig:reoccurrenceTimeIntervals}
It shows that the ecliptic pole regions are generally well covered by time intervals between (approximately) [26, 40], [56, 70], [86, 105],  [121, 135] days, which e.g. allows for efficient detection of decaying SN light curves.
The ecliptic plane region never receives very high recurrence counts of a particular time interval, but is however sampled at a very large range of time-intervals which is useful for e.g. the study of (non-transient) stochastic sources. There are specific time intervals absent between (approximately) [56, 75] and [116, 136] days due to geometric constraints imposed by the 63~day spin-axis precession period, which rules out the possibility of re-observing sources close to the ecliptic plane.
The earlier mentioned high recurrence intervals around the ecliptic poles concentrate around integer multiples of half the 63~day spin period, again due to the geometry of the scanning law.
Some small regions of the sky deviate from this general description, due to the EPSL and transition NSL (e.g. the X-shaped region around the ecliptic poles and two streaks crossing the ecliptic plane as clearly visible in the [66-70] day interval).

\begin{figure*}[h]
\begin{tabular}{@{}lll@{}}
\setlength{\tabcolsep}{0pt} 
\renewcommand{\arraystretch}{0} 
  \vspace{-0.5cm}\\
  \includegraphics[width=0.3\textwidth]{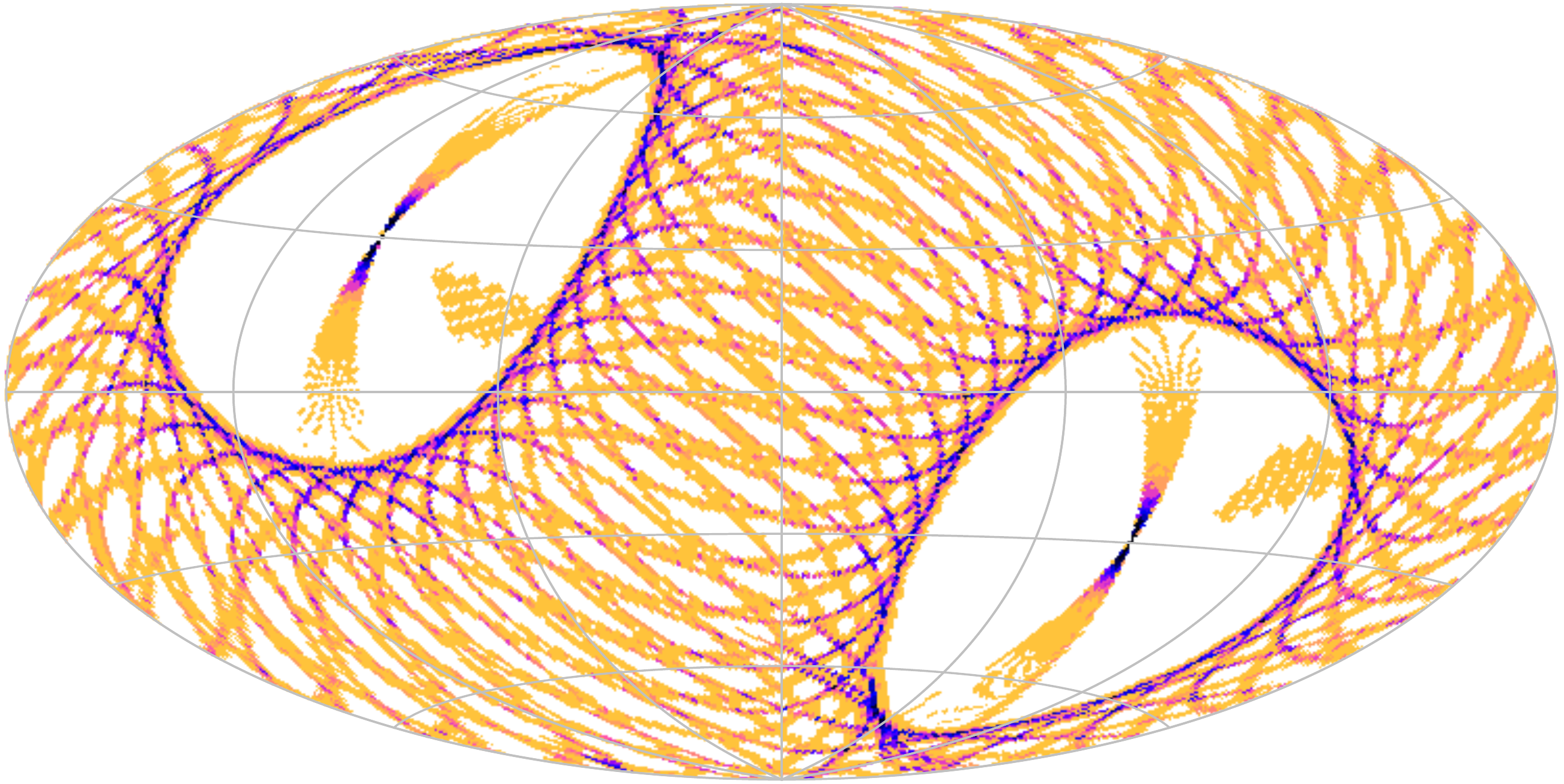}    & \includegraphics[width=0.3\textwidth]{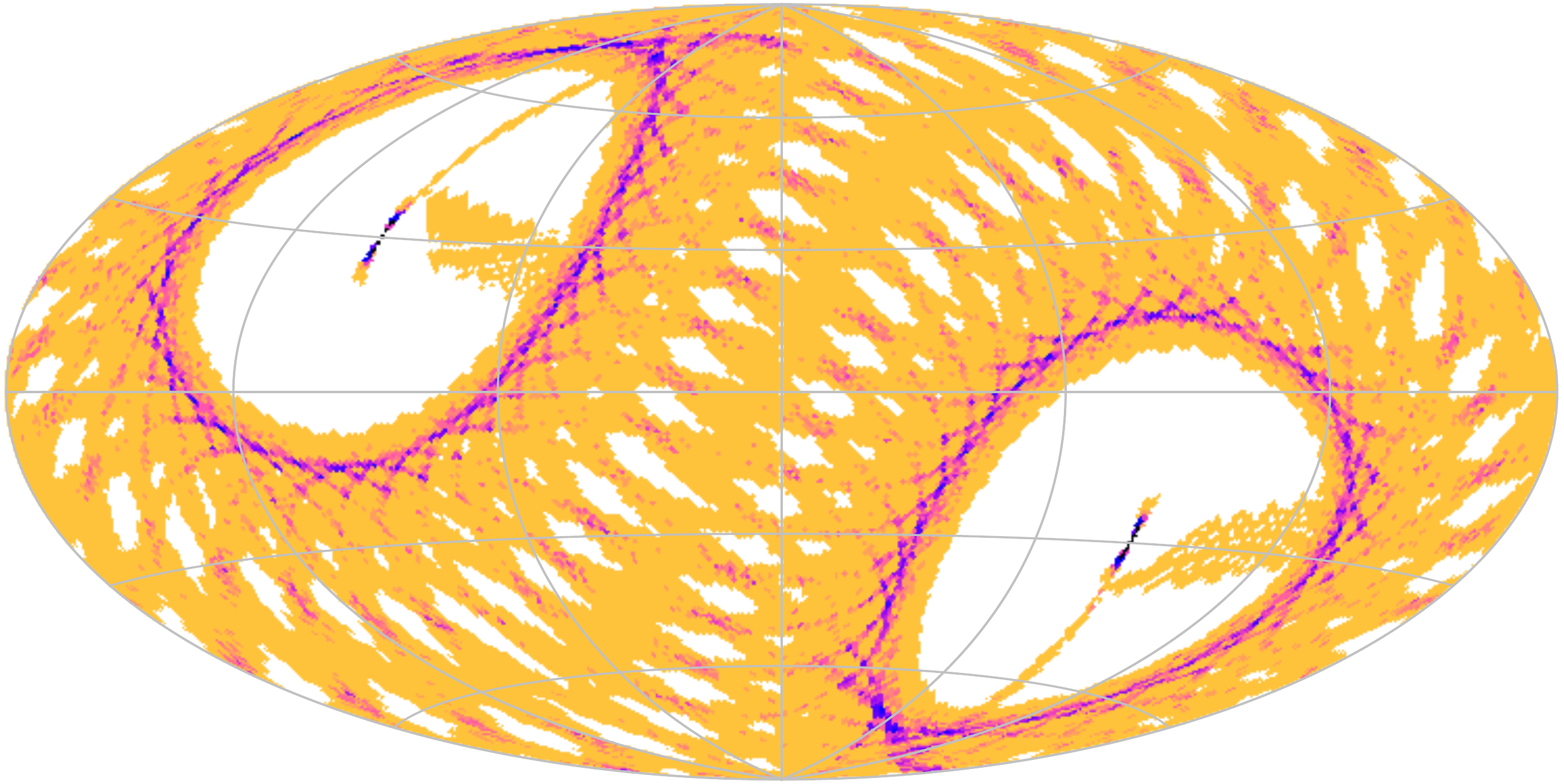}   &  \includegraphics[width=0.3\textwidth]{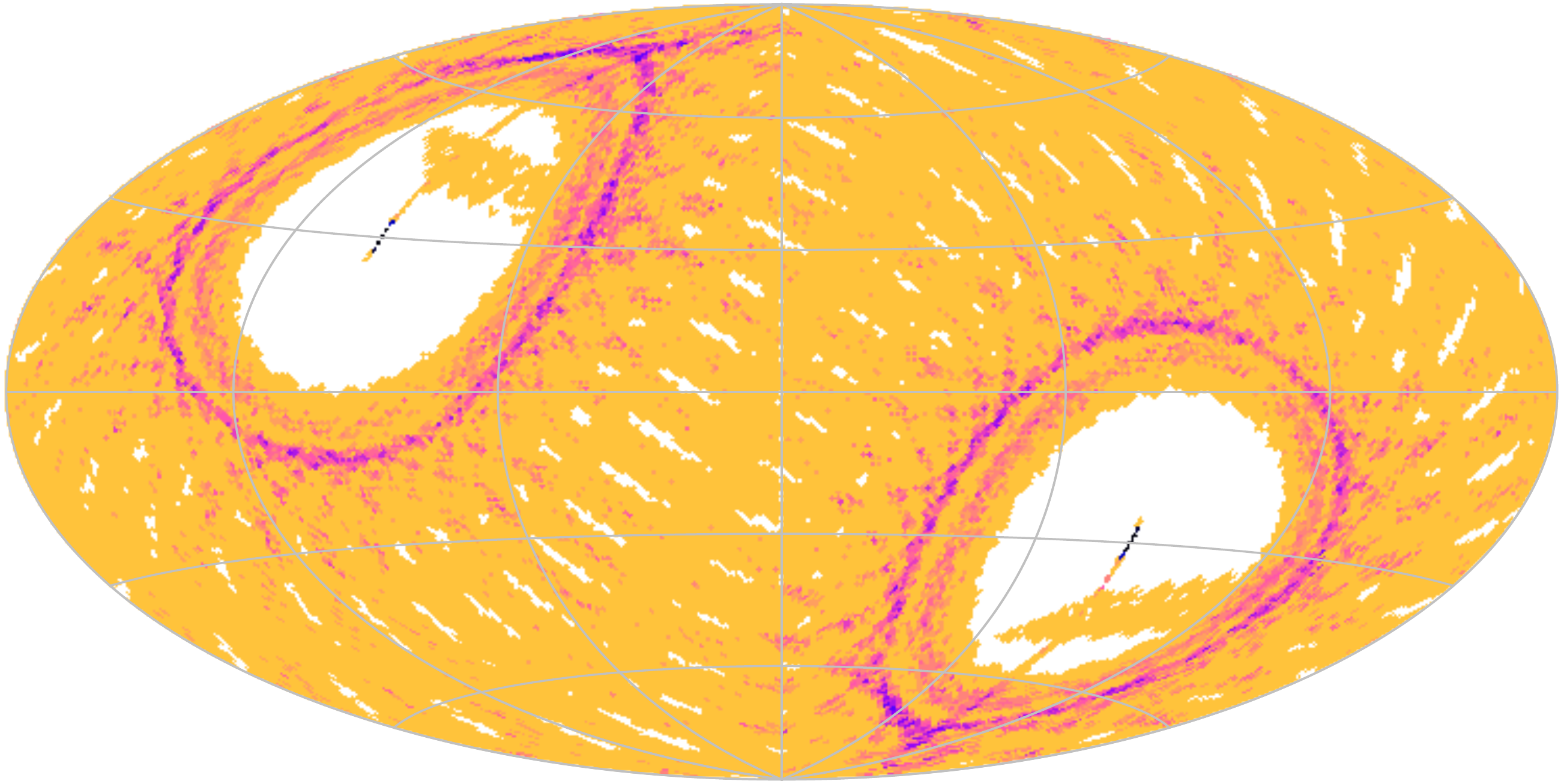}   \vspace{-3cm}\\
2-5~d & 6-10~d & 11-15~d  \vspace{2.4cm} \\[6pt]
 \includegraphics[width=0.3\textwidth]{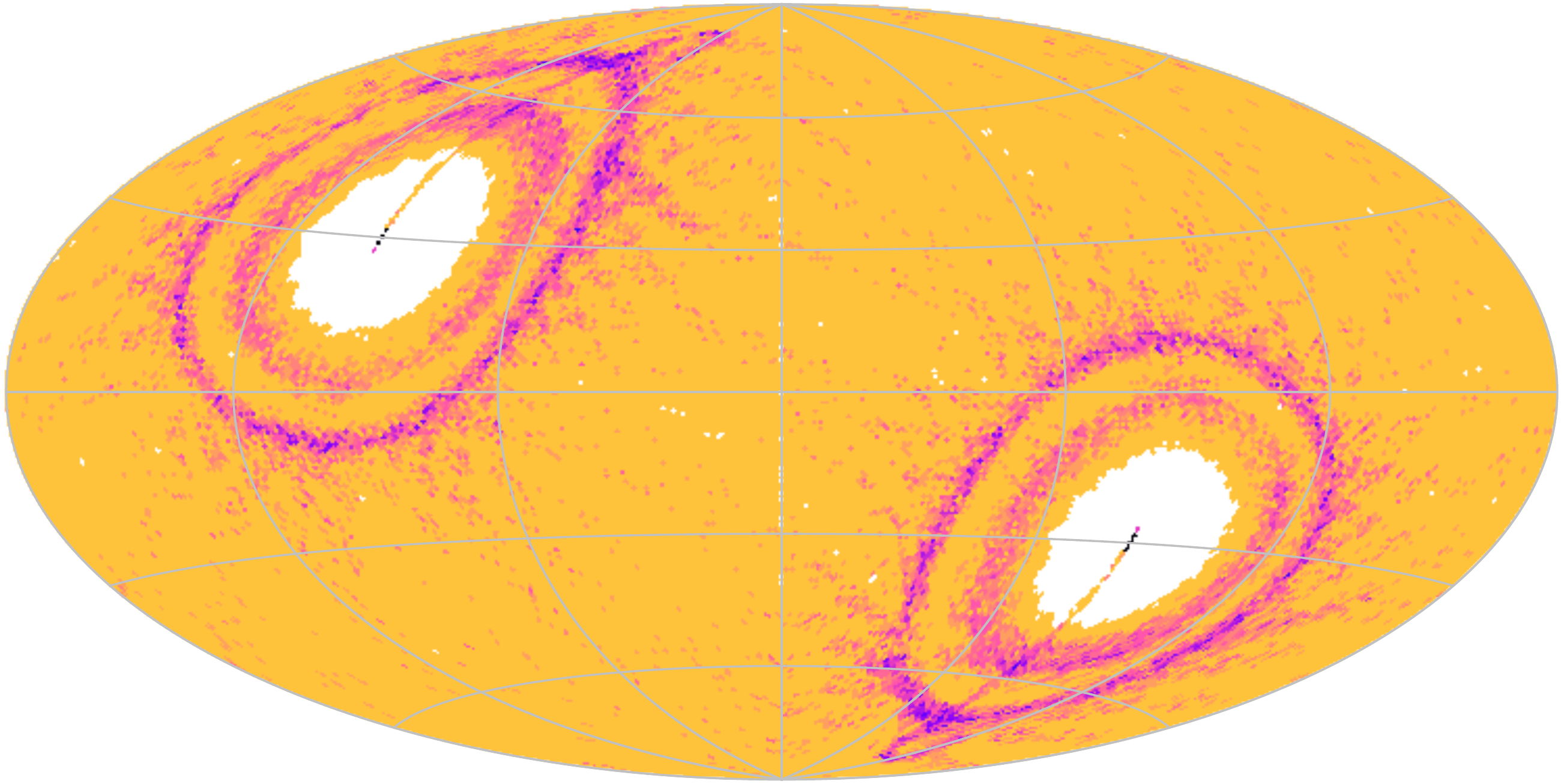}   & \includegraphics[width=0.3\textwidth]{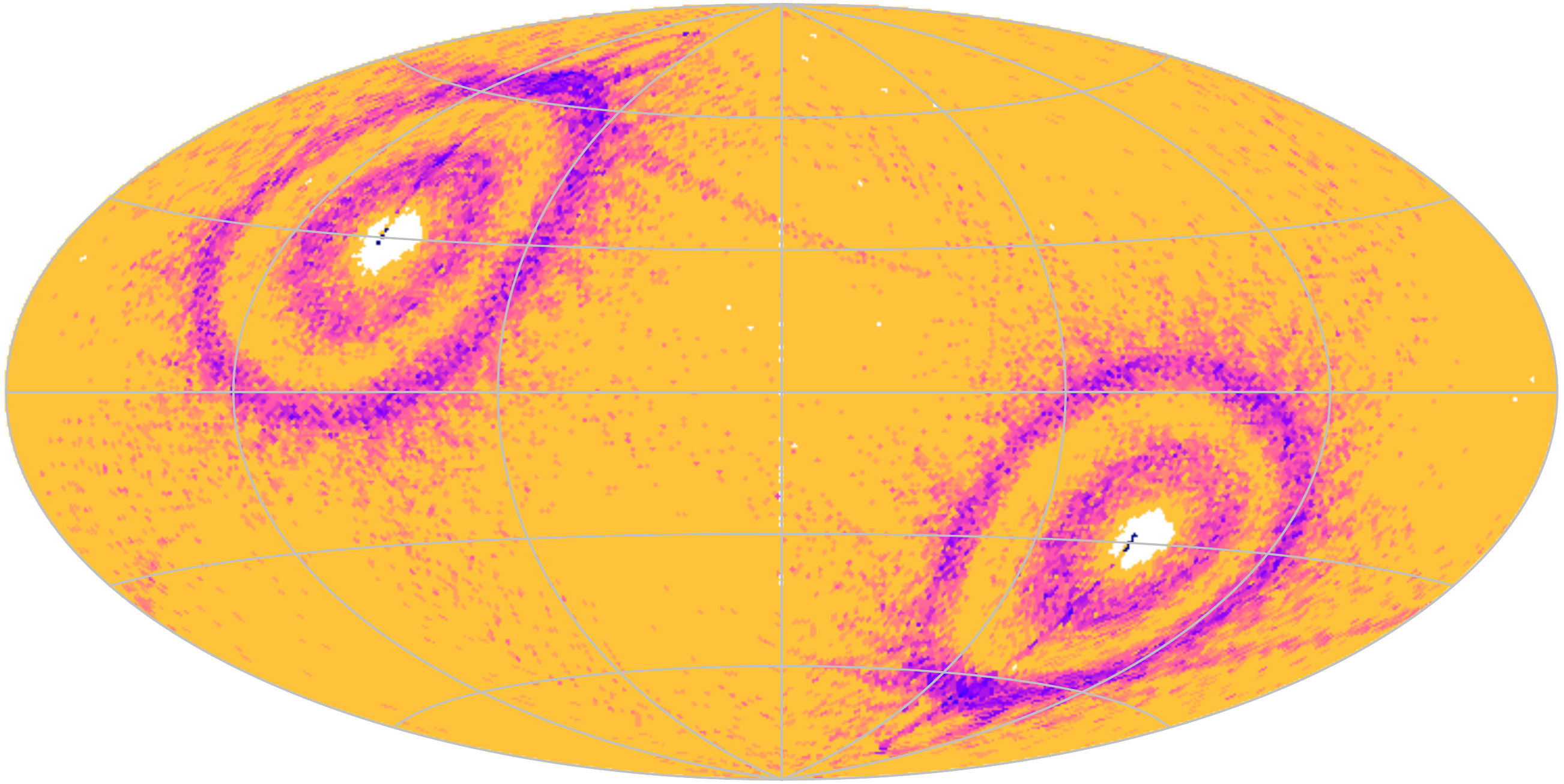}  &  \includegraphics[width=0.3\textwidth]{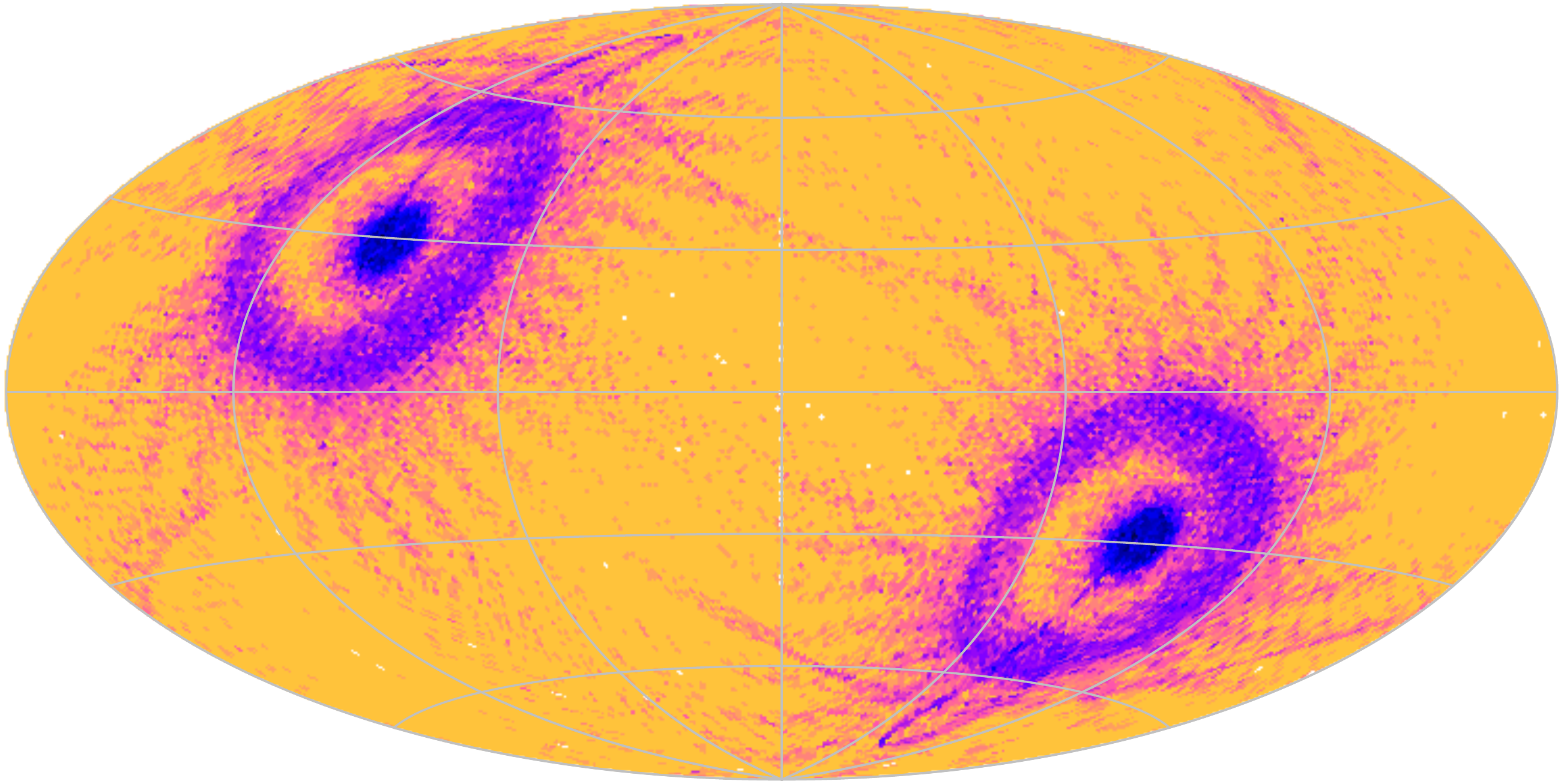}   \vspace{-3cm}\\
 16-20~d & 21-25~d & 26-30~d \vspace{2.4cm}\\[6pt]
  \includegraphics[width=0.3\textwidth]{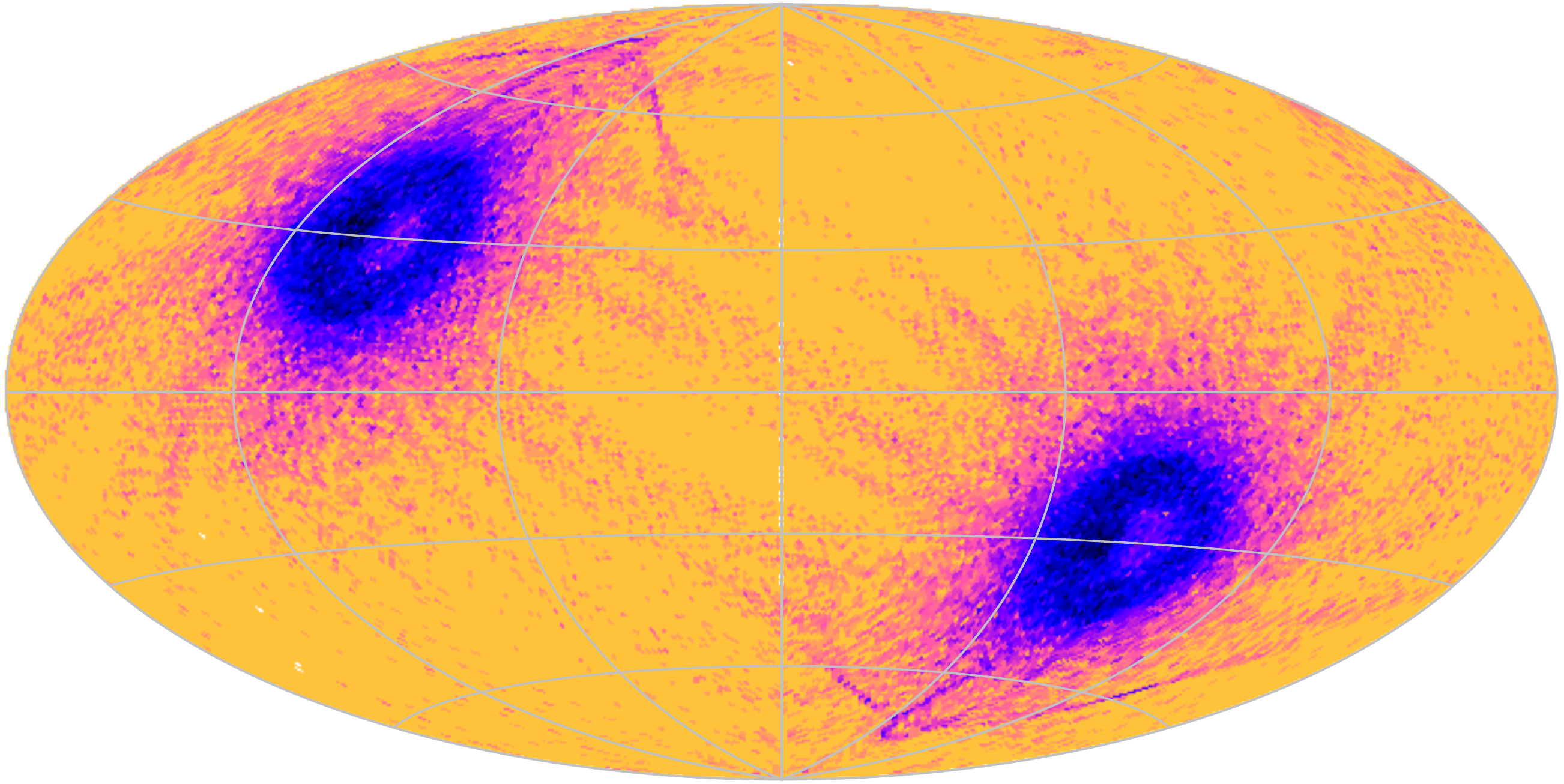}  & \includegraphics[width=0.3\textwidth]{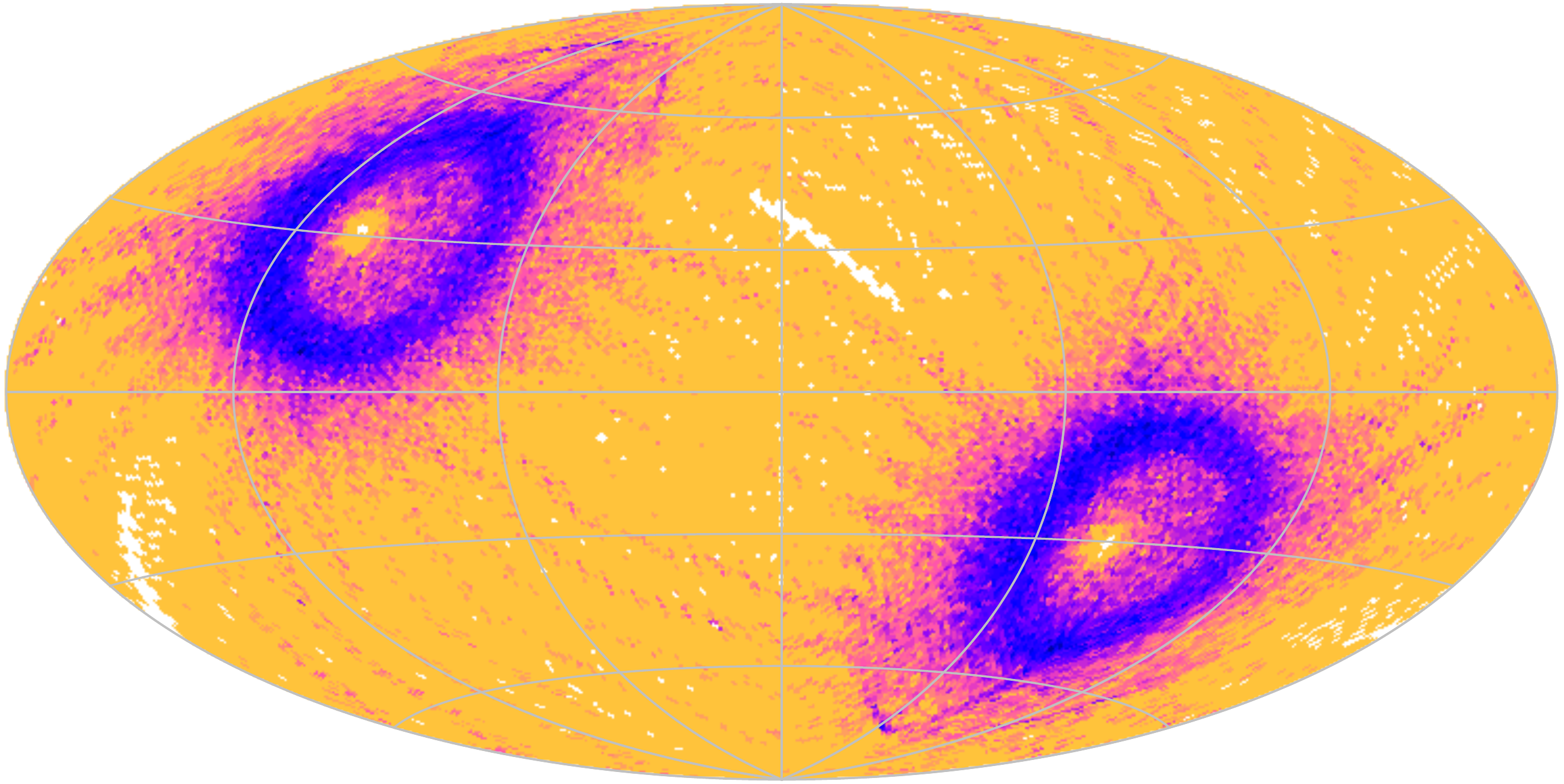}  &  \includegraphics[width=0.3\textwidth]{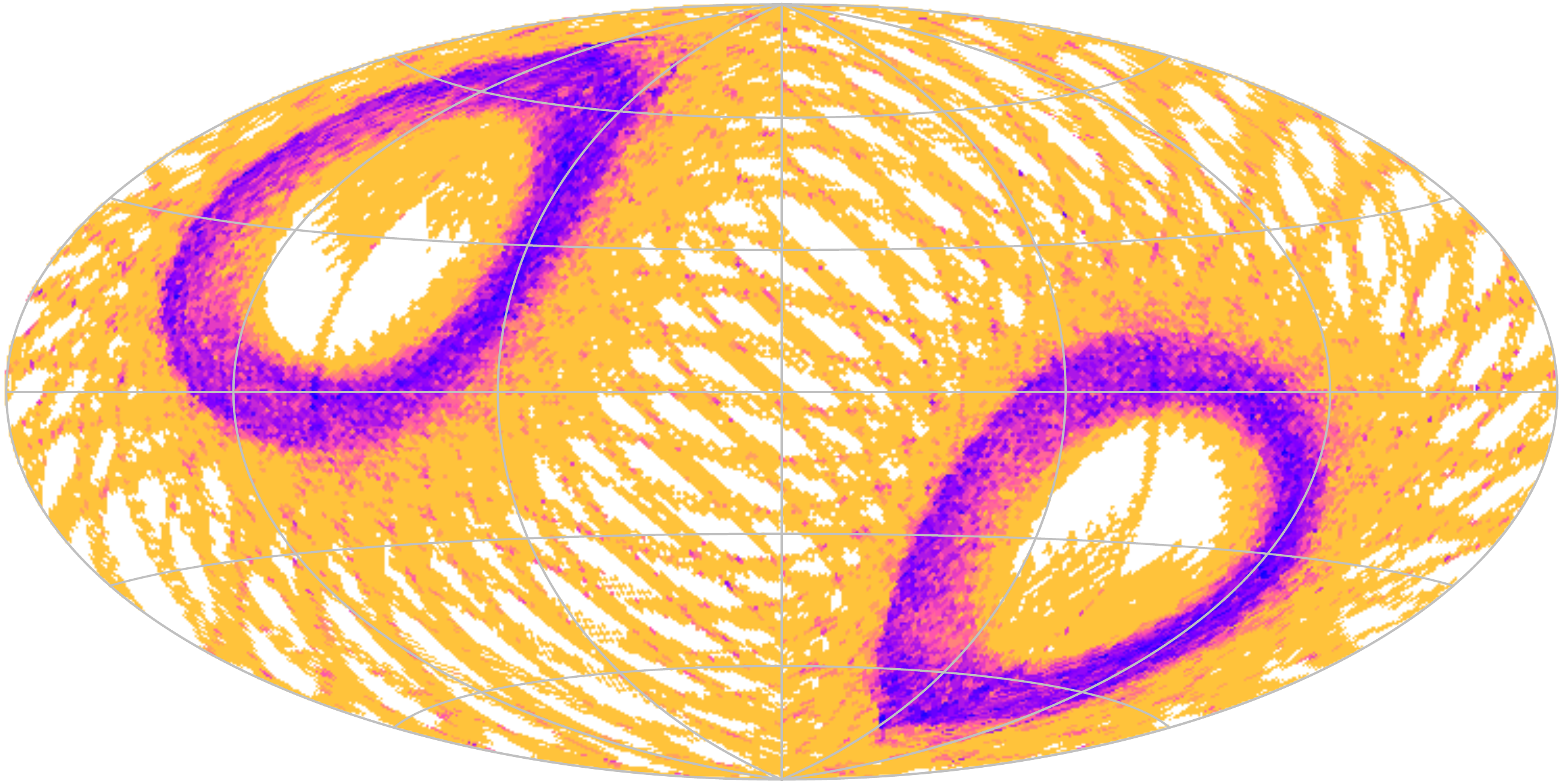}   \vspace{-3cm}\\
 31-35~d & 36-40~d & 41-45~d \vspace{2.4cm}\\[6pt]
  \includegraphics[width=0.3\textwidth]{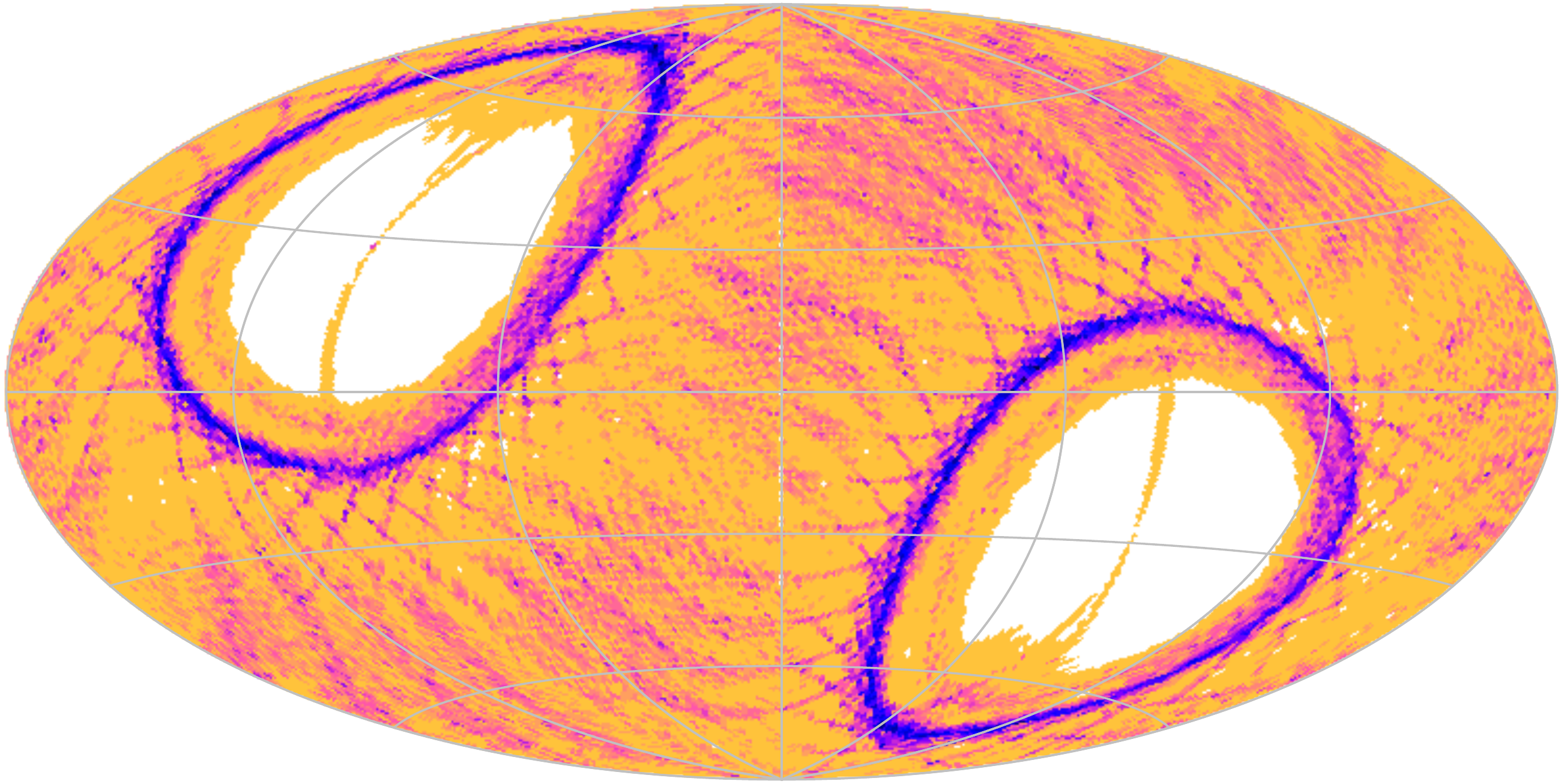}  & \includegraphics[width=0.3\textwidth]{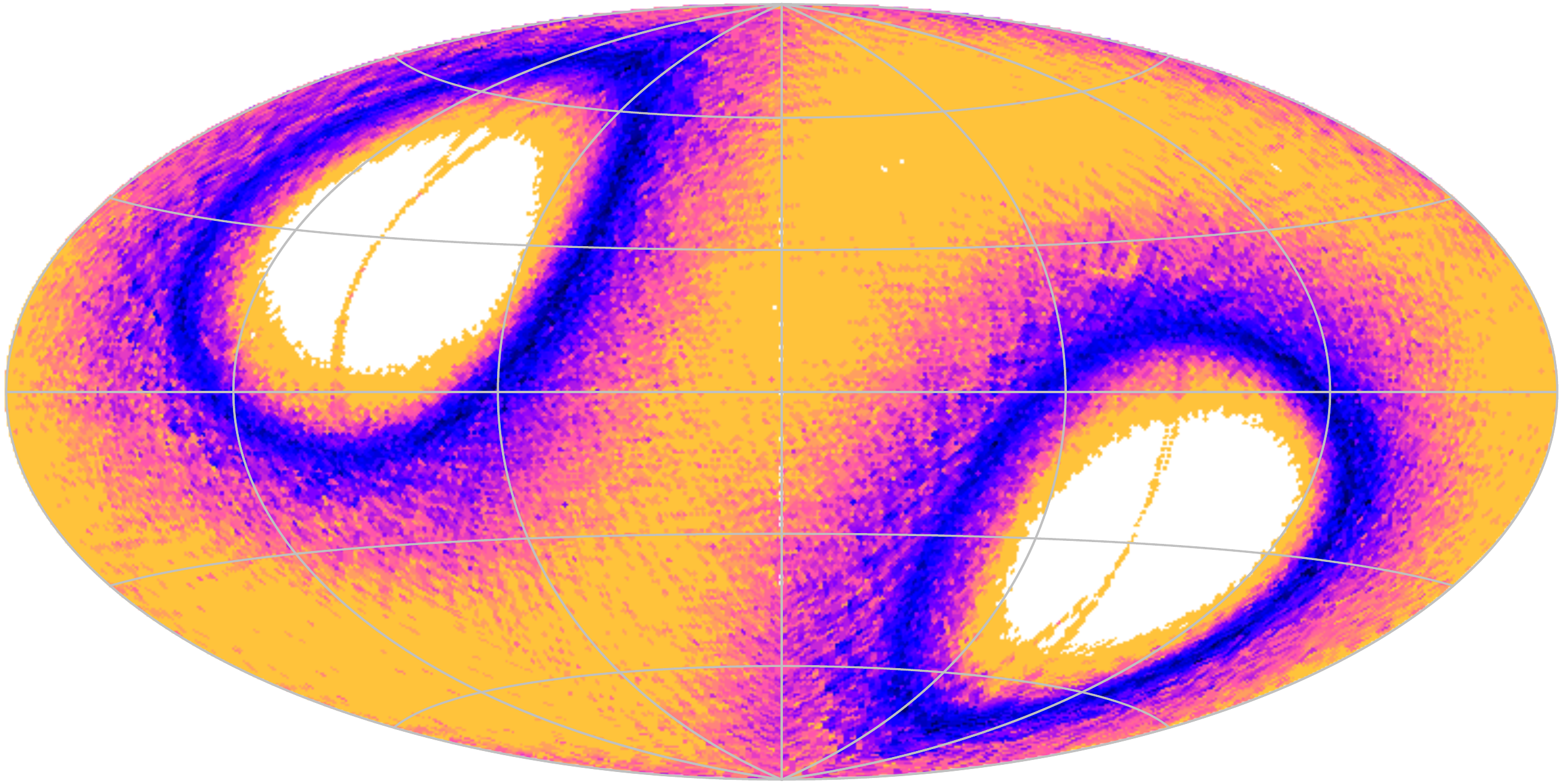}  &  \includegraphics[width=0.3\textwidth]{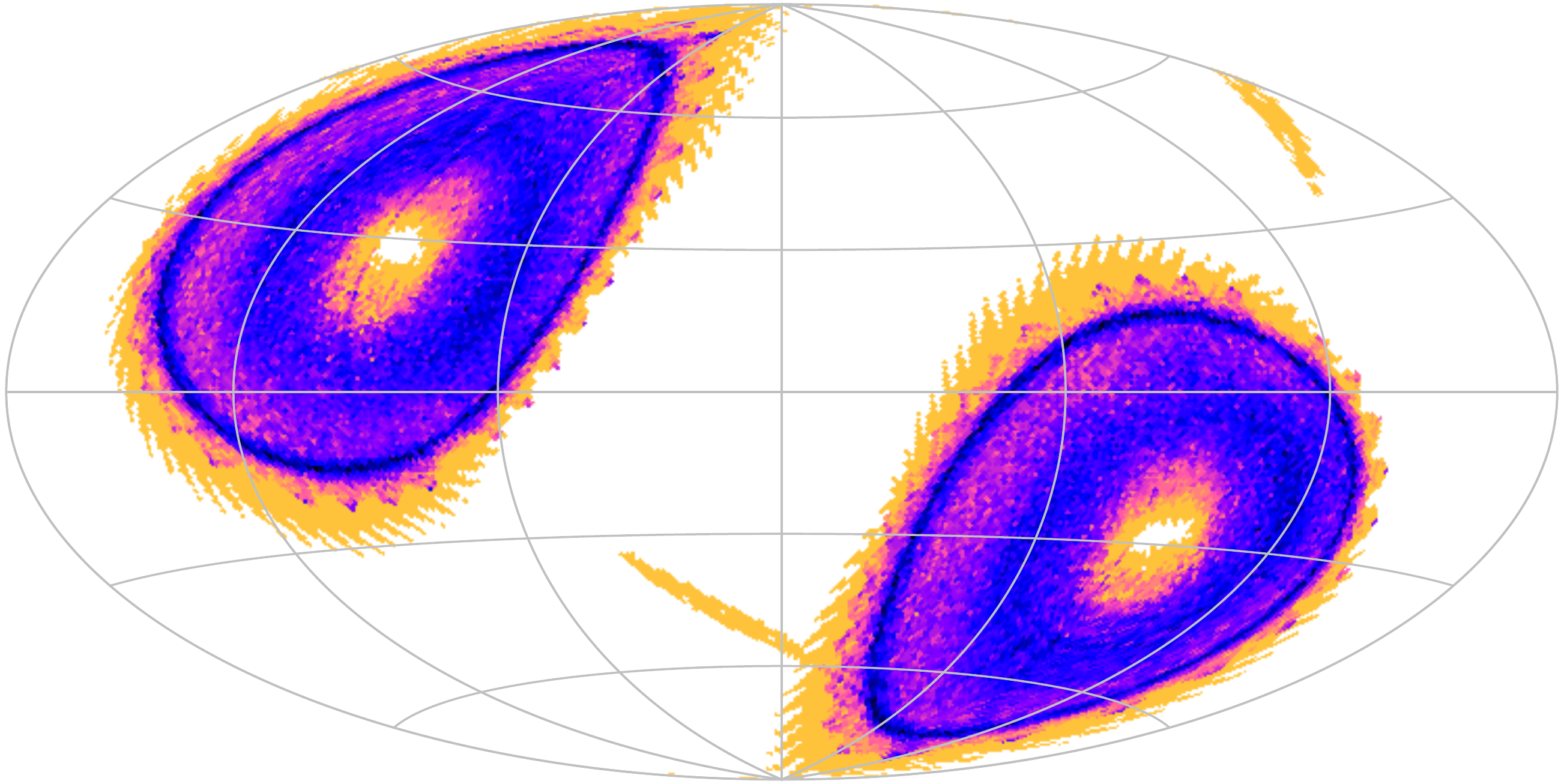}   \vspace{-3cm} \\
  46-50~d & 51-55~d &  56-60~d \vspace{2.4cm}\\[6pt]
  \includegraphics[width=0.3\textwidth]{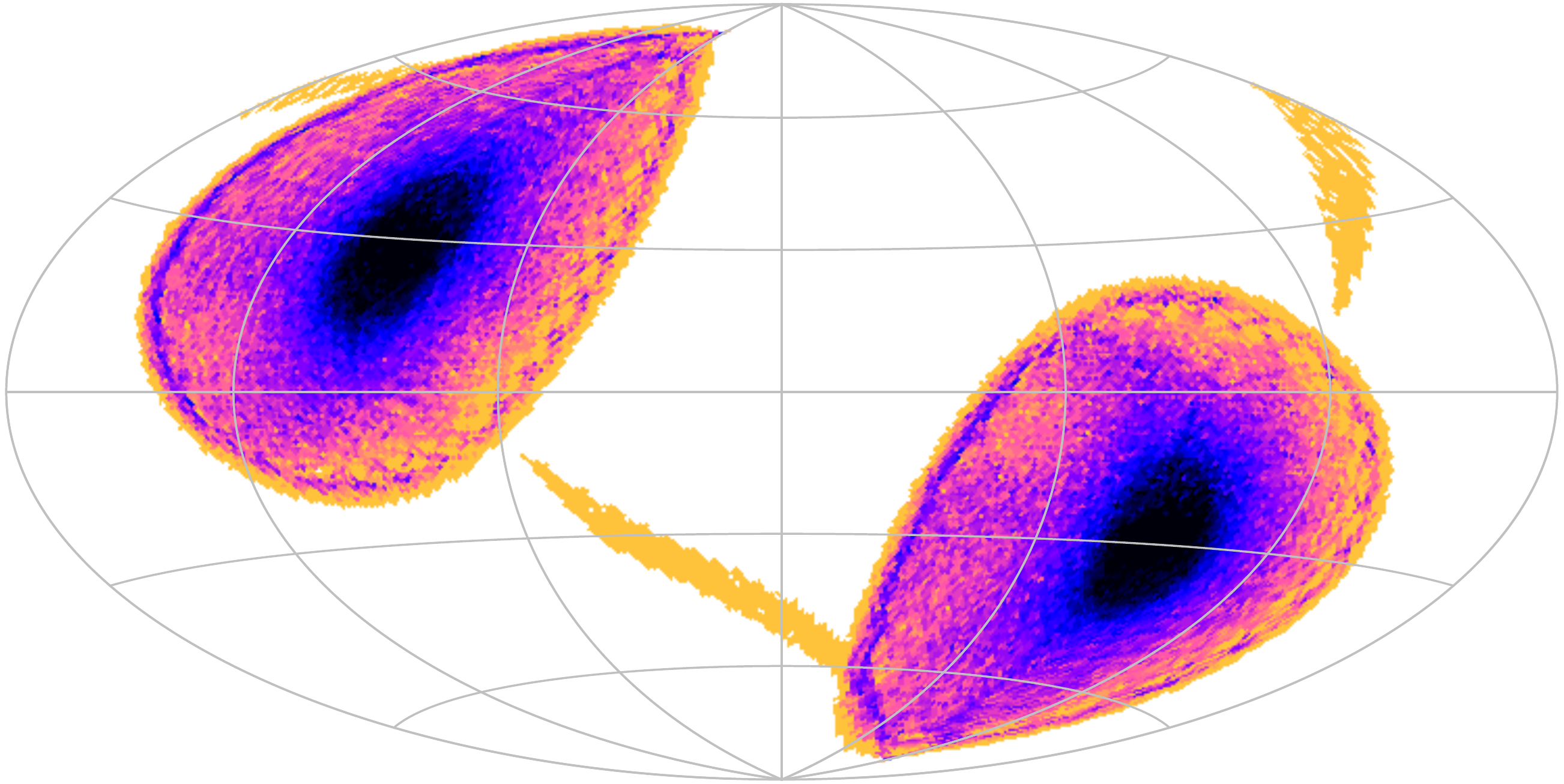}  & \includegraphics[width=0.3\textwidth]{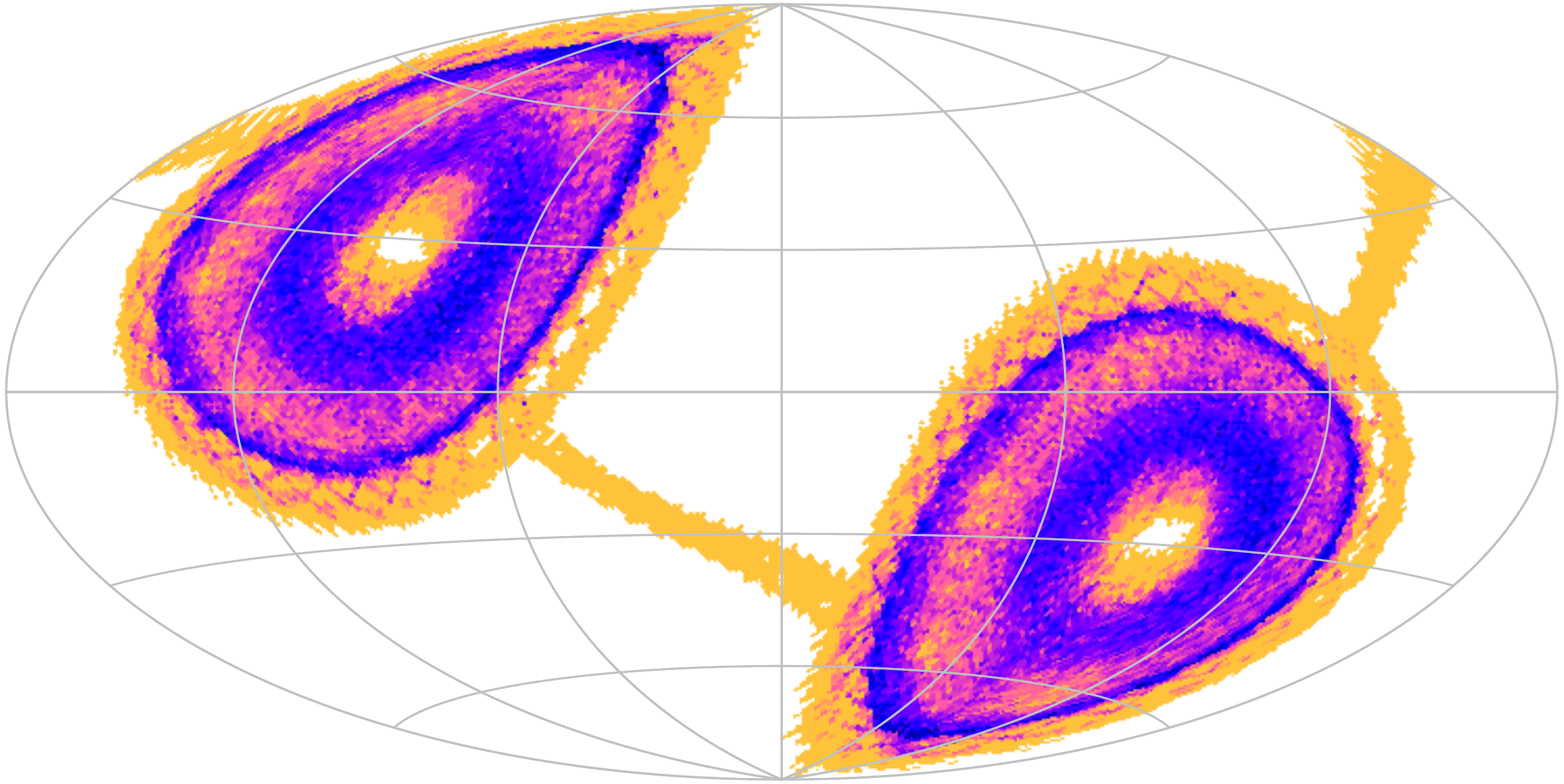}  &  \includegraphics[width=0.3\textwidth]{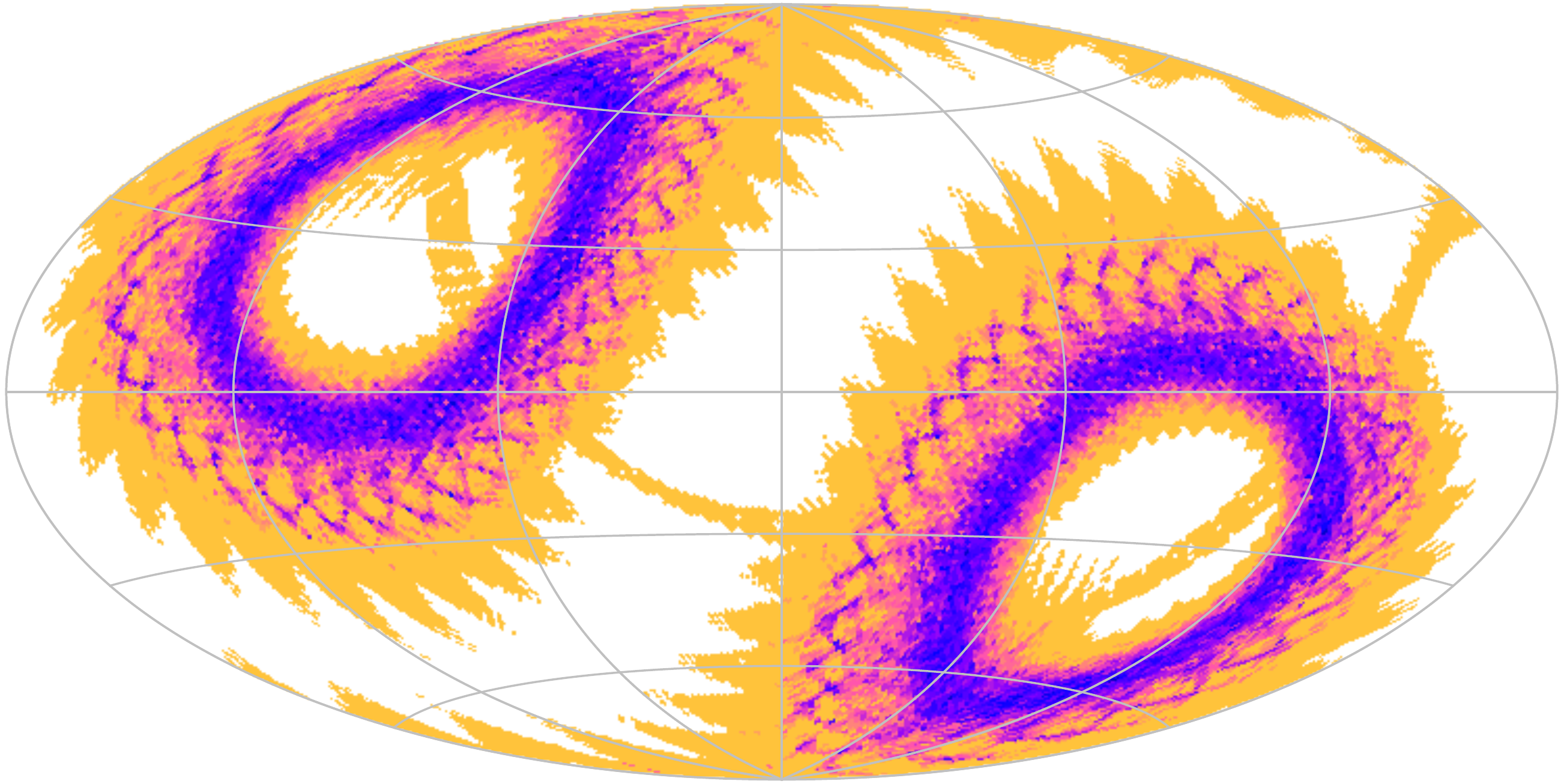}   \vspace{-3cm}\\
 61-65~d & 66-70~d &  71-75~d \vspace{2.4cm}\\[6pt]
  \includegraphics[width=0.3\textwidth]{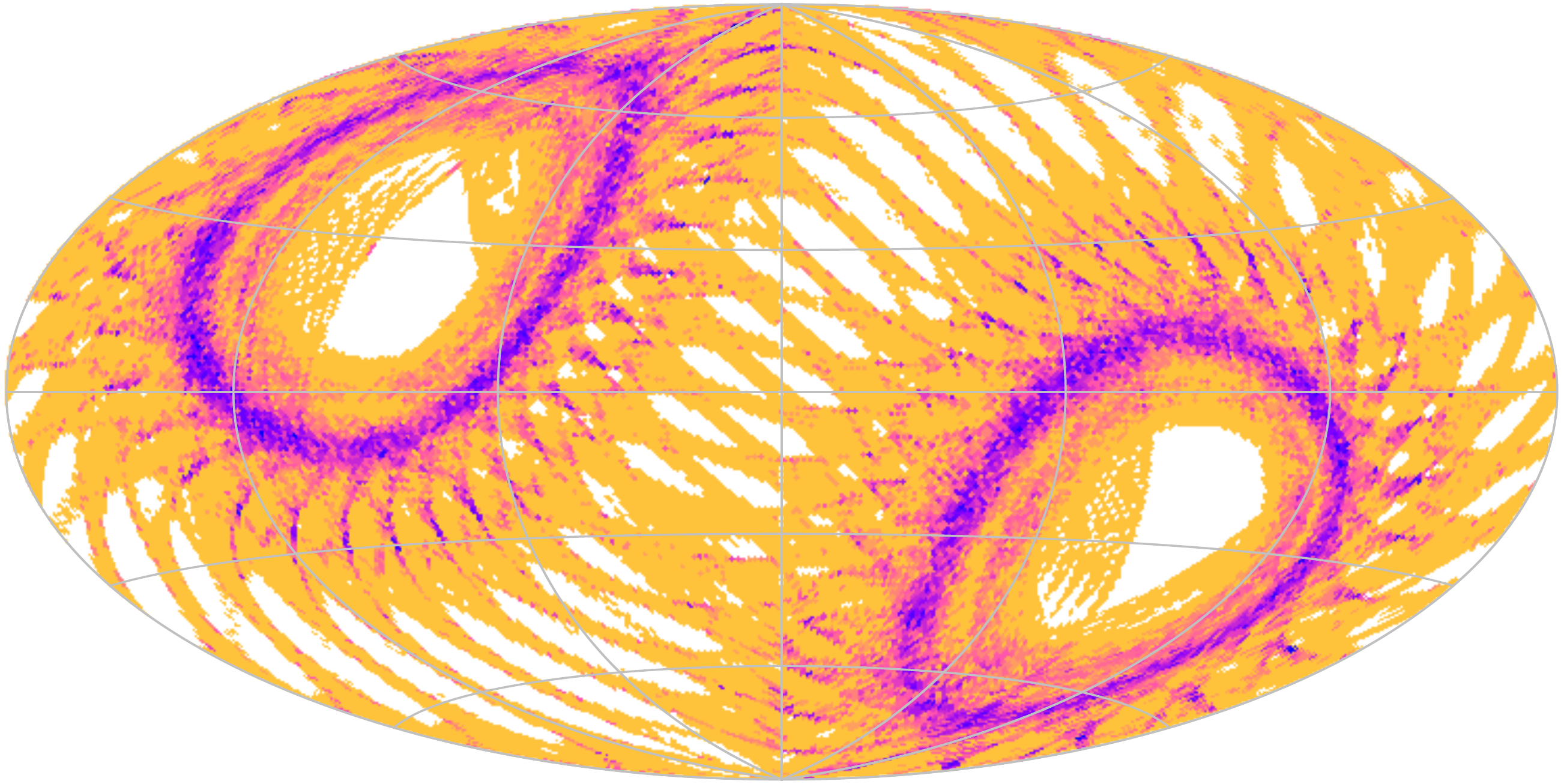}  & \includegraphics[width=0.3\textwidth]{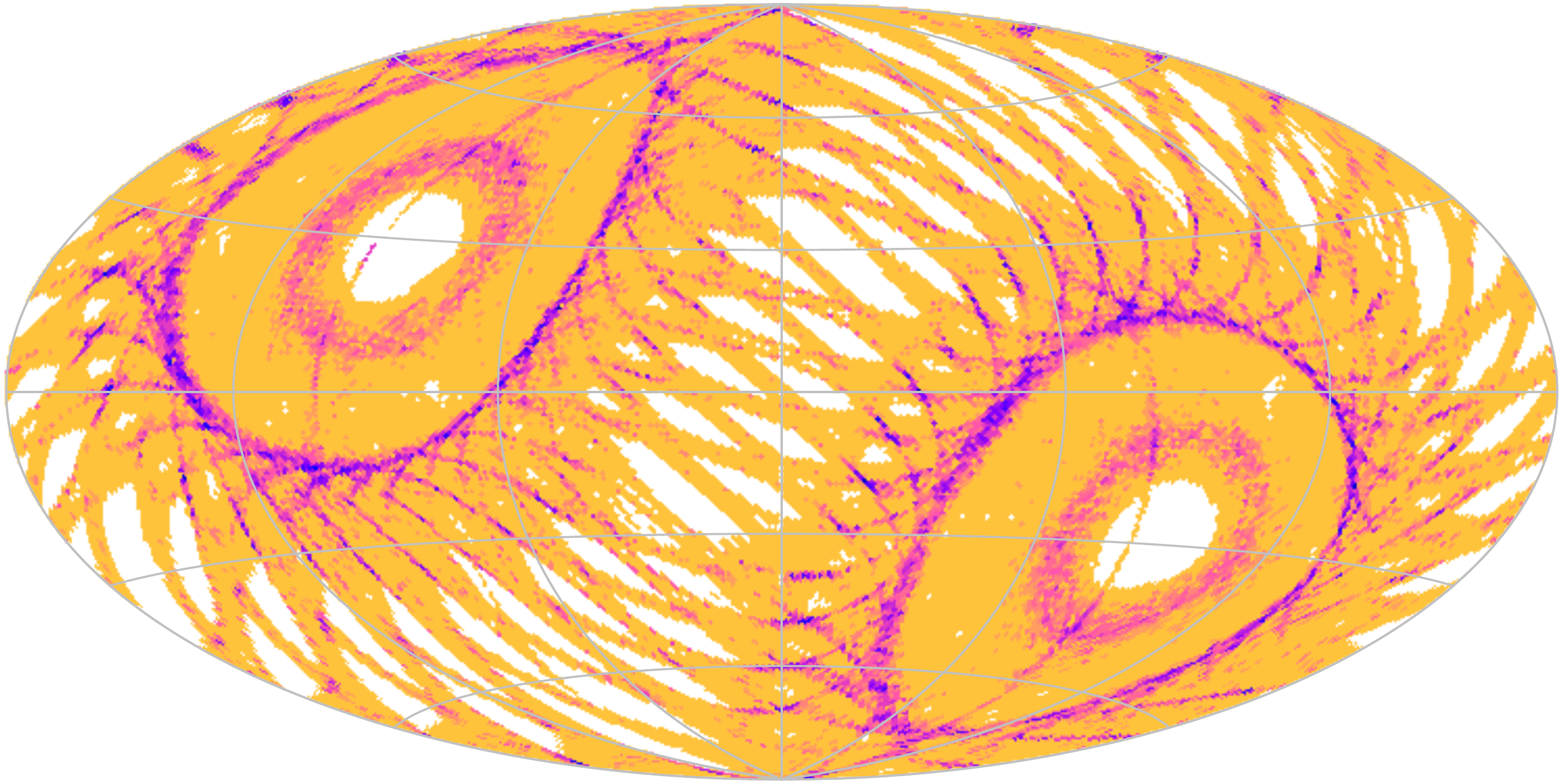}  &  \includegraphics[width=0.3\textwidth]{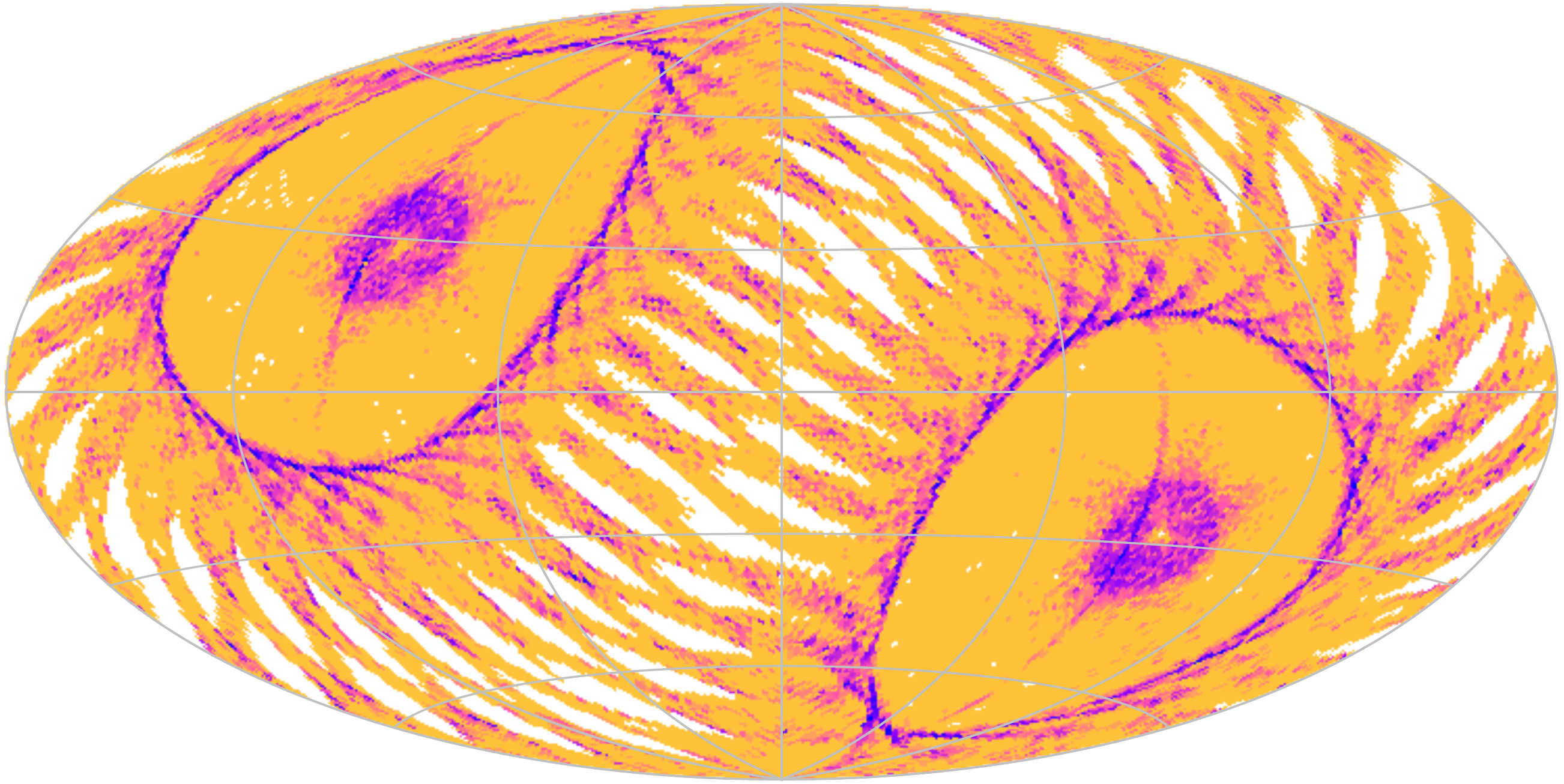}   \vspace{-3cm} \\
   76-80~d & 81-85~d &  86-90~d \vspace{2.4cm}\\[6pt]
  \includegraphics[width=0.3\textwidth]{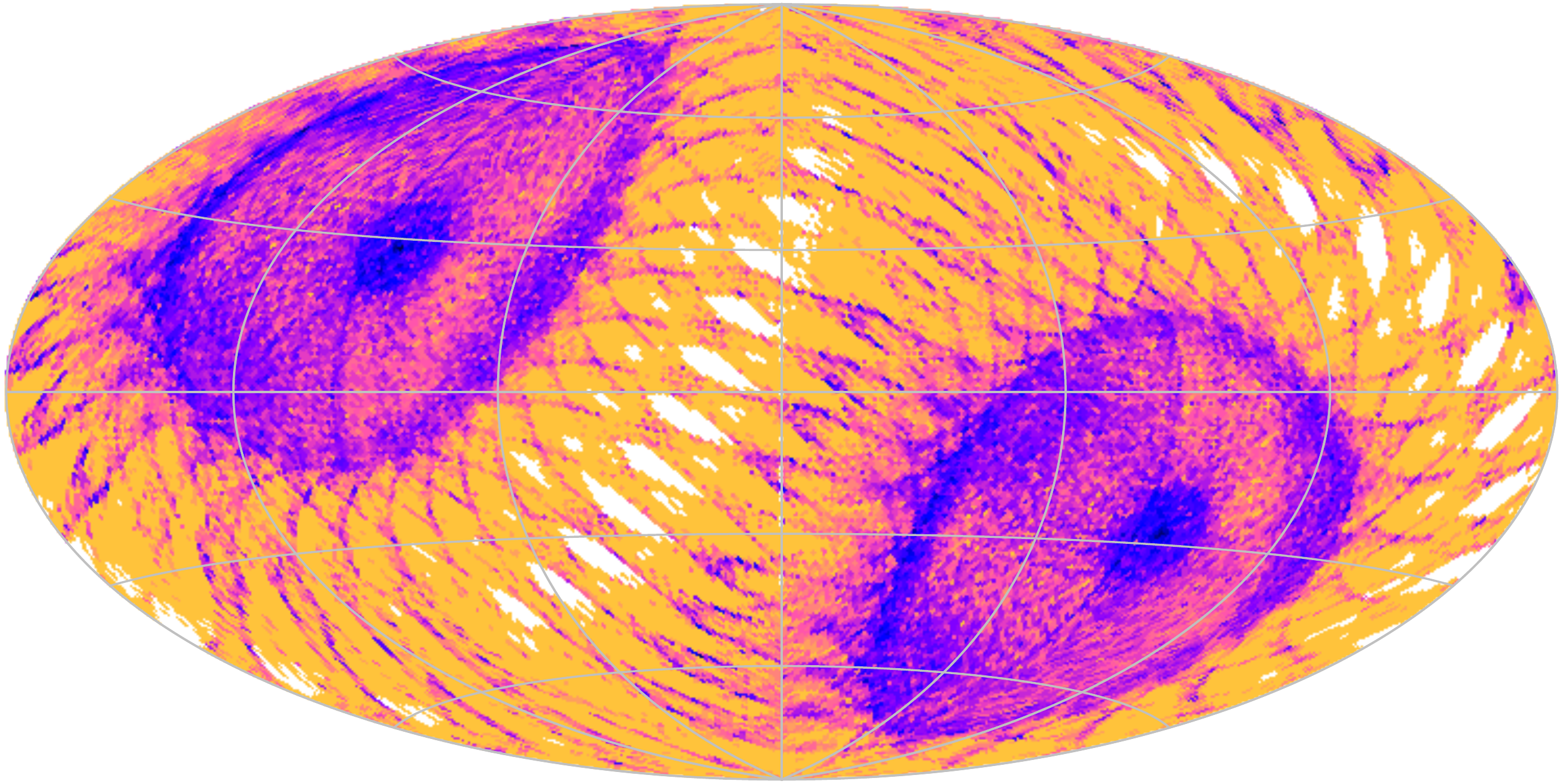}  & \includegraphics[width=0.3\textwidth]{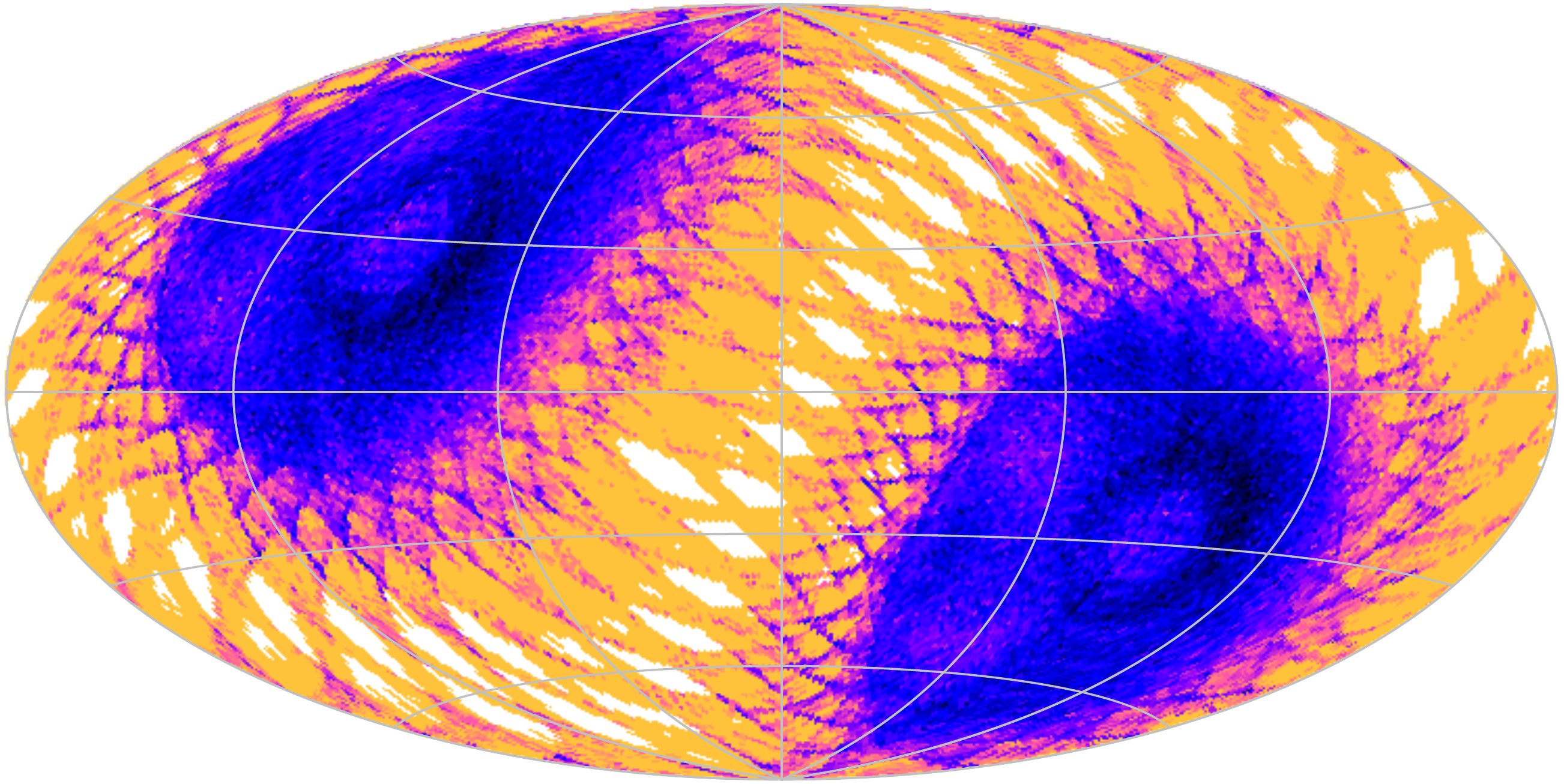} &  \includegraphics[width=0.3\textwidth]{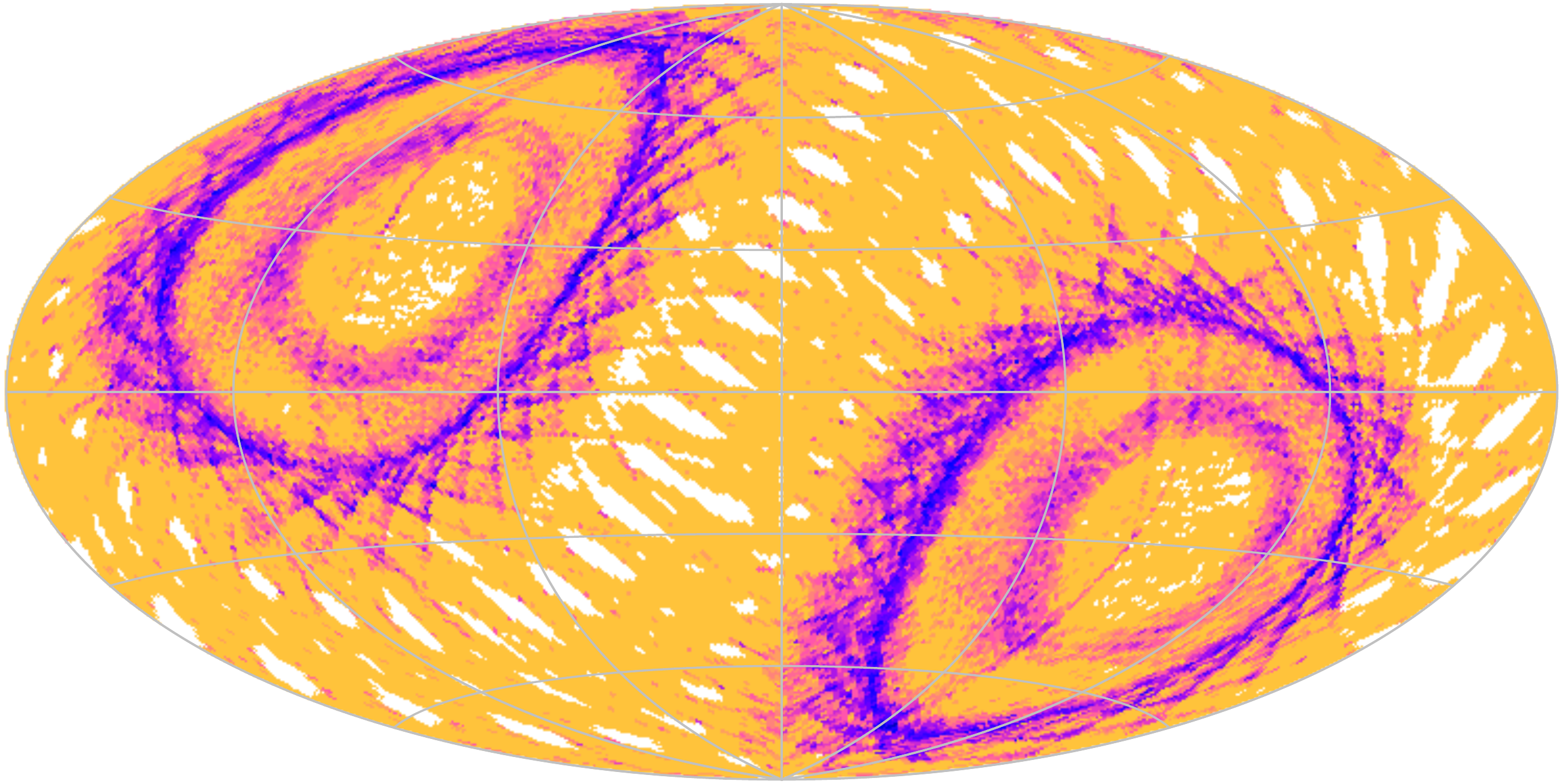} \vspace{-3cm} \\
91-95~d &  96-100~d & 101-105~d \vspace{2.4cm}\\[6pt]
\multicolumn{3}{l}{\includegraphics[width=0.95\textwidth]{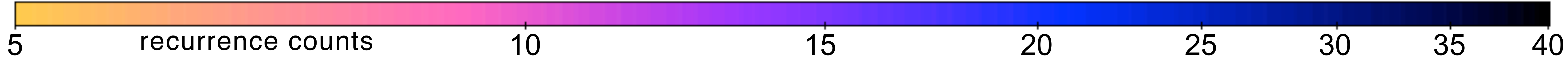} }  \vspace{-0.2cm}\\
\end{tabular}
\caption{Galactic Aitoff projection of the recurrence counts of various time intervals for simulated 5~year scanning law time series, computed following Fig.~\ref{fig:intervalExplanation}. 
Count are clipped between 5 and 40 for easy comparison. Some intervals resemble structures seen in the number of transits of Fig.~\ref{fig:galaxyNslBlend}. Clearly visible are the absence/presence of counts above or below ecliptic latitude $|\beta|=45\degr$ at certain intervals, which is caused by the GAREQ scanning law spin axis precession of 63d at 45$\degr$ around the sun. The symmetric linear structure close to the ecliptic poles (visible in intervals up to 25 days) is due to the 28d ecliptic pole scanning law. Most non-symmetric features are due to the 1~month transition scanning law. See Sect.~\ref{sec:Timesampling} for scanning law details. Galactic map origin $l=b=0$ is at the centre and $l$ increasing to the left. Plots show one sample per 0.84~deg$^2$. }
\label{fig:reoccurrenceTimeIntervals}
\end{figure*} 
\begin{figure*}[h]\ContinuedFloat
\begin{tabular}{@{}lll@{}}
\setlength{\tabcolsep}{0pt} 
\renewcommand{\arraystretch}{0} 
  \vspace{-0.5cm}\\
  \includegraphics[width=0.3\textwidth]{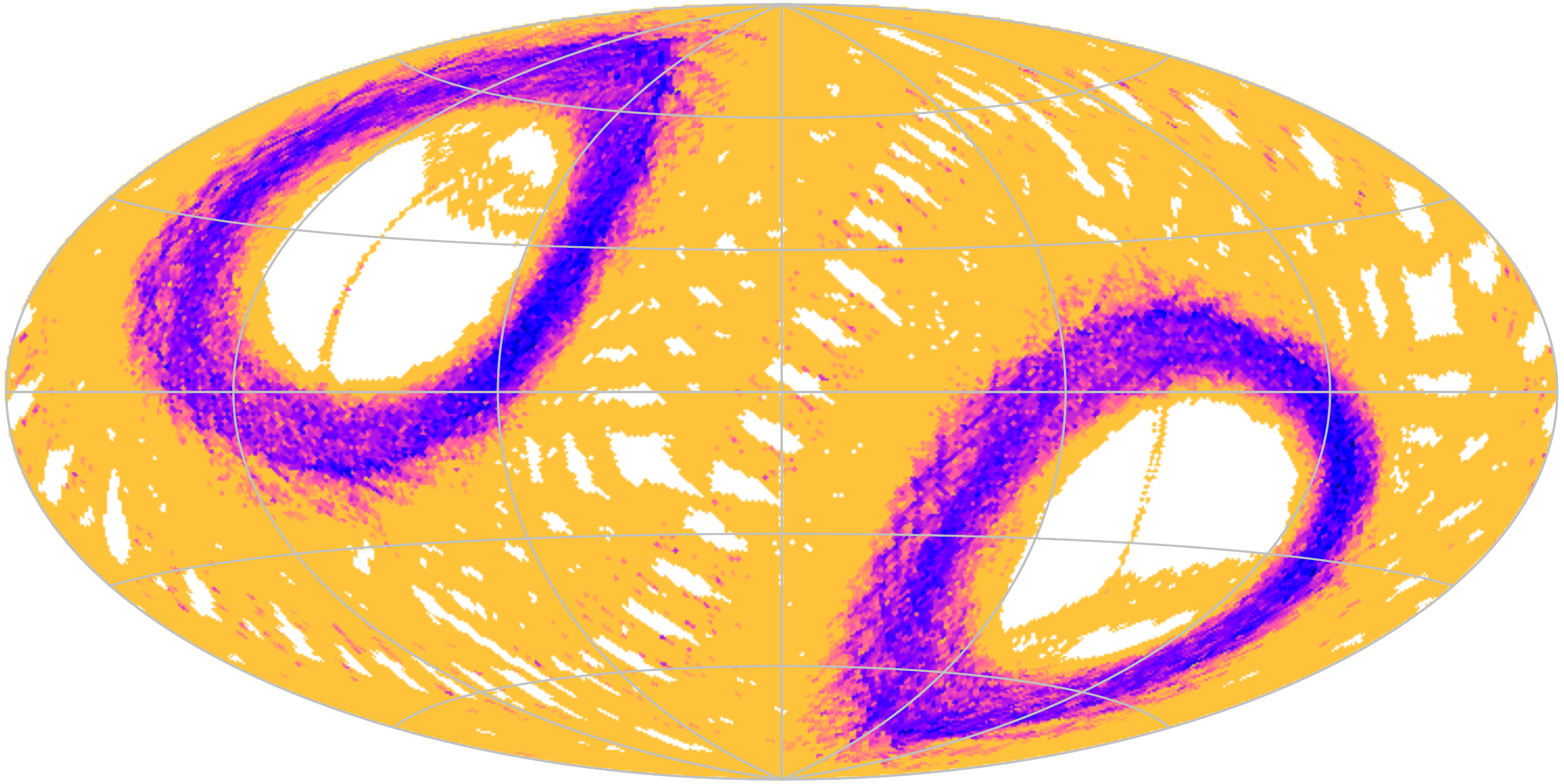} & \includegraphics[width=0.3\textwidth]{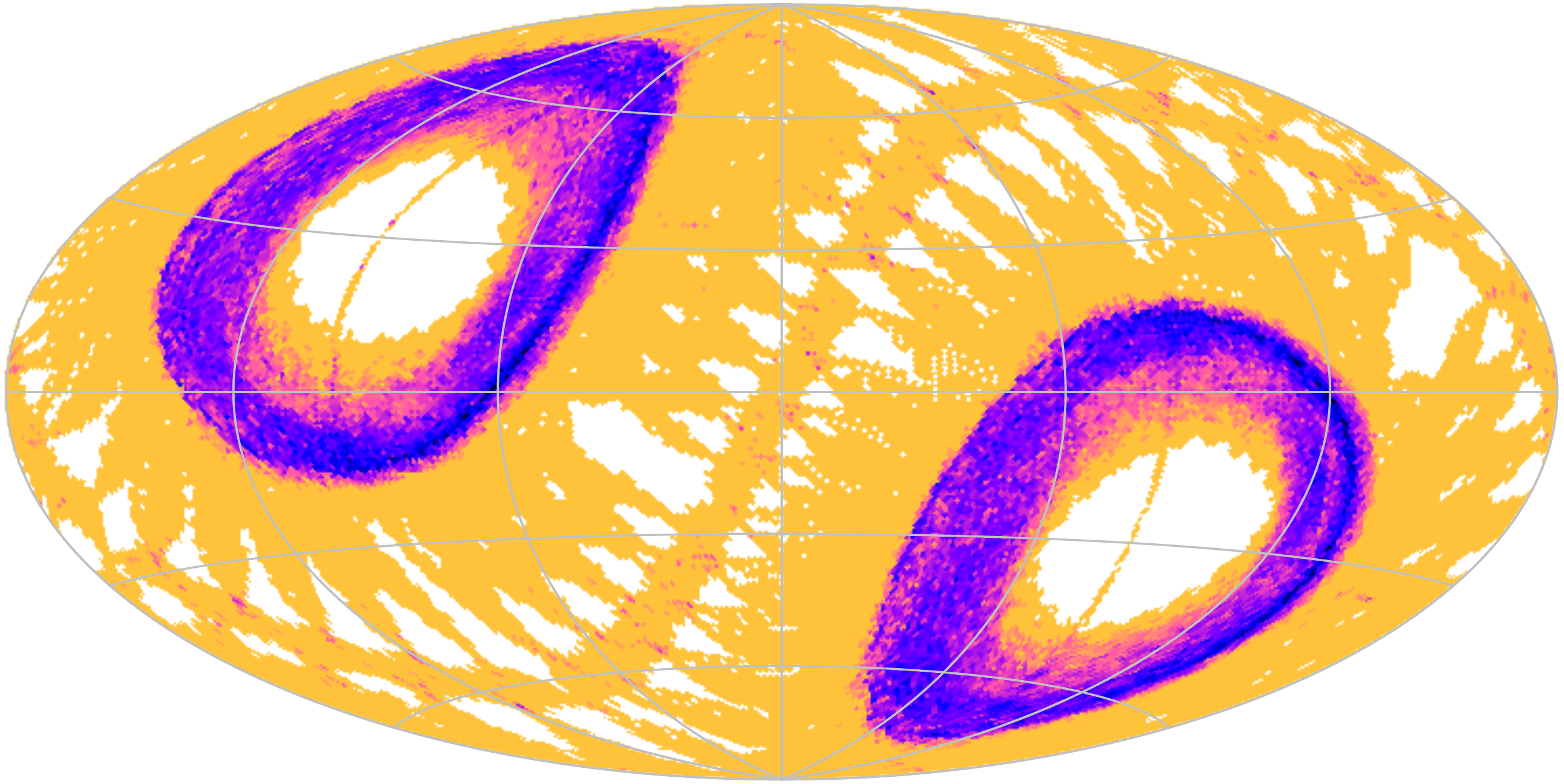} &  \includegraphics[width=0.3\textwidth]{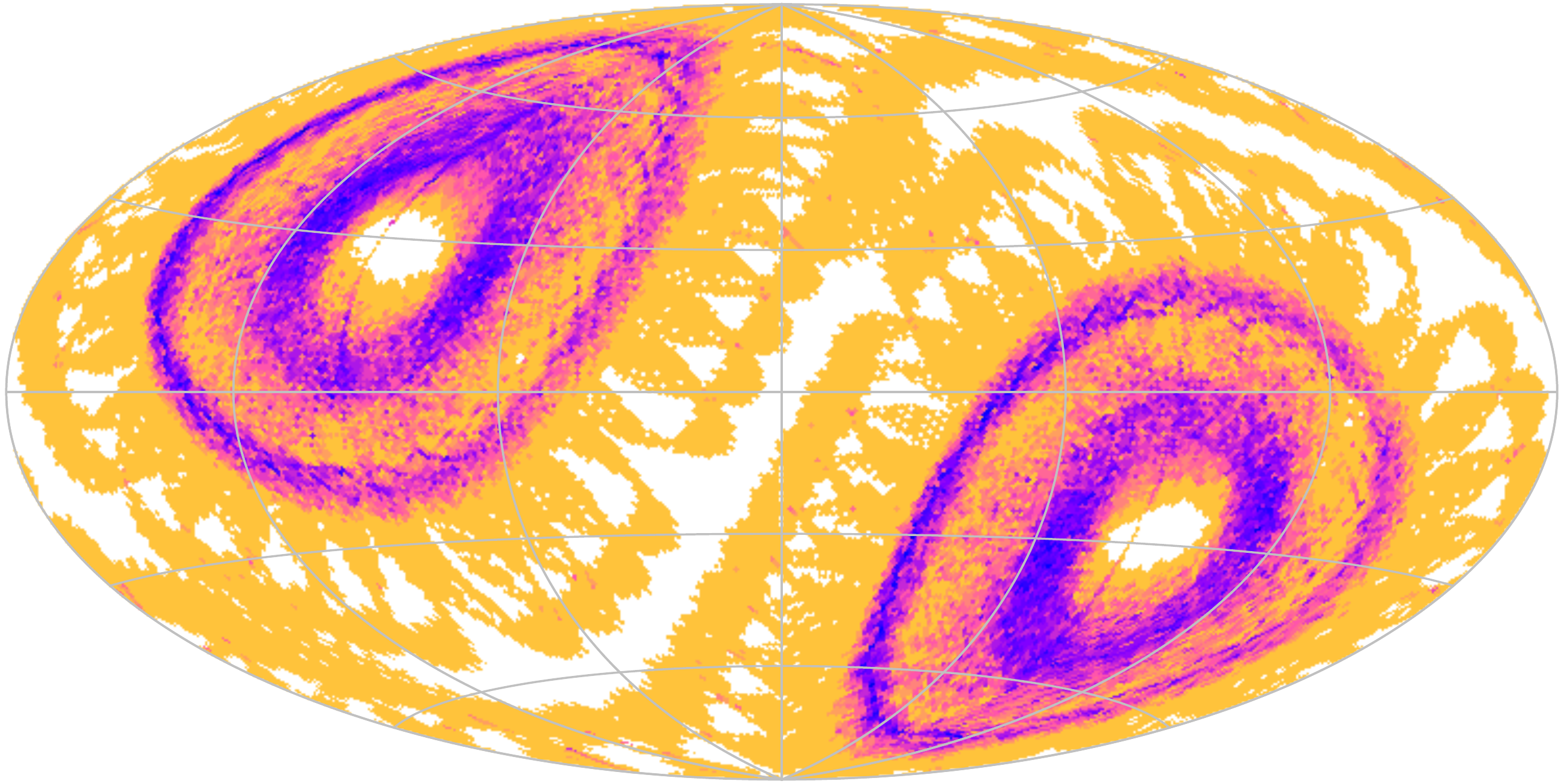}  \vspace{-3cm}\\
 106-110~d & 111-115~d &  116-120~d \vspace{2.4cm}\\[6pt]
  \includegraphics[width=0.3\textwidth]{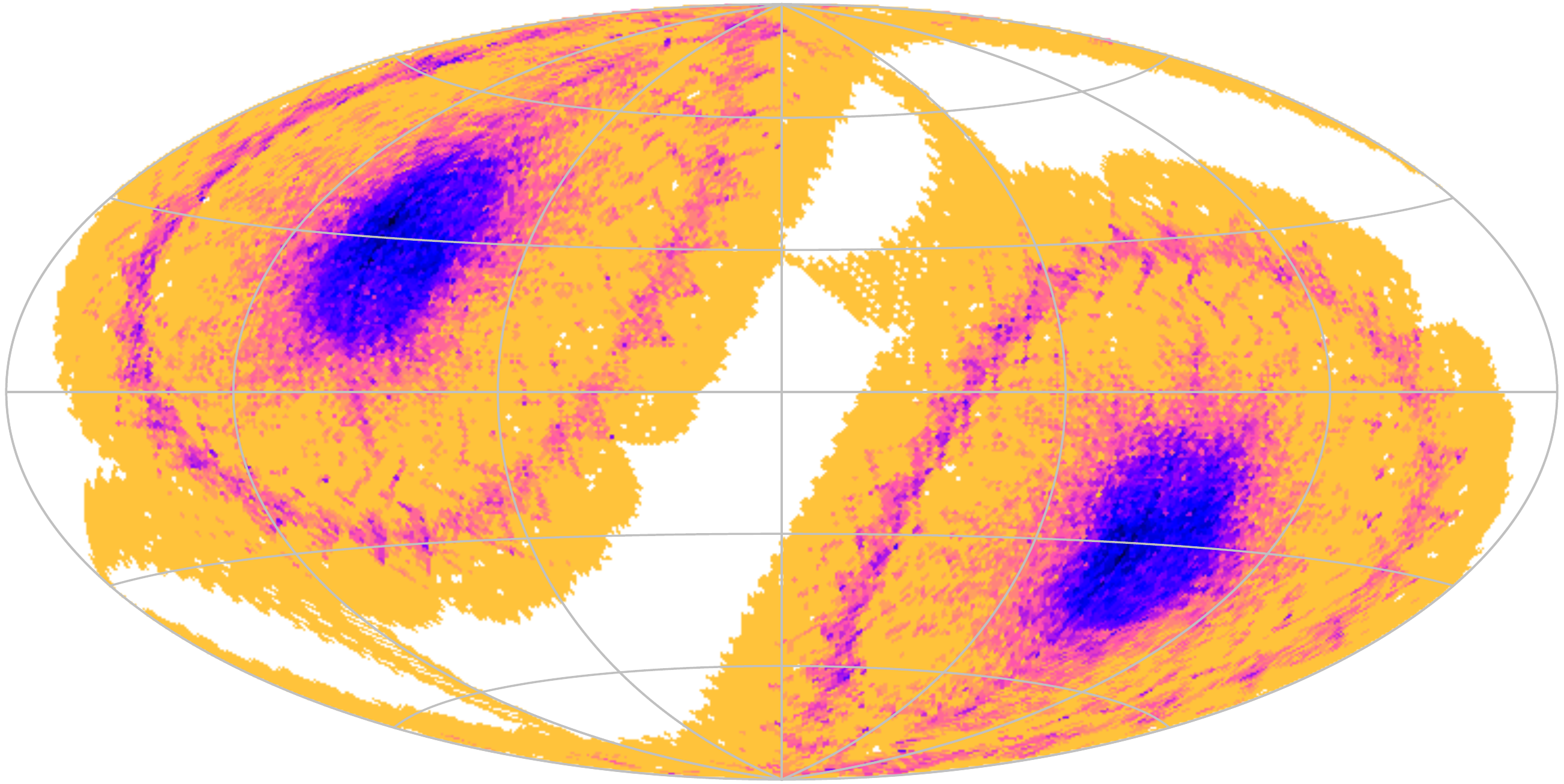} & \includegraphics[width=0.3\textwidth]{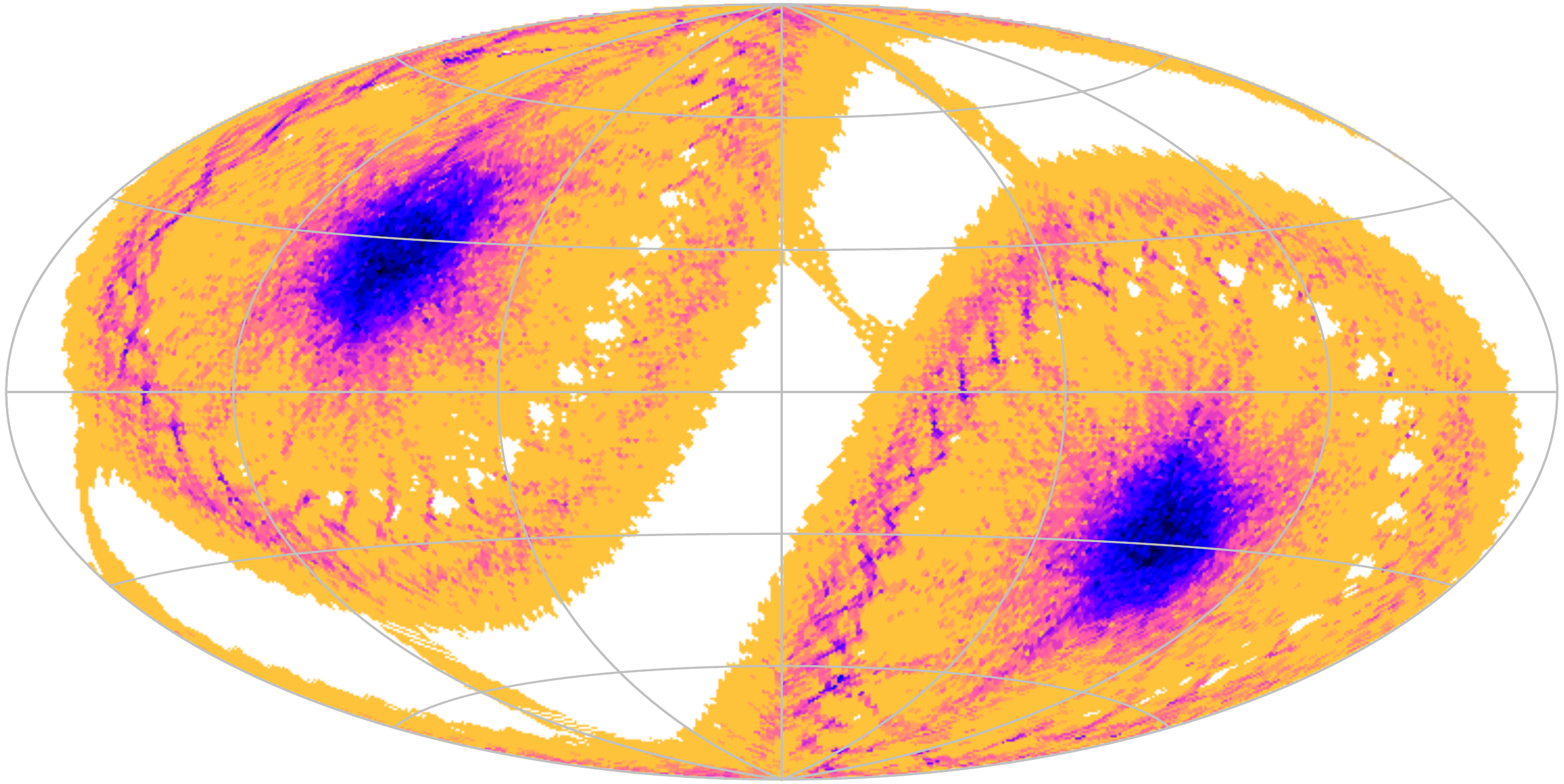} &  \includegraphics[width=0.3\textwidth]{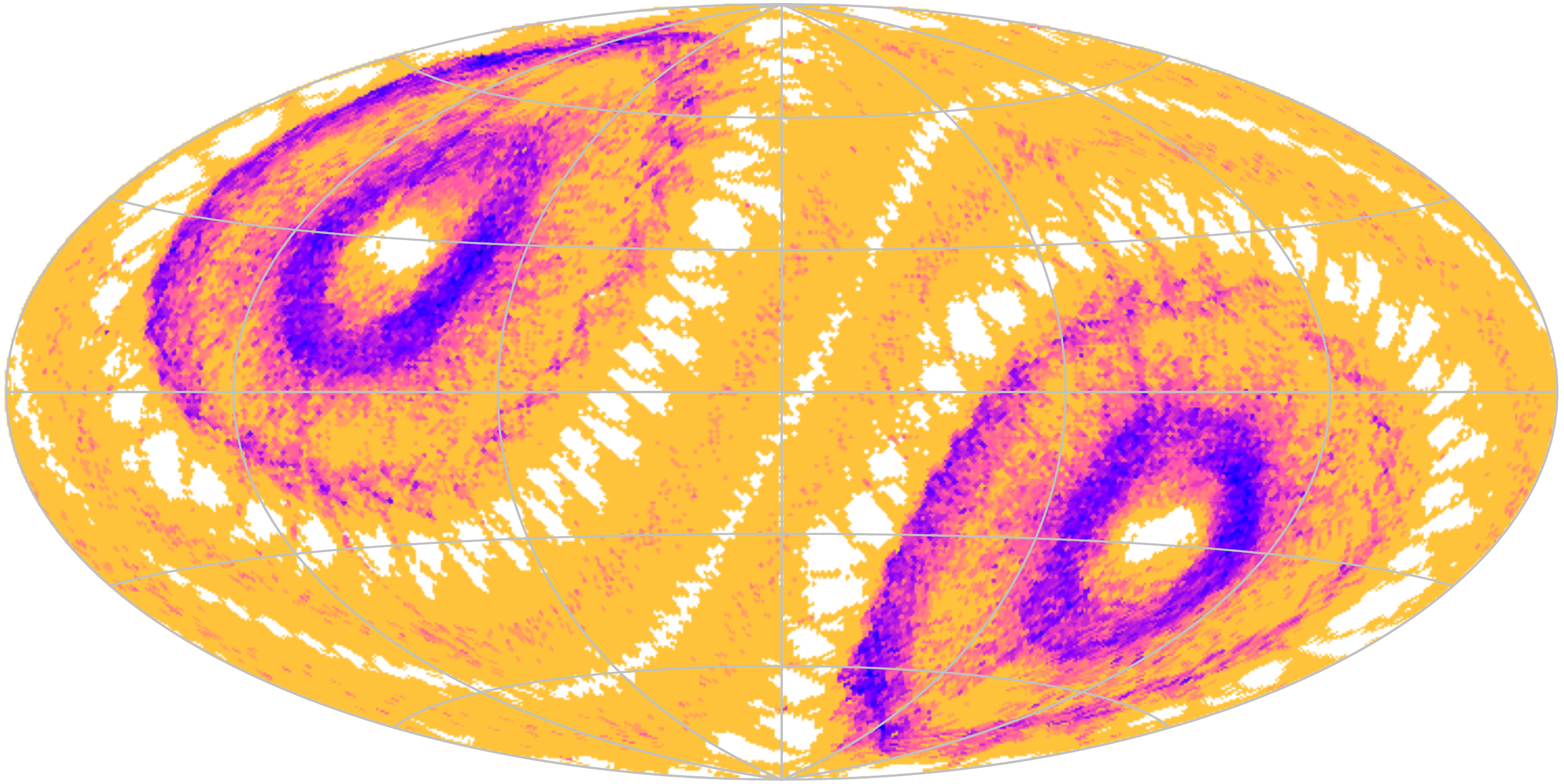}  \vspace{-3cm} \\
   121-125~d & 126-130~d &  131-135~d \vspace{2.4cm}\\[6pt]
\multicolumn{3}{l}{\includegraphics[width=0.95\textwidth]{FigA7ColourBar} }  \vspace{-0.2cm}\\
\end{tabular}
\caption{(continued from previous page).}
\label{fig:reoccurrenceTimeIntervals}
\end{figure*}

\section{Classification}
\label{Appendix:classification}
Distribution of attributes used in the automated supervised classification of the C0 data (not published) as described in Sect.~\ref{sec:C0clasAttributes}.

\begin{figure*}[h]
\centering
\includegraphics[angle=0,width=\textwidth]{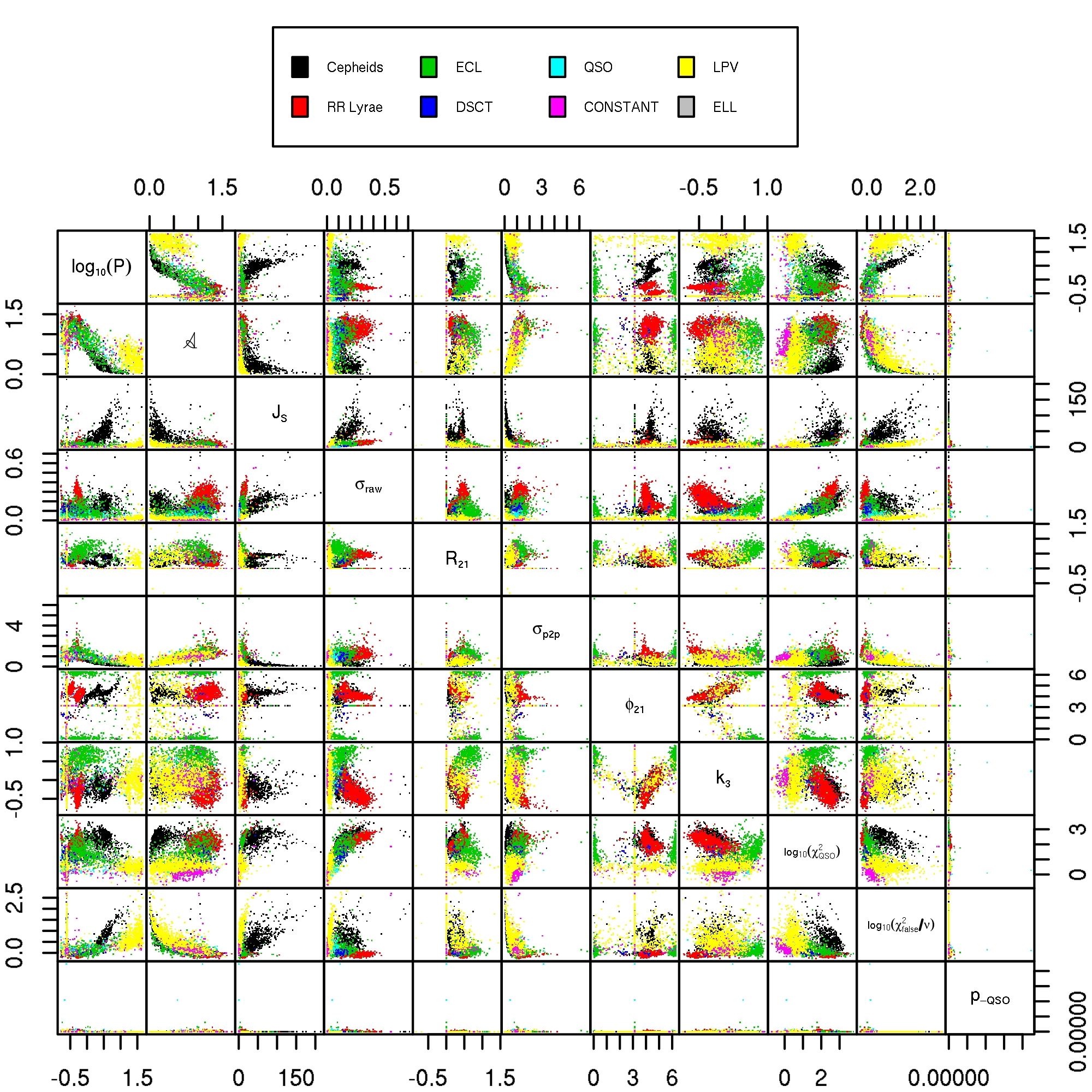}
\caption{The distribution of the attributes of the training set used for the Random Forest classifier. Periods are measured in days, amplitudes, scatters, and phase differences in radians. All other quantities are dimensionless. Note that the distribution of the colour indices ($G_{BP}-G_{RP}$) is omitted due to data publication restrictions. (C0 data)}
\label{fig:trainingsetRF}
\end{figure*}

\begin{figure*}
\centering
\includegraphics[angle=0,width=\textwidth]{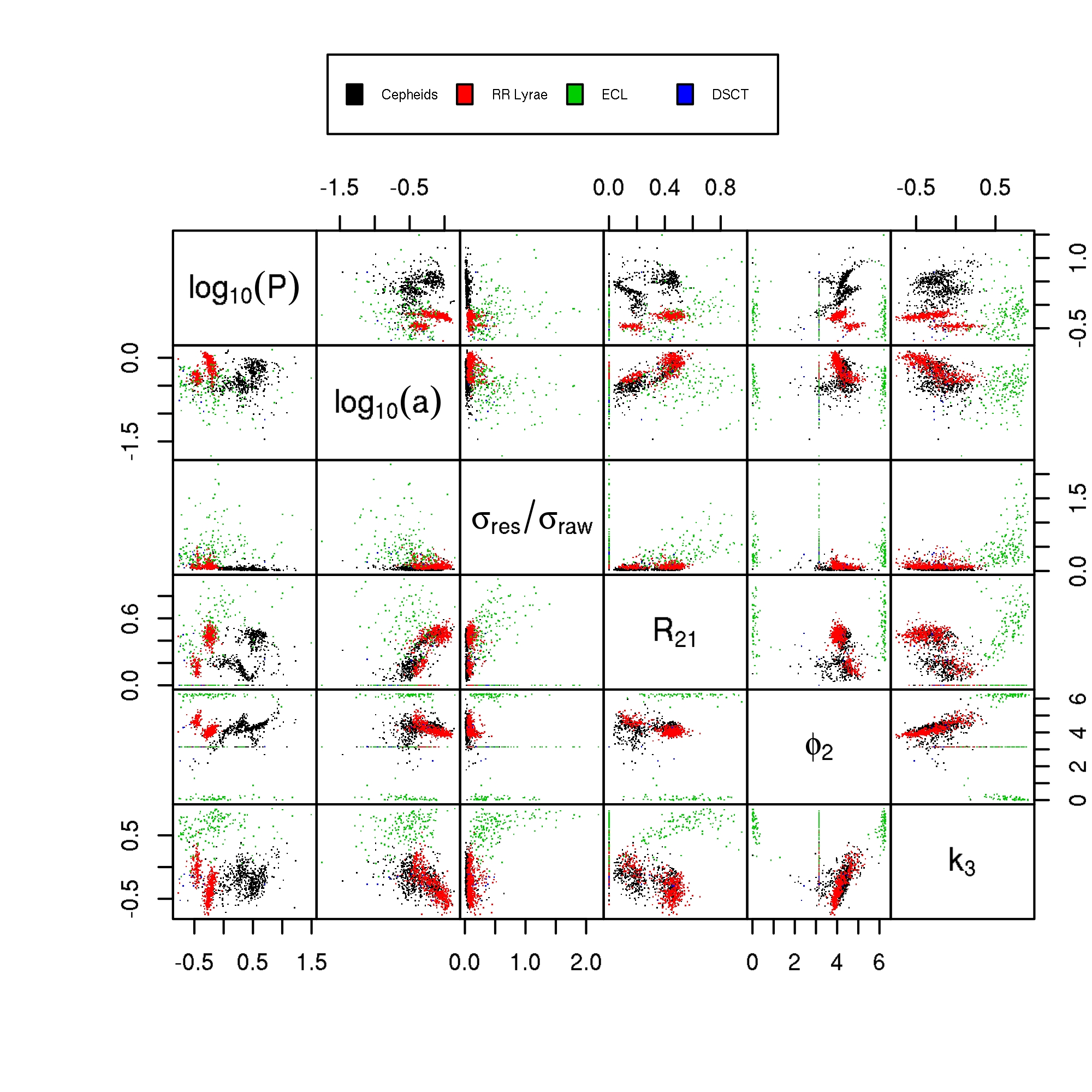}
\caption{The distribution of the attributes of the training set used for the Gaussian Mixture and Bayesian Network classifier. Periods are measured in days, amplitudes, and phase differences in radians. All other quantities are dimensionless. Note that the distribution of the colour indices ($G_{BP}-G_{RP}$) is omitted due to data publication restrictions. (C0 data)}
\label{fig:trainingsetGMBN}
\end{figure*}

\clearpage

\section{Additional PDF comparison}
\label{Appendix:GVS}
Distribution of attributes used in the automated supervised classification of the C0 data (not published) as described in Sect.~\ref{sec:C0clasAttributes}.

\begin{figure*}[h]
\centering
\includegraphics[width=\textwidth]{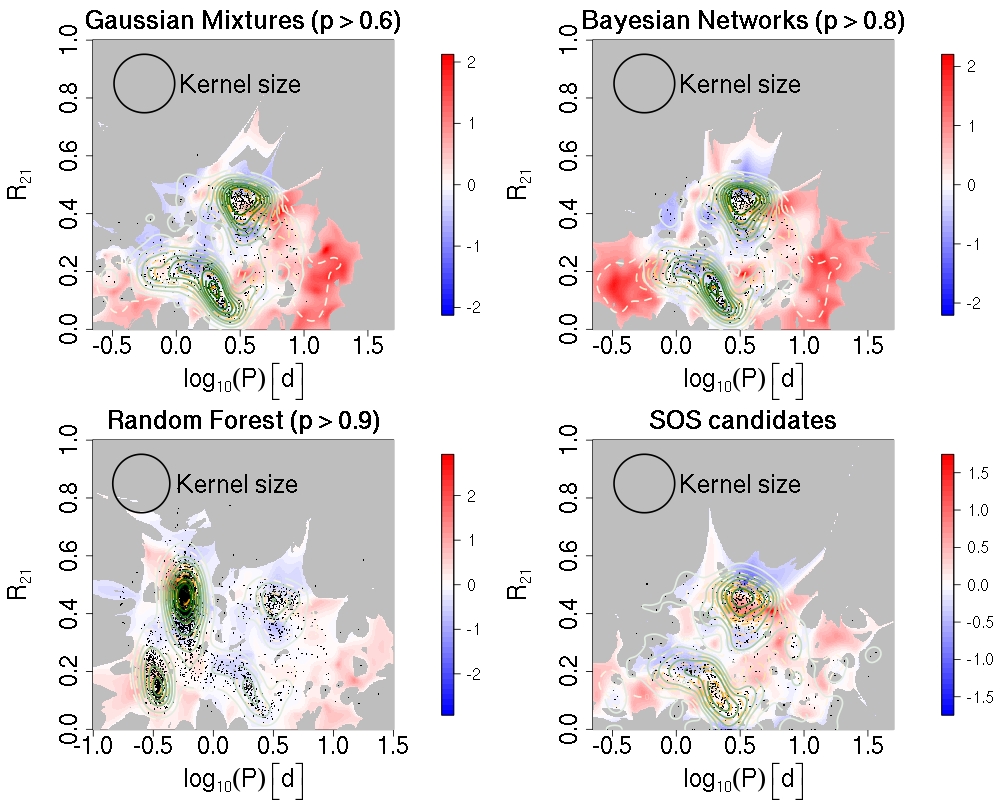}
\caption{Maps of the scaled differences between the kernel density
  estimates of the PDFs that describe the distribution of Cepheids in
  the $\log(P)-R_{21}$ space. The sub-panels correspond to the Gaussian
  Mixture classifier with a membership probability threshold $p > 0.6$
  (left); Boosted Bayesian Networks with $p > 0.8$ (centre left); and
  the Random Forest classifier with $p > 0.9$ for the PDF of Cepheids
  and RR~Lyrae stars combined (centre-right, see Sect.~\ref{DataProcessing:Classification});
  and the final SOS list (right). (C0 data)}
\label{fig:pdfsubs-ceps-r21}
\end{figure*}

\begin{figure*}
\centering
\includegraphics[width=\textwidth]{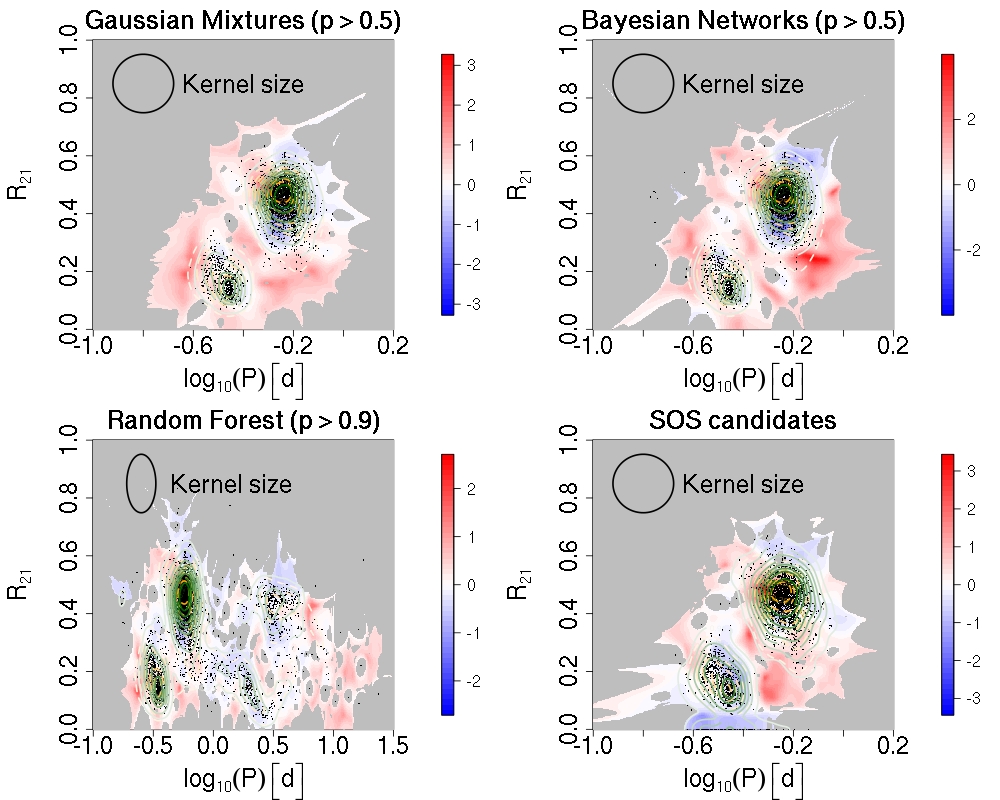}
\caption{Maps of the scaled differences between the kernel density
  estimates of the PDFs that describe the distribution of RR~Lyrae stars in
  the $\log(P)-R_{21}$ space. The sub-panels correspond to the Gaussian
  Mixture classifier with a membership probability threshold $p > 0.5$
  (left); Boosted Bayesian Networks with $p > 0.5$ (centre left); and
  the Random Forest classifier with $p > 0.9$ for the PDF of Cepheids
  and RR~Lyrae stars combined (centre-right, see Sect.~\ref{DataProcessing:Classification});
  and the final SOS list (right). (C0 data)}
\label{fig:pdfsubs-rr-r21}
\end{figure*}

\begin{figure*}
\centering
\includegraphics[width=\textwidth]{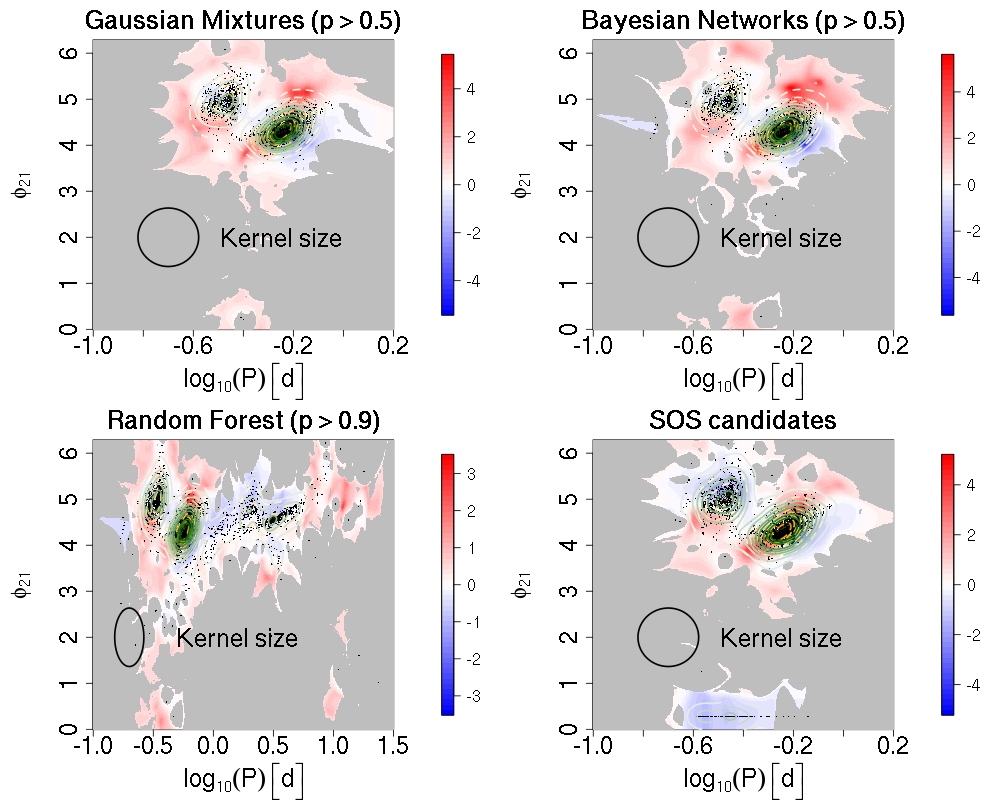}
\caption{Maps of the scaled differences between the kernel density
  estimates of the PDFs that describe the distribution of RR~Lyrae stars in
  the $\log(P)-\phi_{21}$ space. The sub-panels correspond to the
  Gaussian Mixture classifier with a membership probability threshold
  $p > 0.5$ (left); Boosted Bayesian Networks with $p > 0.5$ (centre
  left); and the Random Forest classifier with $p > 0.9$ for the PDF
  of Cepheids and RR~Lyrae stars combined (centre-right, see Sect.~\ref{DataProcessing:Classification});
  and the final SOS list (right). (C0 data)}
\label{fig:pdfsubs-rr-phi21}
\end{figure*}

\begin{figure*}
\centering
\includegraphics[width=\textwidth]{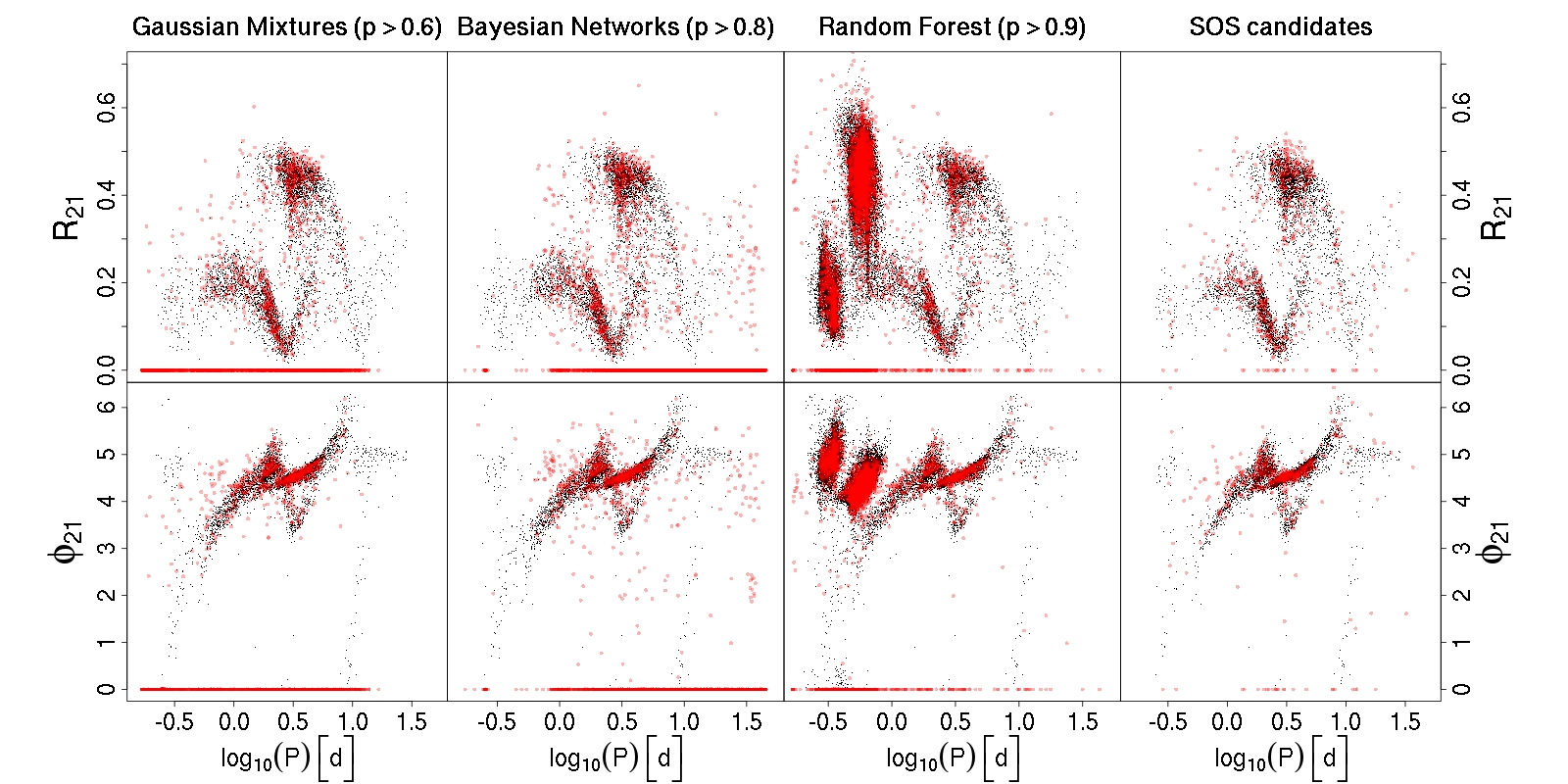}
\caption{Scatter plot of the OGLE IV catalogue of Cepheids (classical,
  type-II and anomalous) and the various {\it Gaia} samples in the
  $\log(P)-R_{21}$ (top) and $\log(P)-\phi_{21}$ (bottom) spaces.
  Black dots correspond to the OGLE reference catalogue and transparent
  red circles represent the different {\it Gaia} samples. The sub-panels
  correspond, from left to right, to the Gaussian Mixture classifier
  with a membership probability threshold $p > 0.6$ (left); Boosted
  Bayesian Networks with $p > 0.8$ (centre left); the Random Forest
  classifier with $p > 0.9$ for the PDF of Cepheids and RR~Lyrae stars
  combined (centre-right, see Sect.~\ref{DataProcessing:Classification});
  and the final SOS candidates (right). (C0 data)}
\label{fig:gvs-scatter-cep}
\end{figure*}

\begin{figure*}
\centering
\includegraphics[width=\textwidth]{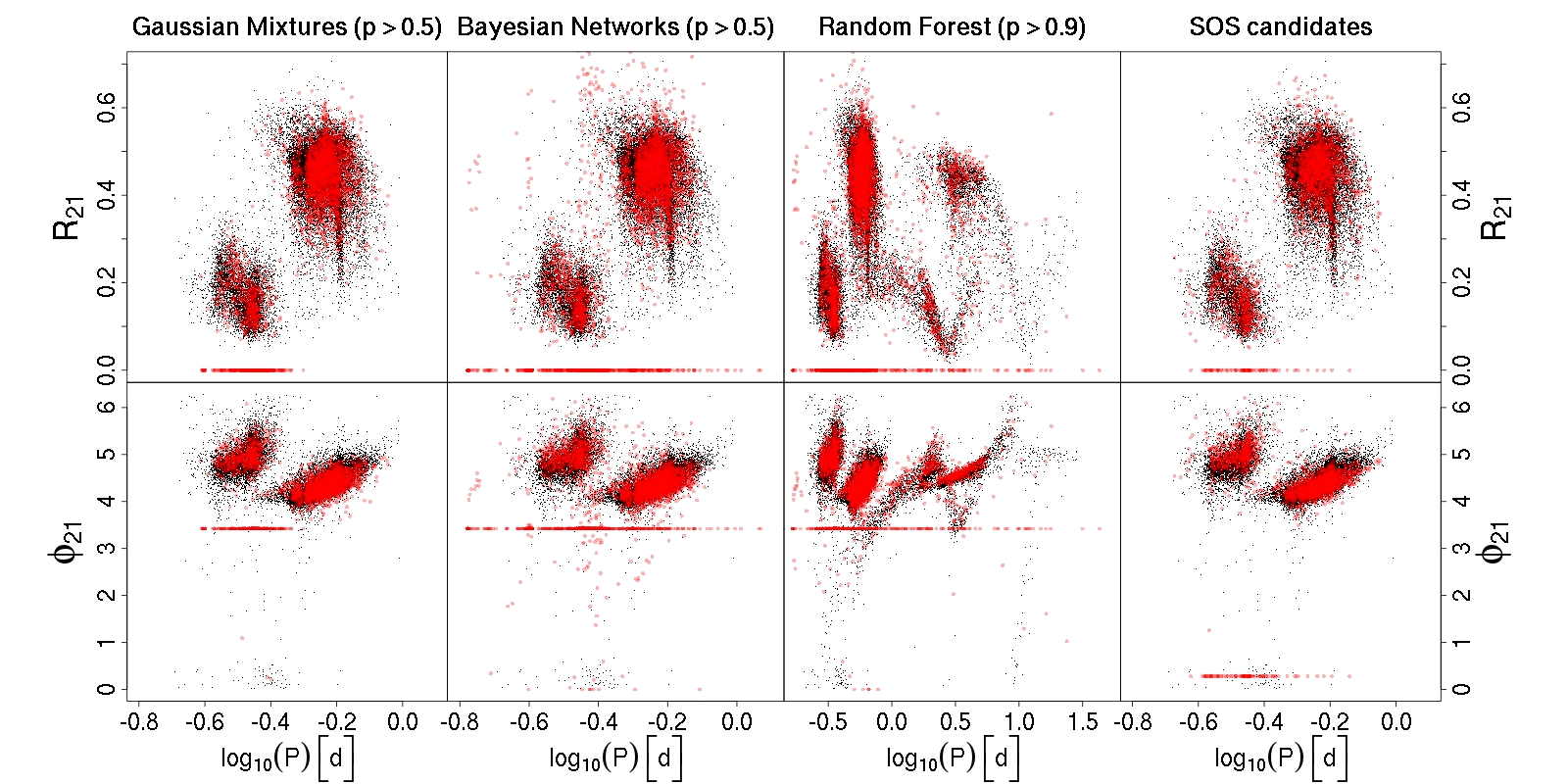}
\caption{Scatter plot of the OGLE IV catalogue of RR~Lyrae stars (fundamental
  and first overtone only) and the various {\it Gaia} samples in the
  $\log(P)-R_{21}$ (top) and $\log(P)-\phi_{21}$ (bottom) spaces.
  Black dots correspond to the OGLE reference catalogue and transparent
  red circles represent the different {\it Gaia} samples. The sub-panels
  correspond, from left to right, to the Gaussian Mixture classifier
  with a membership probability threshold $p > 0.5$ (left); Boosted
  Bayesian Networks with $p > 0.5$ (centre left); the Random Forest
  classifier with $p > 0.9$ for the PDF of Cepheids and RR~Lyrae stars
  combined (centre-right, see Sect.~\ref{DataProcessing:Classification});
  and the final SOS candidates (right). (C0 data)}
\label{fig:gvs-scatter-rr}
\end{figure*}

\clearpage

\section{Data Release 1 Variability catalogue (C1)} \label{sec:Gdr1varCatalog}
The properties of the 3194 Cepheids and RR~Lyrae stars, together with their time series and associated statistics, are published in the first {\it Gaia} data release, which is available through \url{http://archives.esac.esa.int/gaia/} and other distribution nodes. Note that only a subset of the information presented in this paper is available in the released data, i.e., all of the C0 results described in Sect.~\ref{sec:C0sepResults} (C0 statistics, variability information for all sources, and classification attributes and results) as well as  the Characterisation periods and models on C1 data (only the period and model from the SOS Cep\&RRL pipeline are published) are excluded.

Herein, we present a condensed overview of the \textit{tables} related to the variability information in this first data release. In the published data tables, each field has an extensive description which is not be repeated here. We also do not list the unique {\it Gaia} source identifier \texttt{source\_id} and \texttt{solution\_id} present in each table.

Table \texttt{gaia\_source} contains a summary of all 1.2~billion sources published in {\it Gaia} DR1, with the field
\texttt{phot\_variable\_flag} set to \texttt{VARIABLE} for the 3194 published sources
 and to \texttt{NOT\_AVAILABLE} in all other cases.
 
 Table \texttt{variable\_summary} contains a summary of each of the 3194 published variable sources, with fields
\texttt{classification} set to \texttt{RRLYR} or \texttt{CEP}, and \texttt{phot\_variable\_fundam\_freq1} 
being the pulsation frequency transformed from the period field of the \texttt{cepheid} or \texttt{rrlyrae} tables mentioned below.

Table \texttt{phot\_variable\_time\_series\_gfov} contains the \textit{G}-band FoV averaged photometry for each of the 3194 published sources, with fields \texttt{g\_flux} and \texttt{g\_flux\_error} for the flux and associated error in the C1 dataset provided by CU5, \texttt{g\_magnitude} which is transformed from flux using the zero point in table \texttt{ext\_phot\_zero\_point} (also described in Sect.~\ref{sec:time_and_conversions}), \texttt{observation\_time}, whose  definition is discussed in Sect.~\ref{sec:time_and_conversions}, and the flag \texttt{rejected\_by\_variability\_processing} indicating which observations are flagged as outliers in our analysis (see Sect.~\ref{sec:obsFiltering}).

Table \texttt{phot\_variable\_time\_series\_gfov\_statistical\_parameters} contains various statistics computed on the time series available in table \texttt{phot\_variable\_time\_series\_gfov}, excluding all observations for which the \texttt{rejected\_by\_variability\_processing} flag is true. The available fields are: 
\texttt{num\_observations\_processed} (the number of non-rejected observations processed by CU7), 
\texttt{maximum}, 
\texttt{minimum},
\texttt{mean}, 
\texttt{median}, 
\texttt{median\_absolute\_deviation}, 
\texttt{std\_dev},
\texttt{skewness}, 
\texttt{kurtosis}, 
\texttt{abbe}, 
\texttt{iqr}, 
\texttt{range}, 
\texttt{mean\_obs\_time},  
\texttt{time\_duration}. 

Table \texttt{cepheid} contains the SOS Cep\&RRL results of the 599 published Cepheid stars. These fields are discussed in detail in \cite{DPACP-13}, and we only list the available fields and the content of the text fields:
\texttt{type\_best\_classification} (one of \texttt{DCEP}, \texttt{T2CEP} or \texttt{ACEP}),
\texttt{type2\_best\_sub\_classification} (in case of type \texttt{T2CEP}, it specifies 
\texttt{BL\_HER}, \texttt{W\_VIR} or \texttt{RV\_TAU}), 
\texttt{mode\_best\_classification} (in case of types  \texttt{DCEP} or \texttt{ACEP}, it can be 
\texttt{FUNDAMENTAL}, \texttt{FIRST\_OVERTONE}, \texttt{SECOND\_OVERTONE} or \texttt{UNDEFINED}, while for \texttt{T2CEP} the value is set to \texttt{NOT\_APPLICABLE}), 
\texttt{p1},
\texttt{p1\_error},
\texttt{num\_harmonics\_for\_p1},
\texttt{int\_average\_g},
\texttt{int\_average\_g\_error},
\texttt{peak\_to\_peak\_g},
\texttt{peak\_to\_peak\_g\_error},
\texttt{epoch\_g},
\texttt{epoch\_g\_error},
\texttt{phi21\_g},
\texttt{phi21\_g\_error},
\texttt{r21\_g}, and
\texttt{r21\_g\_error}.

Table \texttt{rrlyrae} contains the SOS Cep\&RRL results of the  2595 published RR~Lyrae stars. These fields are discussed in detail in \cite{DPACP-13}, and we only list the available fields and the content of the text fields:
\texttt{best\_classification} (one of \texttt{RRC} or \texttt{RRAB}), 
\texttt{p1},
\texttt{p1\_error},
\texttt{num\_harmonics\_for\_p1},
\texttt{int\_average\_g},
\texttt{int\_average\_g\_error},
\texttt{peak\_to\_peak\_g},
\texttt{peak\_to\_peak\_g\_error},
\texttt{epoch\_g},
\texttt{epoch\_g\_error},
\texttt{phi21\_g},
\texttt{phi21\_g\_error},
\texttt{r21\_g}, and
\texttt{r21\_g\_error}.

\section{Acronyms}
\label{Appendix:acronyms}
The following table contains the acronyms and their definition, and where appropriate, according to the online \textit{Gaia} acronym list.

\begin{longtable}{ll}
\caption{\label{tab:acronyms}Acronyms used in this paper.}\\
\hline\hline
Acronym & Description\\
\hline
\endfirsthead
\multicolumn{2}{c}%
{\tablename\ \thetable\ -- \textit{Continued from previous page.}}\\
\hline\hline
Acronym & Description\\
\hline
\endhead
\hline \multicolumn{2}{r}{\textit{Continued on next page.}}\\
\endfoot
\hline
\endlastfoot
AIC 	& Akaike Information Criterion \\
ASAS	& All Sky Automated Survey \\
BIC		& Bayes Information Criterion \\
BLS		& Box Least Squares \\
BN 		& Bayesian Network(s) \\
CCD   	& Charge-Coupled Device (detector) \\
C0 		& Cycle 0 photometry data\\
C1 		& Cycle 1 photometry data\\
CORAVEL	& CORrelation RAdial VELocities \\
CoRoT	& COnvection ROtation and planetary Transits\\
CU 		& Coordination Unit (in DPAC)\\
CU3		& Core Processing Coordination Unit\\
CU4		& Object Processing Coordination Unit\\
CU5		& Photometric Processing Coordination Unit \\
CU6		& Spectroscopic Processing Coordination Unit \\
CU7 	& Variability Processing and Analysis Coordination Unit\\
CU8		& Astrophysical Parameters Coordination Unit\\
CU9		& Archive and Catalog Coordination Unit\\
DFU		& Directed Follow-Up \\
DPAC 	& Data Processing and Analysis Consortium \\
DPCE	& Data Processing Centre ESAC \\
DPCG 	& Data Processing Centre of Geneva \\
DR1 	& \textit{Gaia} Data Release 1 (September 2016) \\
DRC 	& Data Reduction Cycle \\
EPSL 	& Ecliptic Pole Scanning Law \\
EROS	& Expérience pour la Recherche d'Objets Sombres \\
ESA 	& European Space Agency \\
ESAC 	& European Space Astronomy Centre (VilSpa) \\
FAP 	& False Alarm Probability \\
FFoV	& Following Field of View \\
FoV 	& Field of View \\
GAREQ 	& \textit{GAia} Relativistic Experiment on Quadrupole light deflection \\
GATE	& \textit{GAia} Transiting Exoplanets \\
GM 		& Gaussian Mixture(s) \\
GVD 	& General Variability Detection \\
GSEP	& \textit{Gaia} South Ecliptic Pole \\
HAT		& Hungarian Automated Telescope \\
IQR 	& Inter-Quartile Range \\
KL 		& Kullback-Leibler \\
LMC 	& Large Magellanic Cloud \\
LSST 	& Large Synoptic Survey Telescope \\
MACHO	& MAssive Compact Halo Objects \\
MDB 	& \textit{Gaia} Main DataBase \\
NSL 	& Nominal Scanning Law \\
OGLE	& Optical Gravitational Lensing Experiment \\
Pan-STARRS & Panoramic Survey Telescope \& Rapid Response System \\
PDF 	& Probability Density Function \\
PFoV	& Preceding Field of View \\
QSO		& Quasi-Stellar Object \\
RF 		& Random Forest(s) \\
RV 		& Radial Velocity \\
RVS 	& Radial Velocity Spectrometer \\
SDSS	& Sloan Digital Sky Survey \\
SEP 	& South Ecliptic Pole \\
SMC 	& Small Magellanic Cloud \\
SOS 	& Specific Object Studies \\
SVD 	& Special Variability Detection \\
TCB 	& Temps-Coordonnée Barycentrique (Barycentric Coordinate Time) \\
TESS	& Transiting Exoplanet Survey Satellite \\
\end{longtable}

\end{appendix}

\end{document}